\documentclass[12pt,preprint]{aastex}

\shorttitle{Substructure around M31}
\shortauthors{Gauthier, Dubinski \& Widrow}

\begin{document}

\title{Substructure around M31 : Evolution and Effects}
\author{Jean-Ren\'e Gauthier,\altaffilmark{1,3} John Dubinski\altaffilmark{1} and Lawrence M. Widrow\altaffilmark{2}}
\altaffiltext{1}{Department of Astronomy \& Astrophysics, University of Toronto, 60 St. George Street, Toronto, 
Ontario, M5S 3H8, Canada}
\altaffiltext{2}{Department of Physics and Astronomy, Queen's University, Kingston, Ontario, K7L 3N6, Canada}
\altaffiltext{3}{Email : gauthier@astro.utoronto.ca (JRG); dubinski@cita.utoronto.ca (JD);
                         widrow@.astro.queensu.ca (LMW)}

\begin{abstract}
We investigate the evolution of a population of 100 dark matter satellites orbiting in the gravitational potential 
of a realistic model of M31. 
We find that after 10 Gyr, seven subhalos are completely disrupted by the tidal field of the host galaxy. 
The remaining satellites suffer heavy mass loss and overall, 75\% of the mass initially in the subhalo 
system is tidally stripped. 
Not surprisingly, satellites with pericentric radius less than 30 kpc 
suffer the greatest stripping and leave a complex structure of tails and streams of debris 
around the host galaxy. Assuming that the most bound particles in each subhalo are kinematic tracers 
of stars, we find that the halo stellar population resulting from the tidal debris
follows an $r^{-3.5}$ density profile at large radii. 
We construct B-band photometric maps of stars coming from disrupted satellites
and find conspicuous features similar both in morphology and brightness to the observed 
Giant Stream around Andromeda.
An assumed star formation efficiency of 5-10\% in the simulated satellite galaxies results 
in good agreement with the number of M31 satellites, the V-band surface brightness distribution, 
and the brightness of the Giant Stream. 
During the first 5 Gyr, the bombardment of the satellites heats and thickens the disk by a small amount. 
At about 5 Gyr, satellite interactions induce the formation of a strong bar which, in turn, leads to a
significant increase in the velocity dispersion of the disk. 
\end{abstract}


\section{Introduction}

In the current cosmological paradigm, large-scale structure
forms via hierarchical clustering wherein small systems 
composed of dark matter and gas merge to form larger objects. 
These systems originate from the primordial
density fluctuation spectrum of cold dark matter (e.g., \citealt{dav85}). 
The hierarchical clustering hypothesis leads to a picture in which subhalos are incorporated
into larger systems. A number of processes such as tidal stripping and dynamical friction can lead
to the destruction of substructure \citep{whi78}. However, many subsystems survive 
and remain in orbit about the parent halo.
This phenomenon gives birth to a large
population of small satellites orbiting around galaxies and clusters of galaxies 
\citep{moo99,kly99,ghi00,die04,gao04,ben05}. Typically this population consists of several
hundred subhalos and contributes about 10\% of the total dark halo mass in a galaxy \citep{fon01,ghi00}. 

From an observational point of view, there is no clear evidence that such a 
large population of dark matter satellites exists. There are only $\approx$ 40 observed satellites in the 
Local Group out of which 13 belong to the Milky Way \citep{mat98}. This number is an order of magnitude less than predicted
from cosmological simulations because the stellar content and hence surface brightness is below current detection limits.  How then is it 
possible to reconcile the low number of observed satellites with models 
predictions?  There are several possibilities that we examine below.

The satellites may still be there but difficult to observe.
Many of the satellites may be associated with high-velocity clouds \citep{bli99} or 
may not be detected because the stellar component is not important enough and the surface brightness 
of these objects is below current detection limits. 
There are also theoretical reasons why star formation may be suppressed or inefficient in low mass satellites.
Several studies explore  mechanisms that should 
operate in the early stages of galaxy formation to suppress star formation in low-mass satellites. 
Gas ejection by supernovae-driven winds from the shallow potential wells of satellites may
quench star formation (e.g. \citealt{dek86}).
Also a strong intergalactic ionizing background can prevent
the collapse of gas into low-mass systems \citep{tho96} with circular velocities $v_c \lesssim 30$ km $s^{-1}$
as revealed in  numerical simulations. Recently, \citet{min05} claim to have detected an HI source in the Virgo Cluster
that could be a dark halo without an accompanying bright stellar galaxy. If the observed system is bound, 
this implies a dark halo mass up to $10^{11}$ $M_{\odot}$. 
In addition, the census of the Milky Way satellites may be incomplete
at low latitude due to obscuration. \citet{wil04} estimated a 33\% incompleteness in the total number of dwarfs
due to obscuration which represents only a small increase of the total number compared to cosmological predictions. All these possible explanations and 
considerations are plausible solutions to the 
satellite abundance problem. 

A large satellite population may produce a strong dynamical effect and modify 
the structure and kinematics of the galactic disk.
Many studies look at the dynamical response of the host galaxy from
an infalling satellite \citep{tot92,qui93,wal96,hua97,vel99,ben04}.
\citet{mih95} show that the accretion of a low-mass satellite by a disk galaxy can
generate a strong bar. This bar can buckle vertically and eject disk material out of
the disk giving rise to an X-structure similar to structures observed in some bulges (e.g. \citealt{whi88}).
Angular momentum transfer from the satellite to the disk may also occur during the infall
phase. \citet{hua97} demonstrate that satellites with a mass 0.3$M_{\rm disk }$ change the orientation of 
the disk up to $10^{\circ}$ and generate warps as the satellites, under dynamical friction, sink toward the center of the galaxy.
In addition, \citet{qui86} and \citet{qui93} note that the effects on the vertical structure of the disk are not uniform
across the disk. They observe some flaring of the disk at large radii, i.e. particles at
large radius are orbiting at larger $z$ than those at small radius. As suggested by these
authors, a thick disk produced by minor mergers should have a scale height
that increases with radius. Repeated interactions of a large population of satellites with a stellar disk also lead to disk heating.
\citet{qui86} shows that the disk is not heated isotropically by the infall of a satellite. They note that the heating 
is largest near the center of the disk and that most of the kinetic energy is distributed in the disk plane, 
i.e. $\sigma_z$ receives a small fraction of the available energy. 
In fact, most of the radial and azimuthal heating comes from the disk spiral response to an infalling satellite \citep{qui93}. 
In addition, the disk spreads radially due to an angular momentum exchange between 
the satellite and the disk.  
\citet{vel99} demonstrates that the heating and thickening rate of the disk
differs for satellites on prograde and retrograde orbits. The former heat the disk and the second tilt
it. 
\citet{ben04} show that
there are significant differences between heating rates for prograde and retrograde orbits and those
differences are amplified by increasing the mass and concentration of the satellites.
As noted by some authors \citep{ben04,fon01}, only satellites with orbits that bring them close to the center of the 
galaxy affect the disk significantly. 

Most of the results described above were based on controlled numerical experiments where a single satellite interacts
with a "live" disk. 
The exception was the pioneering study by \citet{fon01} who examine the evolution of a stellar disk 
embedded in a system of several hundred subhalos. The work presented here improves upon that study in 
a number of ways. First, the mass resolution of our simulations is a factor of 250 higher than in the 
\citet{fon01} study. This is high enough to adequately follow the evolution of the satellites and 
disk and to suppress two-body effects. Second, the base model for the galaxy is the self-consistent 
equilibrium model of M31 designed to fit the observed photometric and kinematic data for the Galaxy. 
Finally, the assumed mass distribution of the subhalo population and internal structure of individual 
subhalos are motivated by results from cosmological simulations. 

Our aim is to study the effect of tidal stripping on the satellite population. In particular, we attempt
to quantify the number of satellites that survive with a detectable stellar surface brightness. (Our subhalos
have a single component, but we can infer the surface brightness by modelling the stellar content, as described
below.) In addition, we study the stellar halo formed from tidal debris paying particular attention to extended
structures. As we will see, these structures are similar to 
observed features of M31 such as the Giant Stream \citep{iba01}.
 
In \S 2, we describe our
N-body model of M31 and describes a control experiment without satellites designed to test the model's inherent
stability.
In \S 3, we describe
the initial conditions of our satellite system. In \S 4,
we discuss the evolution of the satellite population and follow with a discussion of
the hypothesis that the metal-poor stellar halos arise in the tidal stripping
of satellite galaxies.  Our main conclusions are summarized in \S 5.

\section{An N-body model of M31}

For our experiments, we use a self-consistent numerical model of M31
derived from a composite distribution function \citep{wid05} (hereafter the WD models).
In general, the WD models are axisymmetric equilibrium solutions of the collisionless Boltzmann equation which may be 
subject to non-axisymmmetric instabilities. 
We choose their model M31a which provides a good match to the observational data for M31. Numerical simulations of this 
model assuming a smooth halo find that the system is stable against the formation of a bar for 10 Gyr.
 Therefore any heating or instabilities that are observed in simulations where part of the smooth halo is replaced by 
a subhalo population should be the result of the interactions between the disk and
 the satellite system.

The model consists of an exponential disk, a Hernquist-model bulge and a truncated NFW halo with plausible mass-to-light ratios for the disk and bulge
and simultaneous fits to the surface brightness profile, rotation curve and bulge velocity dispersion and rotation
with an NFW halo that extends to the expected virial radius.
These models contrast with other studies of infalling satellites (e.g \citealt{ben04,vel99}) 
that use disk galaxy models that must be pre-relaxed from an approximate equilibrium initial state \citep{her93}.
The M31 model has the main advantage that the initial state is formally an equilibrium solution to the collisionless Boltzmann equation 
since it is derived from a distribution function constructed from integrals of motion.

Table \ref{m31} contains the physical parameters of our Andromeda model.
The bulge follows the Hernquist profile 
\begin{equation}
\rho_{\rm{H}}(r) = \frac{\rho_b}{\left(r/r_b\right) \left(1 + r/r_b\right)^3} 
\end{equation}
while the halo is modeled according to the NFW density profile \citep{nav96}
\begin{equation}
\rho_{\rm{NFW}} = \frac{\rho_0}{r/r_s \left(1+r/r_s\right)^2} \; . 
\end{equation}
The disk distribution function is taken from \citet{kui95}. In brief, 
the surface density profile follows an exponential with scale radius 
$R_d$ and truncation radius $R_{out}$. The vertical component is given 
by $\rm{sech}^2$($z$/$2z_d$) where $z_d$ is defined as the scale height. 

\subsection{Control Experiment Results}

We first run a control experiment of the M31 system for 5 Gyr 
using a parallel treecode \citep{dub96}. The softening length is 15 pc. We use 10,000 equal time steps and the total energy is conserved to within $0.7 \%$. 
We take the 4.25 Gyr snapshot of the 35M-particle control model as the galaxy initial conditions for our
run with satellites. Most of the initial transient spiral instabilities have died away by that time and the increase in velocity dispersion
after that point is very small. Starting at this initial time minimizes the contamination of disk heating by initial transient spirals.
We also run other control experiments at lower resolution (350K and 3.5M particles) to look for convergence. The simulations 
are run using CITA's McKenzie cluster \citep{dub03}.

\begin{figure}
\centering{
\includegraphics[angle=-90,scale=.90]{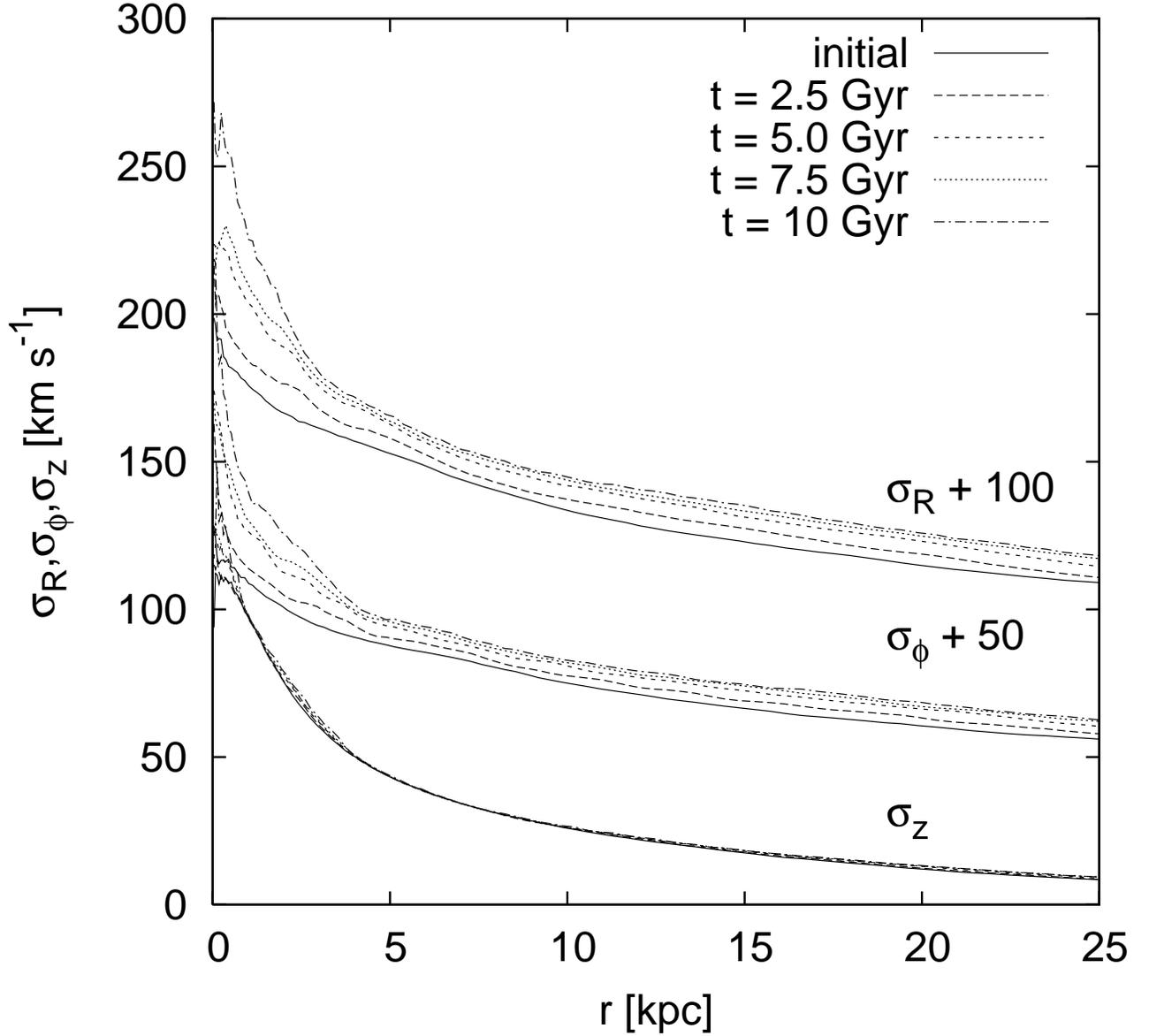}}
\caption{Evolution of the disk velocity dispersion for the 35M control simulation. 
From bottom to top : $\sigma_z$, $\sigma_{\phi} + 50$ km s$^{-1}$ and $\sigma_r + 100$ km s$^{-1}$. 
Note that there is no vertical heating over 10 Gyr. Most of the heating in the 
radial and azimuthal directions occurs in the first 5 Gyr and is due to transient spiral features present early in the 
simulation.}
\label{veldisp35M}
\end{figure}

\begin{figure}
\centering{\includegraphics[angle=-90,scale=.90]{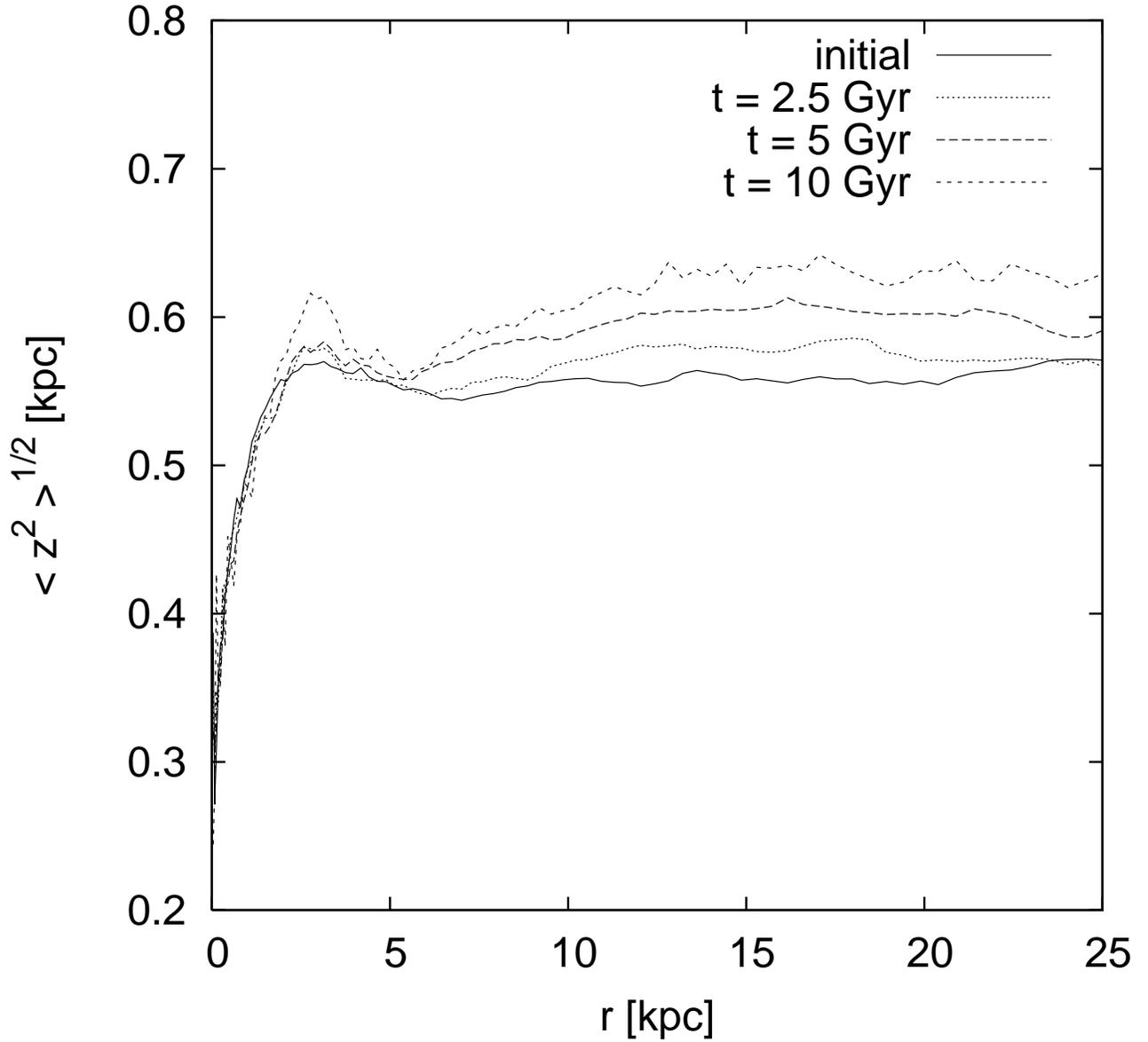}}
\caption{Disk scale height versus radius for the control experiment. Scale height is taken to be 
the variance of $z$ averaged over circular rings in the disk. 
Note that there is only minor growth of the scale height with time reflecting the quiet equilibrium of the disk DF and adequate resolution 
of the simulation.}
\label{scaleheight35M}
\end{figure}

After taking the 4.25 Gyr snapshots as initial conditions, we let the galaxy evolve for 10 Gyr.
Figure \ref{veldisp35M} shows the evolution of the disk velocity dispersion for the 35M model. 
Outside 5 kpc, $\sigma_r$ and $\sigma_{\phi}$ increase by 5-10 km $\rm{s}^{-1}$ while $\sigma_z$ is virtually unchanged. 
Inside 5 kpc, all components increase significantly. In fact, most of the heating occurs in the first 
few billion years where spiral patterns arising from swing-amplified noise
heat the radial and azimuthal components of the velocity ellipsoid. Nevertheless, the heating is quite small 
and negligible in the $z$ direction. The disk scaleheight evolution for the same 
model is shown in Figure \ref{scaleheight35M} and found to be minimal in the inner 
region of the disk. There is some flaring on the edges, but the increase is only around 100 pc.
The velocity dispersion ellipsoid ($\sigma_r$,$\sigma_{\phi}$,$\sigma_z$) at an equivalent solar radius of $R =2.5 R_d$
changes from (17.8, 24.6, 18.9 ) km $\rm{s}^{-1}$ to (20.5, 29.0, 19.1) km $\rm{s}^{-1}$ after 2.5 Gyr. 
In comparison, \citet{fon01} let their disk relax for
3.5 Gyr before introducing the satellite population and their velocity dispersion ellipsoid 
changes from (31, 27, 26) km $\rm{s}^{-1}$ to (43, 30, 33) km $\rm{s}^{-1}$. Because we use a large number of particles 
for the galaxy (20M - halo, 10M - disk, 5M bulge), the mass of the halo particles are quite small 
and artificial disk heating due to halo particles bombarding the disk is almost negligible.

\section{Initial conditions of the satellite population}

The properties of the initial satellite population are motivated from the latest results of $\Lambda$CDM simulations of cluster and galaxy formation.
Since the properties and distribution of subhalos do not differ significantly on galaxy and cluster size scale 
(\citet{moo99}, L. Gao private communication), the different relations at cluster size scale (number density profile, slope of 
the mass function, total mass ratio, etc.) can be rescaled down to the galactic scale.

\subsection{The mass distribution function}

\begin{figure}
\centering{
\centering{\includegraphics[angle=-90,scale=.50]{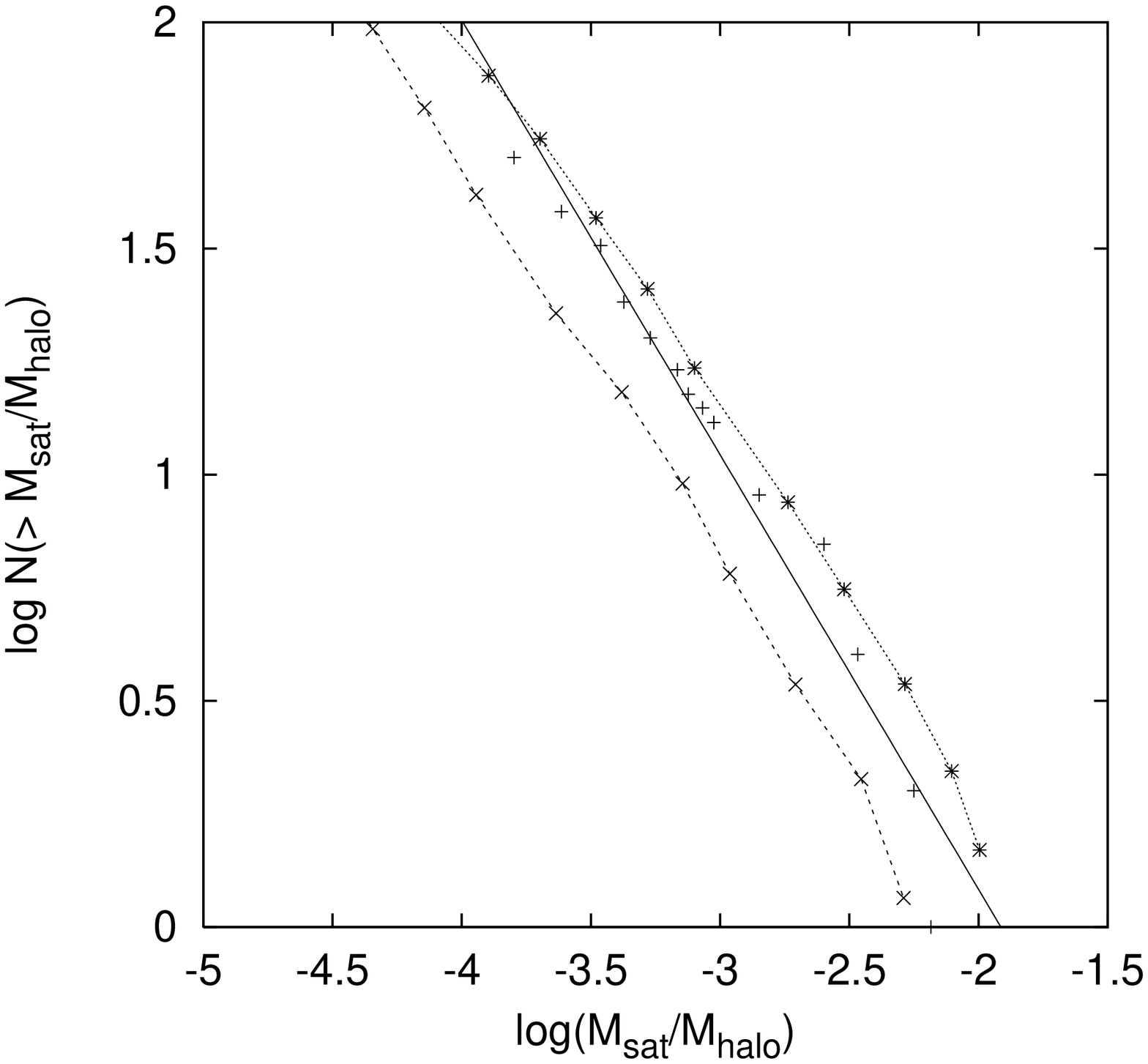}}
\centering{\includegraphics[angle=-90,scale=.50]{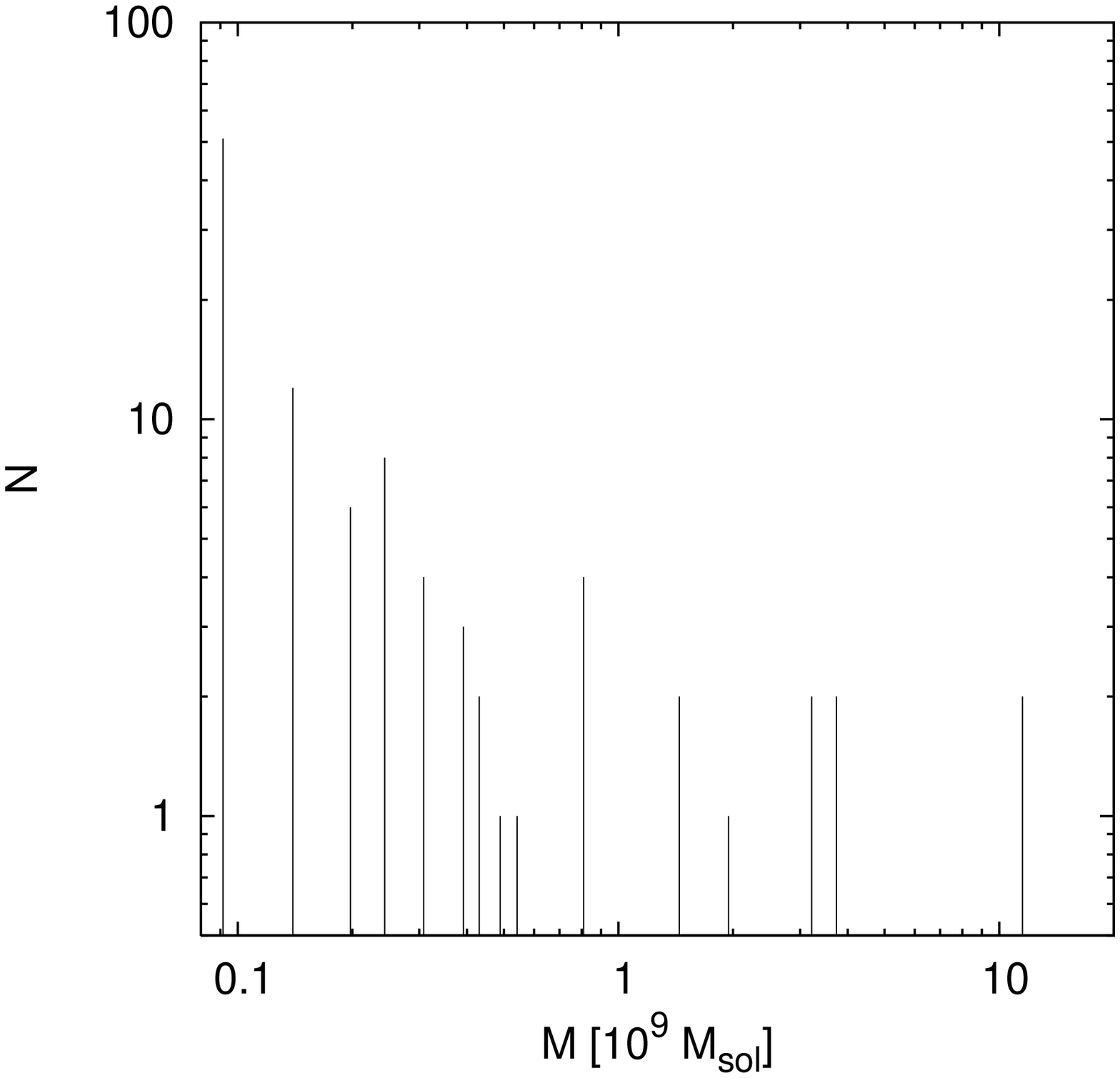}}
}

\caption{Satellite cumulative mass function (top) and mass function (bottom). 
For the top figure, the \emph{plus} symbols are the data points for our mass function  
and the \emph{filled} line is the best $dn/dM_{\rm sat} \propto M_{\rm sat}^{-1.9}$ 
fit to the data. Note that the mass function we use is normalized in such a way that 
in the interval 1.5 $\times$ $10^{-4}$ $<$ $M_{\rm{sat}}/M_{\rm{halo}}$ $<$ 0.02 the total mass in 
subhalos is 0.1 $M_{\rm{halo}}$. The other lines with symbols are taken from \citet{gao04}.
The \emph{X} symbols and \emph{short-dashed} line represent the subhalo mass function of a 
high resolution 2 $\times$ $10^{12}$ $M_{\odot}$ halo (GA3n run). The \emph{star} symbols and the 
\emph{dotted} line refer to an average of all the cluster-type simulations.  
}
\label{mdf}
\end{figure}

We use the following mass function in agreement with recent  
$\Lambda$CDM simulations by \citep{gao04} : 
\begin{equation}
\frac{d\cal {N}}{dM} = 10^{-3.2}\left (\frac{M_{\rm{sat}}}{h^{-1}M_\odot}\right )^{-1.9}
\left (\frac{1}{h^{-1}M_\odot}\right )^2 \; .
\label{massfunction}
\end{equation}
Here $\cal{N}$ represents the number of satellites per unit parent mass. We distribute the satellites between $M_{\rm{sat}}/M_{\rm{halo}} = 1.5 \times 10^{-4}$
to $M_{\rm{sat}}/M_{\rm{halo}} = 0.02$ in 15 different logarithmic mass bins. With $M_{\rm{halo}} = 5.8$ $\times$ $10^{11}$ $M_{\odot}$, we get $n \simeq 28$ and a mass fraction of 0.024.   
In \citet{gao04}, Figure 12 suggests that most of the satellites for a M31-sized galaxy are in place between $z=0.7$ and 1.1. 
Their Figure 13 shows that satellites accreted at $z=1$ lose of order 70-80\% of their mass by $z=0$. With these results in mind, we start our simulation with a total mass fraction of 0.1 and 100 satellites. 
The sample of 100 satellites is large enough to provide the correct stochastic treatment of disk heating and will provide a smooth distribution of
tidal debris.   
Figure \ref{mdf} shows the comparison between the expected
satellite cumulative mass function, the integral 
of Equation (\ref{massfunction}) normalized to 0.1 for 1.5 $\times$ $10^{-4}$ $<$ $M_{\rm{sat}}/M_{\rm{halo}}$ $<$ 0.02, 
and the data points generated for our population composed of 100 satellites. 

\subsection{Spatial Distribution of Satellites}

We assume that 
the spatial distribution of satellites in the parent halo is spherically symmetric. 
Recent cosmological simulations (e.g \citealt{die04,gao04}) suggest that the number density profile of satellites is 
less centrally concentrated than the dark matter halo. \citet{gao04} found that the cumulative number
density profile of subhalos in a Milky Way size halo is well fitted by 
\begin{equation}
N( < x) = N_{\rm tot} x^{2.75} \frac{\left( 1+0.244x_s\right)}{\left(1+ 0.244x_sx^{2}\right)} 
\end{equation}    
where $x = r/r_{200}$, $x_s = r_s/r_{200}$ and $N_{tot}$ is the total number of satellites within $r_{200}$. 
We sample that cumulative number density profile according to this formula with the further assumption that the spatial distribution of satellites 
is independent of the satellite mass. 

\subsection{Satellite orbital distribution}

According to cosmological simulations, satellite orbits are isotropic in the center and  slightly radially biased in outer regions (e.g. \citet{ben05,die04}). 
The anisotropy parameter 
\begin{equation}
\beta \equiv 1 - \frac{\sigma_{\theta}^2 + \sigma_{\phi}^2}{2\sigma_r^2}
\end{equation}
varies with radius as $\beta(r/r_{200}) \approx 0.35 r/r_{200}$ for the satellite population \citep{die04}. We use a constant value 
of $\beta = 0.3$ over the range $ 0 < r/r_{200} < 1$ as a good approximation. The velocity components $\left(\sigma_r,\sigma_{\theta},\sigma_{\phi}\right)$
are determined using the Jeans equation \citep{bin87} 
\begin{equation}
\frac{d(n \overline{\sigma_r^2})}{dr} + \frac{n}{r}\left[ 2\overline{\sigma_r^2} - \left(\overline{\sigma_{\theta}^2} + \overline{\sigma_{\phi}^2} \right) \right] = -n \frac{d\Phi}{dr}
\label{jeans}
\end{equation}
where $n$ is the satellite number density profile and $\Phi$ is the gravitational potential. 
Here we assume that the gravitational potential of the host galaxy can 
be approximated by an NFW density profile and the gravitational potential 
generated by the satellite population is a small perturbation of the host gravitational potential, i.e. we neglect the contribution 
of the satellites. The NFW approximation is valid everywhere, except within $\simeq$ 30 kpc  where the bulge and disk modify 
the potential significantly ($r \lesssim 30$ kpc). 
We solve equation (\ref{jeans}) numerically for $\sigma_r(r)$ and assign the velocity $(v_r, v_\theta, v_\phi)$ at position $r$ 
assuming they are random variates with dispersions $(\sigma_r,\beta \sigma_\theta, \beta \sigma_\phi)$.

\begin{figure}
\centering{\includegraphics[angle=-90,scale=0.9]{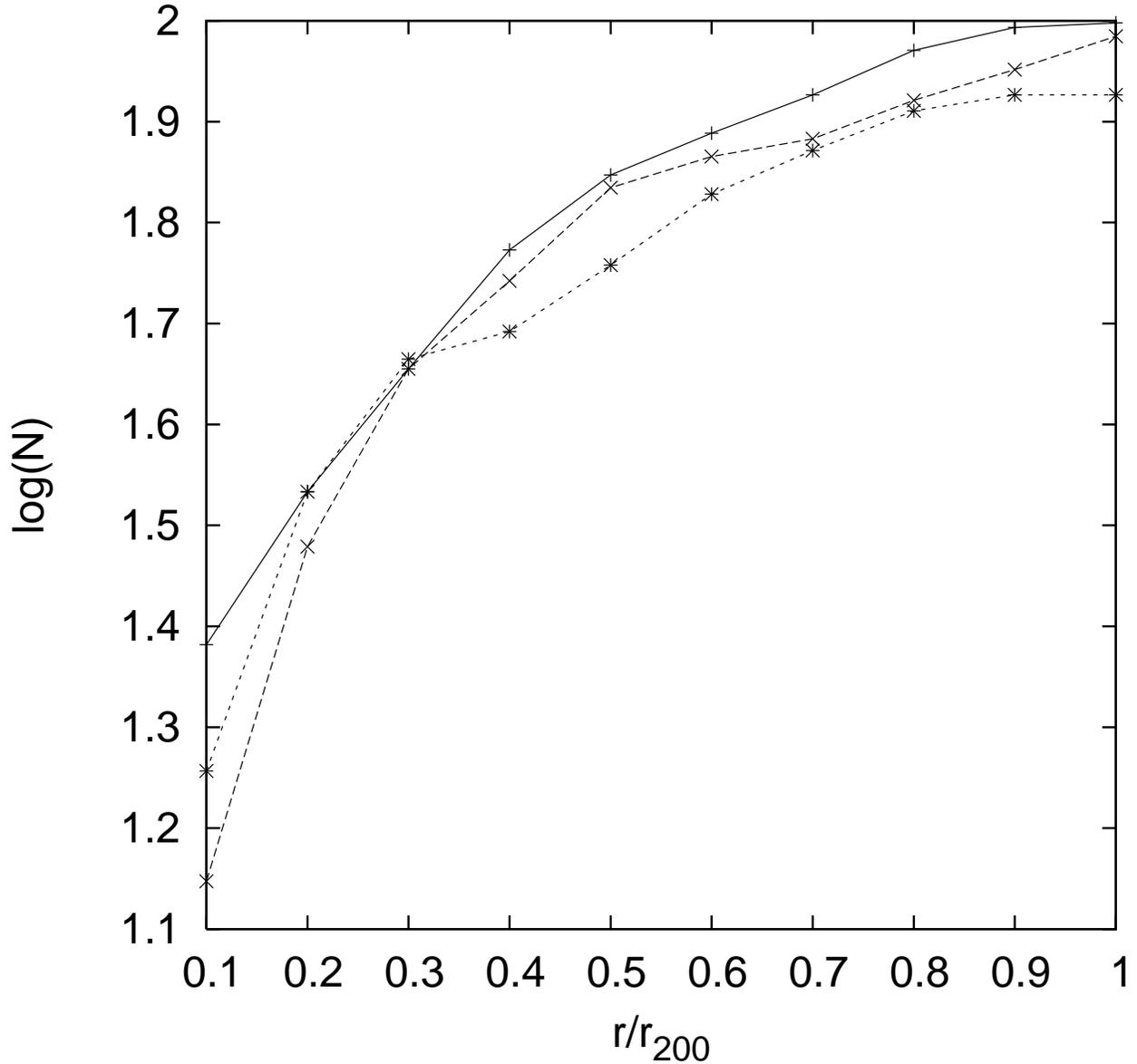}}
\caption{Evolution of the satellite cumulative number density function using solid test particles in the rigid M31 potential (no dynamical
friction/tidal stripping). The different lines with symbols are for : \emph{plus-filled} 0 Gyr, \emph{X-dashed} 5 Gyr, and \emph{stars-dotted} 10 Gyr snapshot.  The cumulative number density function evolves slightly over 10 Gyr reflecting the small number statistics
and the fact that spherical NFW potential is only an approximation to the full potential for the galactic model.
}
\label{evoldistribution}
\end{figure}

To validate the equilibrium of the satellite distribution within the M31 model galaxy, we replace the satellites by
test particles and compute the forces assuming a rigid potential. 
So long as the satellites are test particles neither 
dynamical friction nor tidal stripping come into play. 
In Figure \ref{evoldistribution}, we show the evolution of the cumulative number density function of satellites.
There is an evolution of the distribution, but this may be attributed to small number of satellites and the
inexactness of using the Jeans equations dispersion profiles to approximate the distribution function.
Despite this slight evolution, we believe that 
our initial distribution could represent a population in quasi-equilibrium with the host galaxy and is therefore suitable
for our experiments.

\subsection{The internal structure of the satellites}

Individual satellites are modeled as single cuspy NFW halos truncated by the tidal field of the galaxy model.
The WD model for an isolated NFW halo has an adjustable parameter $\gamma$ that can introduce an arbitrary tidal radius akin to the
the tidal radius of a King model.  Satellite models can therefore be created that have an $r^{-1}$ cusp with NFW behavior at
the extremities and a self-consistent cut-off.
Many previous studies used the King model for the satellites (e.g. \citealt{ben04,vel99,hua97}) in satellite-disk interactions. 
Since NFW satellites are cuspy, we might 
expect them to be more robust to tidal interactions than King-model satellites and therefore 
produce more damage to the disk. 
We use 15 different satellite models to cover the mass range $1.5 \times 
10^{-4} < M_{\rm sat}/M_{\rm halo} < 0.02$ logarithmically. This mass range corresponds to  $8.75 \times 10^{7} < M_{\rm sat}/M_{\odot} < 1.17 \times 10^{10}$. As a comparison, the Sagittarius dwarf and Large Magellanic Cloud, 
our galactic companions,
have estimated masses of (2-5) $\times$ $10^{8}$ $M_{\odot}$ \citep{law05,iba95} and 2 $\times$ $10^{10}$ $M_{\odot}$ \citep{sch92}. Note that the mass of the LMC corresponds to the upper limit of the mass function. M31's satellites M32 and 
NGC205 have masses of 2.1 $\times$ $10^{9}$ and 7.4 $\times$ $10^8$ $M_{\odot}$ respectively \citep{mat98}. Our mass
range includes most of the observed satellites in the Local Group.  

For simplicity, the truncation of each satellite is determined by the length of the tidal radius  
at the mean apocentric radius of the satellite system ($r = 50$ kpc). Here we assume that the tidal radius 
of the satellites $r_s$, is defined by the Jacobi approximation \citep{bin87}. The radial extent of the satellites
is determined by comparing the mean density of the satellite inside the tidal radius and the mean density of the 
halo inside $r=50$ kpc : 
\begin{equation}
\overline{\rho_s}(r_t) \approx 3\overline{\rho_h}(r=50 \rm{ kpc})
\end{equation}
where $\overline{\rho_s}(r_t)$ is the satellite mean density inside the tidal radius $r_t$ and $\overline{\rho_h}(r=50 \rm{ kpc})$
is the halo mean density inside 50 kpc. 
Satellites within $r=50$ kpc will over fill their tidal radii while those beyond will lie well within it. We determine
that satellites initially located inside 50 kpc 
will lose a cumulative $5.3 \times 10^8$ $M_{\odot}$ due to the tidal radius overestimate. This accounts 
for an ``artificial" tidal stripping at the beginning of the simulation and corresponds to less than 1\% of 
the total satellite mass. 
A true satellite system is never really in equilibrium since dynamical friction leads to continual tidal erosion so our
choice of a mean satellite orbital radius for estimating the tidal radius is a compromise.  We see below that there is indeed
a transient startup where satellites that overfill their tidal radii are quickly stripped but by an amount that is
small (less that 1\%) compared the total stripping over the course of the satellite system evolution.

Table \ref{15models} contains the main characteristics of each satellite model. 
We model our satellites assuming a shallow power law distribution for the concentration 
of the satellites
\citep{nav96,nav97}. 
The internal properties of the satellite halos found in cosmological 
simulations are not well determined, especially for the low-mass end where the concentration 
and density profile are not well constrained because of the poor mass resolution. 

\subsection{Evolution of individual satellites}

To compute the mass, center-of-mass position and density profile evolution of each infalling satellite, we
use the technique described in \citet{ben04}. This algorithm identifies the particles that are bound to the
satellites, computes the center-of-mass of the system, and iterates until the total mass converges. Typically, the criterion we use for convergence varies by less than 0.5\% in mass and most of the steps
require only a few iterations. The method is straightforward and offers a good alternative to the friends-of-friends algorithm
which is not appropriate when the satellites are numerous and characterized by different scale lengths.

\section{Results}

We perform two simulations at low and high resolution to test for numerical convergence. 
In the low resolution run, each of the one hundred satellites is represented by 1000 particles and put in orbit
around an M31 galaxy model with 100K disk particles, 50K bulge particles and 200K halo particles. In the high resolution simulation, particle numbers are increased by a factor of 100.
The population of satellites for both simulations have the same mass function, spatial distribution and orbital velocities.
Both simulations are run for 10 Gyr (20,000 equal time steps) using a parallelized treecode \citep{dub96} with
opening angle parameter $\theta = 0.9$ (quadrupole order) and $\epsilon = 25$ pc ($\epsilon = 15$ pc) for the 450K (45M) run. For the 45M-particle run, the energy is conserved to within 0.4 \%. Unless otherwise 
stated, all the results presented in this section will be for the high resolution version. 

\subsection{Galaxy Evolution in Three Acts}

\subsubsection{Act I. The First Orbit}

During the first few billion years, the satellites accomplish their first complete orbits and leave 
behind an interwoven web of tidal streams. Some of these 
streams extend beyond the virial radius of the host galaxy.
One of the most interesting features of the first two billion years is the presence 
of shell structures with clearly defined edges similar to those seen in some elliptical galaxies. 
These shells are produced via phase wrapping \citep{qui84,her87} and tend to disappear quickly due to phase
mixing. During this period, the disk remains quiet with only a small amount of heating. 

\subsubsection{Act II. Bar Formation}

At about 5 Gyr, a stunning phenomenon appears in the 
disk : the unexpected formation of a bar. We can associate its creation with the interaction of the disk and 
the satellite system since no bar formed in the control run after 10 Gyr.
The strong bar is responsible for most of the disk heating after 5 Gyr and creates a puffy vertical structure.

\subsubsection{Act III. Anticlimax : The Quiet End}

Following bar formation, the disk and satellite system evolution is more gradual. After two or three pericenter passages, 
the satellites are stripped significantly. The galactic halo is enriched with tidal debris and the initial 
streams become more mixed. The final state of the debris is that of a spheroid of about the size of the Galactic stellar 
halo. 
During that time, the bar suffers a bending instability that gives rise to the buckling mode \citep{rah91}. This process 
generates a conspicuous X-design in the disk when viewed edge-on and may point to the origin of the peanut-shaped bulges
observed in many galaxies \citep{bur99,kui95b,whi88}. 
During the final two billion years, the bar remains the dominant feature of the disk but slowly spins down. 
By the end of the simulation, a typical satellite completes five to seven orbits and loses a significant fraction of its initial mass. 
Snapshots of the satellite population and the host galaxy are shown in Figure \ref{snapshots} (see also the animation of the disk galaxy evolution at http://www.cita.utoronto.ca/\~{}jgauthier/m31). 

In Figures \ref{veldisp45M} and 
\ref{scaleheight45M} we show the evolution of the disk velocity dispersion ellipsoid and 
scaleheight for the high-resolution run. As noted in Figure \ref{veldisp45M}, 
the formation of the bar around 5 Gyr produces a sudden jump in velocity dispersion for disk 
particles. The same event triggers the scaleheight increase in Figure \ref{scaleheight45M}.
Further details of the disk evolution will be discussed in a forthcoming paper. 

\subsection{Evolution of the Satellite System}

The satellites are initially distributed 
isotropically in space between approximately 50 kpc to 250 kpc ($\approx$ $r_{200}$). Although most of them 
survive the strong tidal interactions in the vicinity of the disk, they create extended tidal streams
and shell structures. These streams and shells are particularly obvious in the first 
few billion years, but tend to lose their sharp edges as phase-mixing proceeds.
In Figure \ref{ind_sats}, we show the evolution of the two most massive and three least massive satellites. 
The plots, which show the evolution of the satellites position and 
density profile demonstrate that dynamical friction has little effect in bringing the satellites
close to the center of the galaxy. Figure \ref{number_density_profile} shows that there is no significant change 
to the number density profile of satellites after 10 Gyr. Because of the very steep slope of the mass function, most of our satellites have a mass of $1.5 \times 10^{-4}$ $M_{\rm halo}$ and do not feel the effects of dynamical friction. In fact, plots of the satellite 
orbital decay indicate that the value of the apocenter radius decreases very slowly for 
almost all satellites, except for the very massive ones (see Figure \ref{ind_sats}). 

The inner slope of the satellite density profile remains the same over the simulation (see Figure \ref{density_profiles.sats}).
As noted by \citet{hay03}, it is possible to describe the structure of a stripped halo by modifying the NFW profile : 
\begin{equation}
\rho(r) = \frac{f_t}{1+(r/r_{\rm te})^3}\rho_{\rm NFW}
\label{hayashi}
\end{equation}
where $f_t$ is interpreted as a reduction in the central density of the profile and $r_{\rm te}$ is an effective 
tidal radius that describes the cutoff due to tidal forces. Comparisons are hard to make with the work done by \citet{hay03} because
our satellites suffer an initial truncation to avoid a divergent mass profile. In some way, our initial satellites density profiles look quite similar to the final profiles of their subhalos. 
Nevertheless, we show, in Figure \ref{density_profiles.sats} the best fit of equation (\ref{hayashi}) for typical profiles of 
three different satellites mass bins. Equation (\ref{hayashi}) provides a good fit of the final profile especially for the low-mass 
satellites with $M_{\rm sat}/M_{\rm halo} \lesssim 5.5 \times 10^{-4}$. For the more massive ones, the fit tends to overestimate 
the mass loss at large radii. 

\begin{figure}
\centering{\includegraphics[angle=-90,scale=.90]{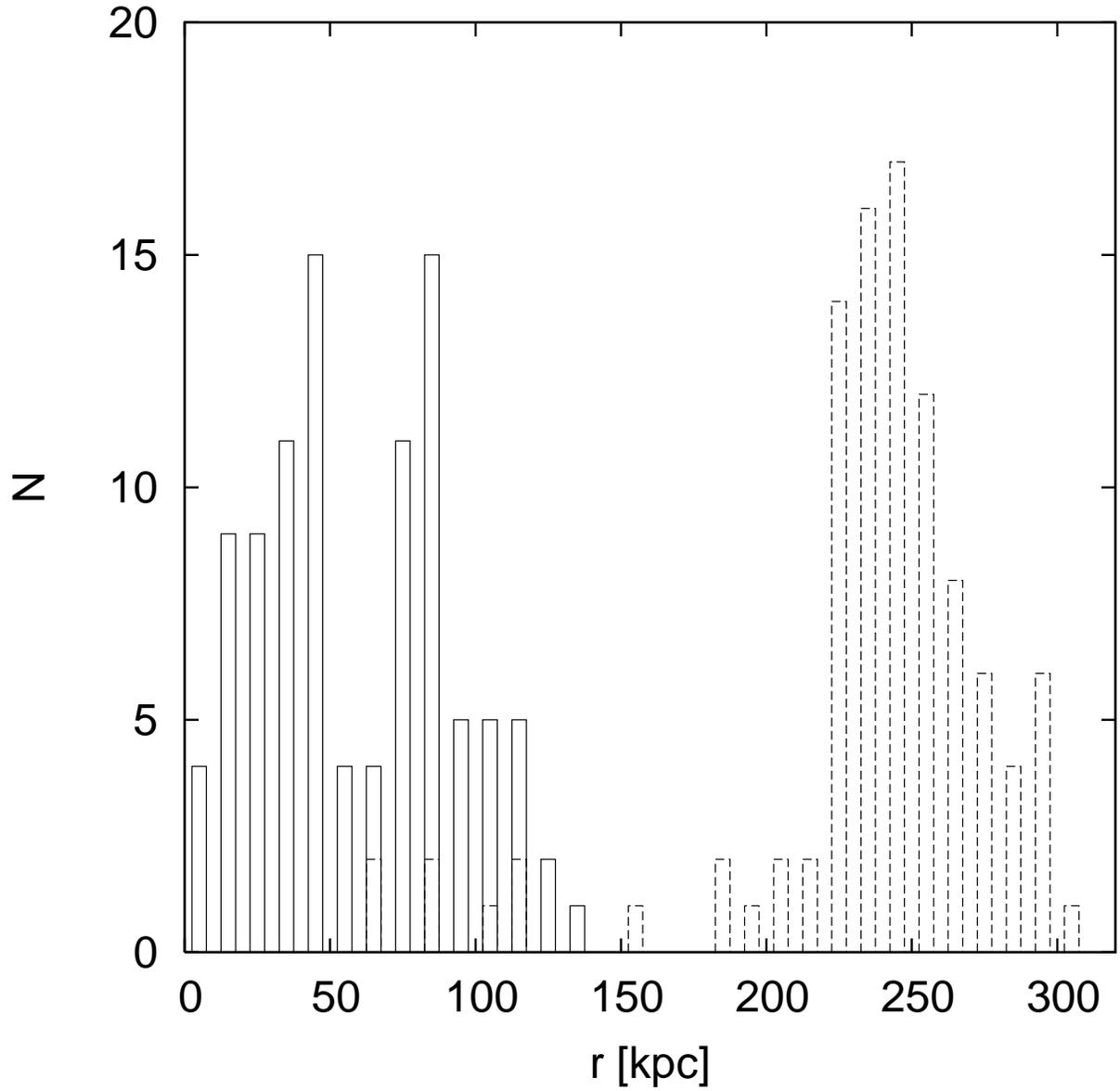}}
\caption{Distribution of the initial pericenter and apocenter passages radii for the satellites population.\emph{Filled} line :
the pericenter radii; \emph{dashed} line : apocenter radii. The pericenter distribution peaks at 50 kpc and the apocenter one
at 250 kpc. The ratio $r_a/r_p$ $\approx$ 5 is typical of cosmological simulations (e.g. \citet{moo99}).}
\label{apoperi}
\end{figure}

\begin{figure}

\centering{\includegraphics[angle=0,scale=0.8]{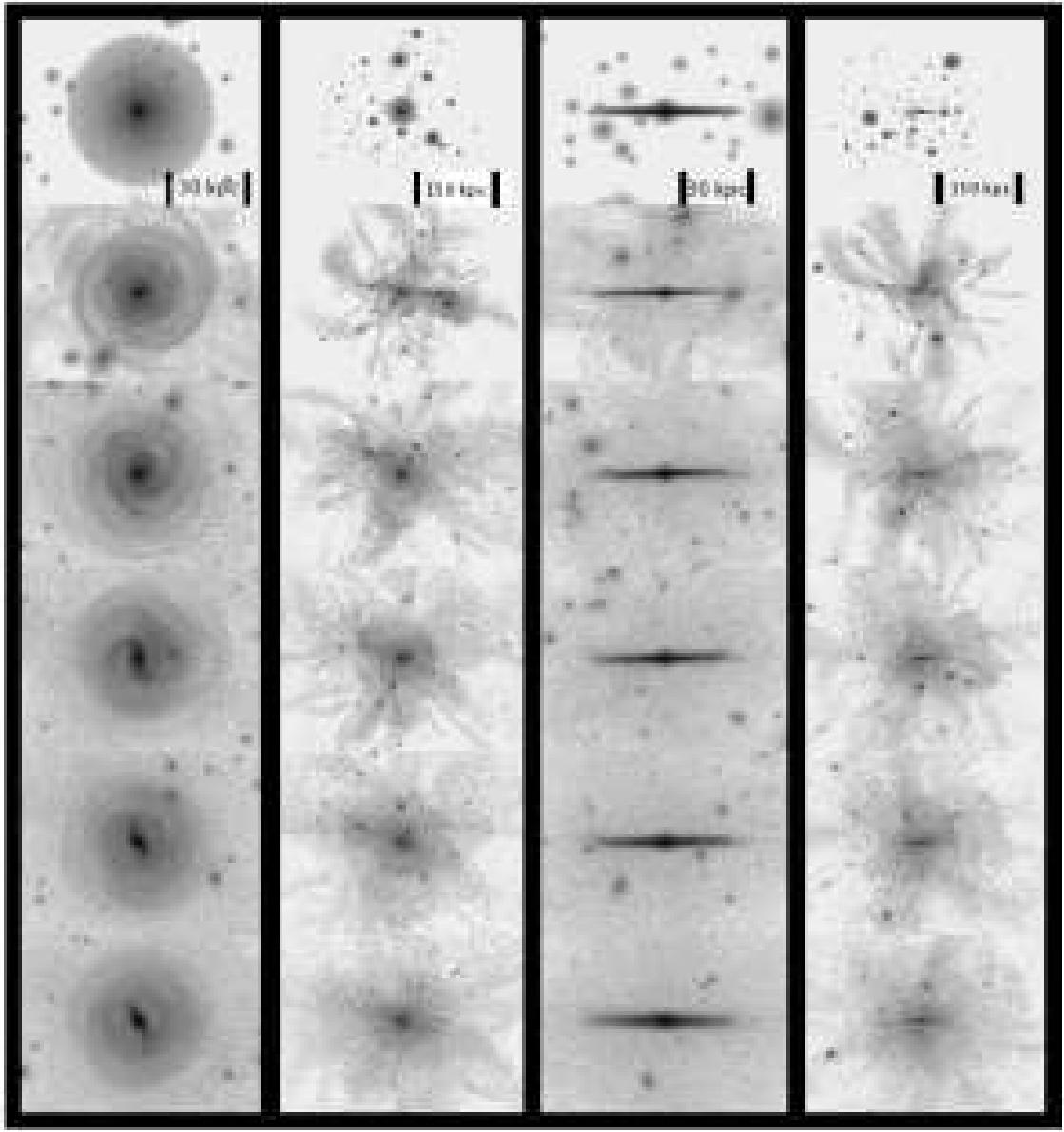}}
\caption{Snapshots of the 100-satellite simulation. 
From top to bottom, t=0, 2, 4, 6, 8, and 10 Gyr. One can see 
that a strong bar is formed between 4 and 6 Gyr. Several conspicuous shell structures are visible, especially 
in the first few billion years.(\emph{See high-resolution color snapshots at http://www.cita.utoronto.ca/\~{}jgauthier/m31})}
\label{snapshots}
\end{figure}

\begin{figure}
\centering{\includegraphics[angle=-90,scale=.90]{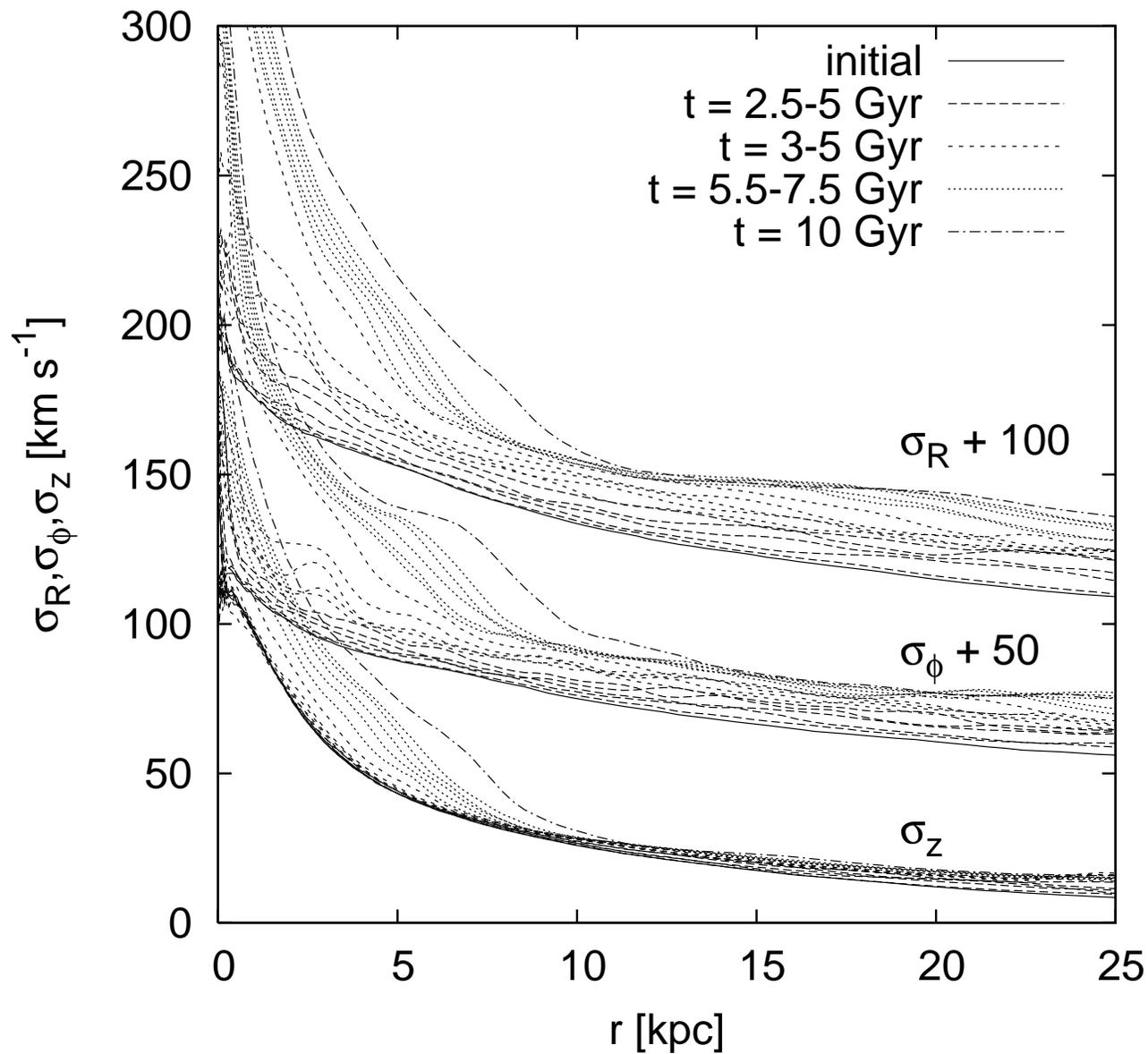}}
\caption{Evolution of the disk velocity dispersion for the simulation with subhalos. From bottom to top : $\sigma_z$, $\sigma_{\phi} + 50$ km s$^{-1}$
and $\sigma_r + 100$ km s$^{-1}$. Between 4 and 5 Gyr, a bar forms. This explains the sudden increase of $\sigma_r$ and $\sigma_{\phi}$ in the 
inner region of the galaxy between 3 and 5 Gyr.}
\label{veldisp45M}
\end{figure}

\begin{figure}
\centering{\includegraphics[angle=-90,scale=.90]{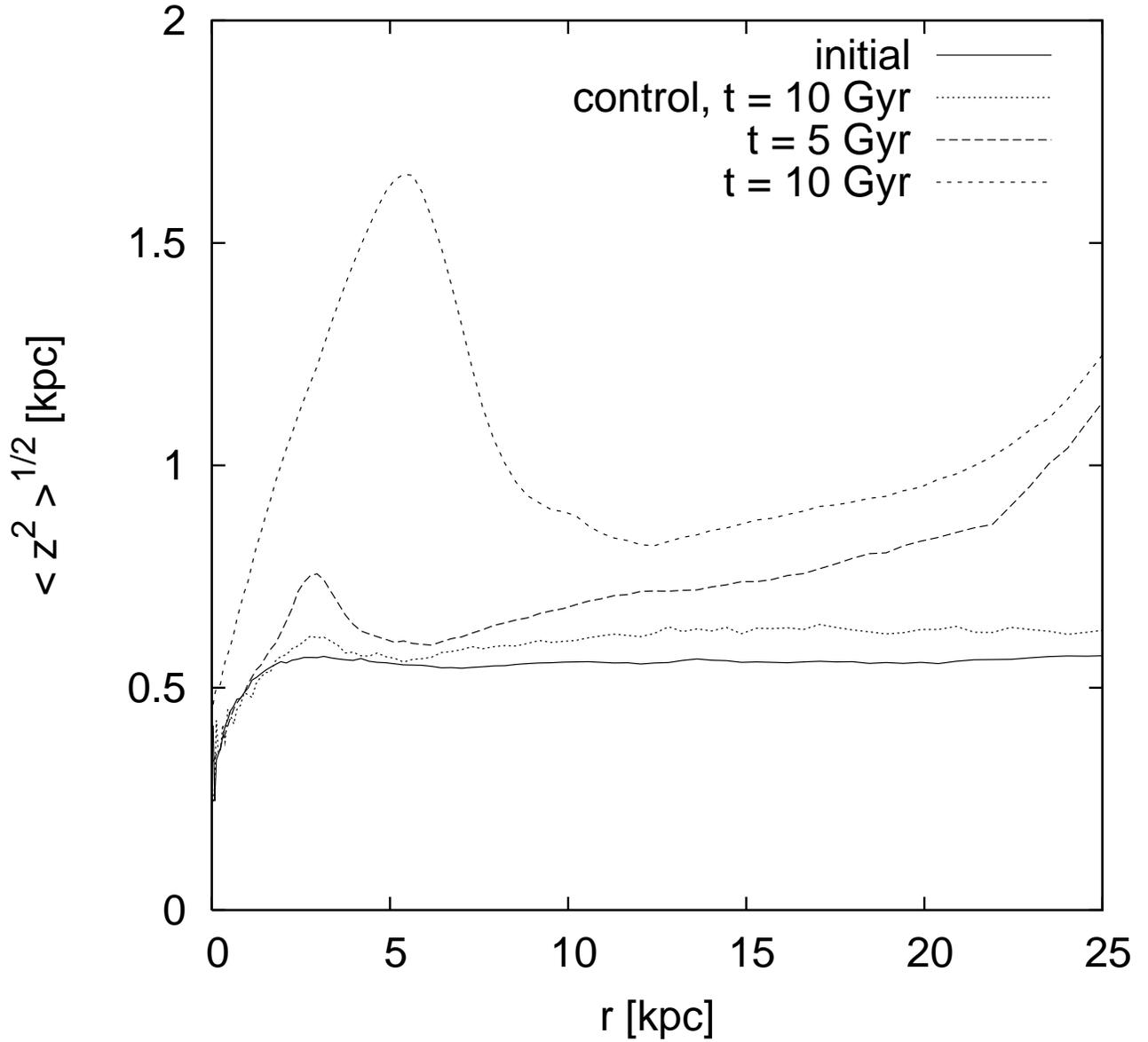}}
\caption{Same as Figure \ref{scaleheight35M} but with satellite population. The scale height only grows slightly during the first 4 billion years reflecting only mild heating from the satellite population. 
The rapid heating thereafter is due to the formation of the bar.}
\label{scaleheight45M}
\end{figure}

\begin{figure}
%

 \vspace{7pt}
  \centerline{\hbox{\hspace{0.0in}
\includegraphics[angle=-90,scale=.22]{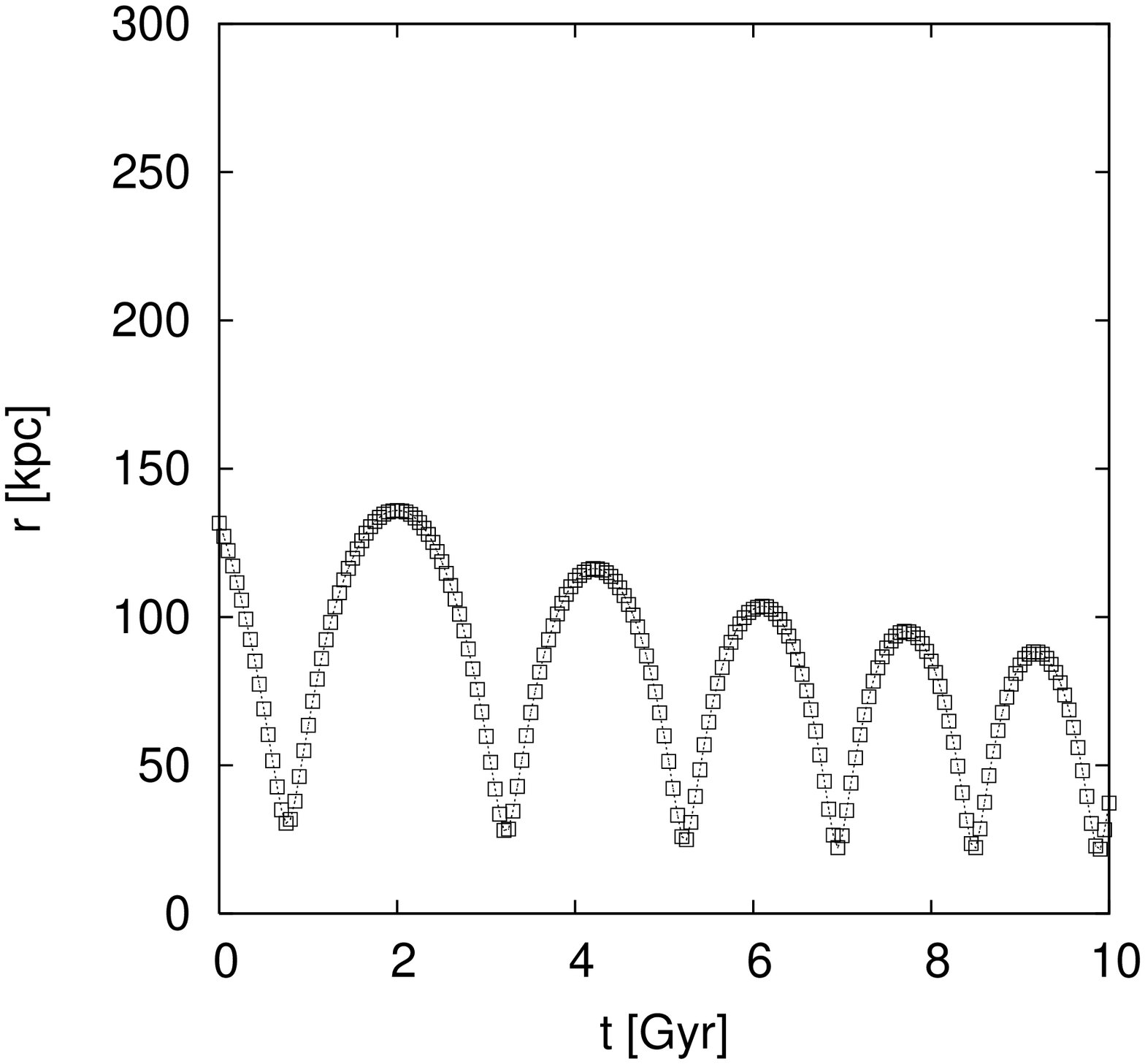}
    \hspace{0.1in}
\includegraphics[angle=-90,scale=.22]{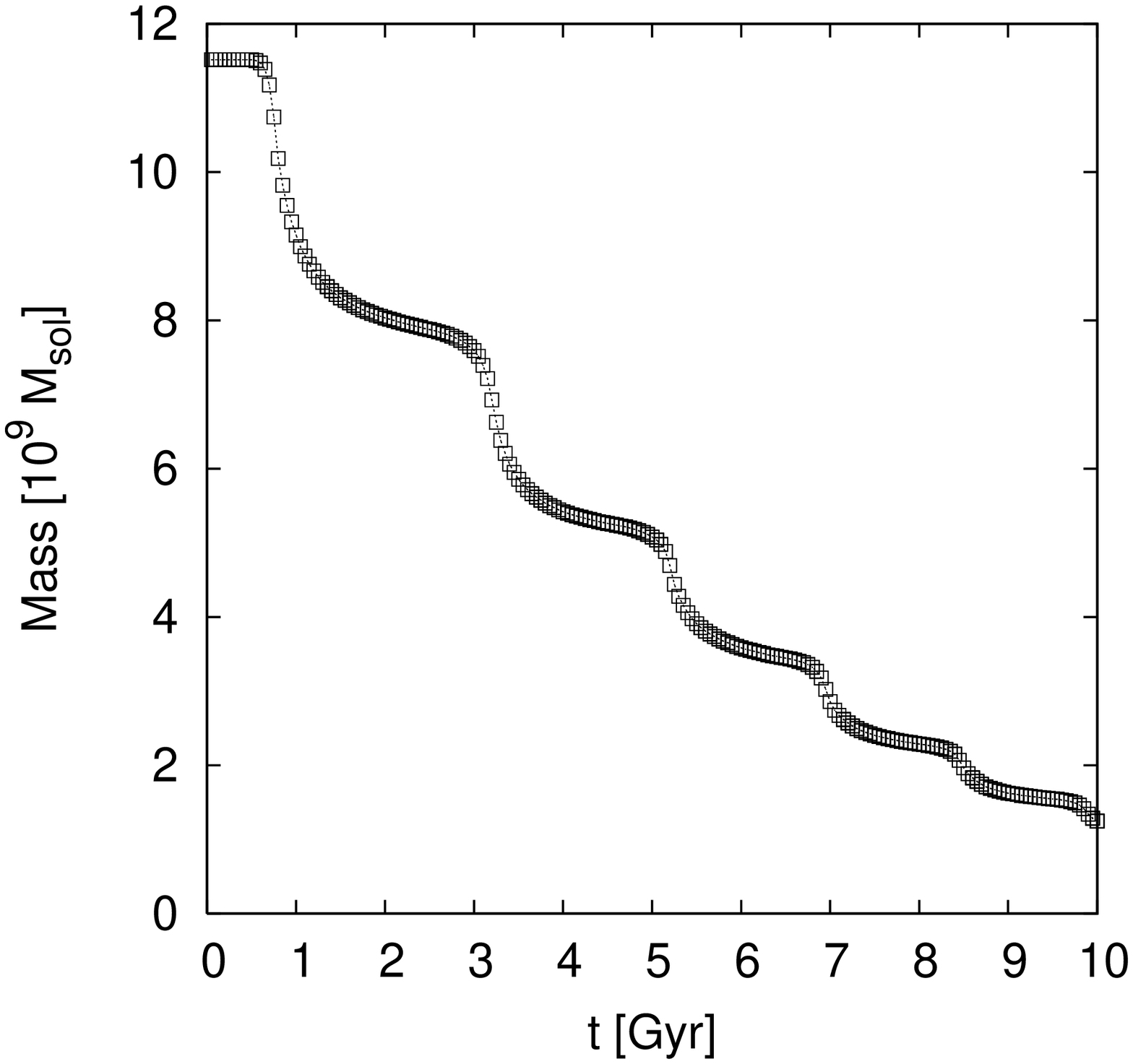}
    \hspace{0.1in}
\includegraphics[angle=-90,scale=.22]{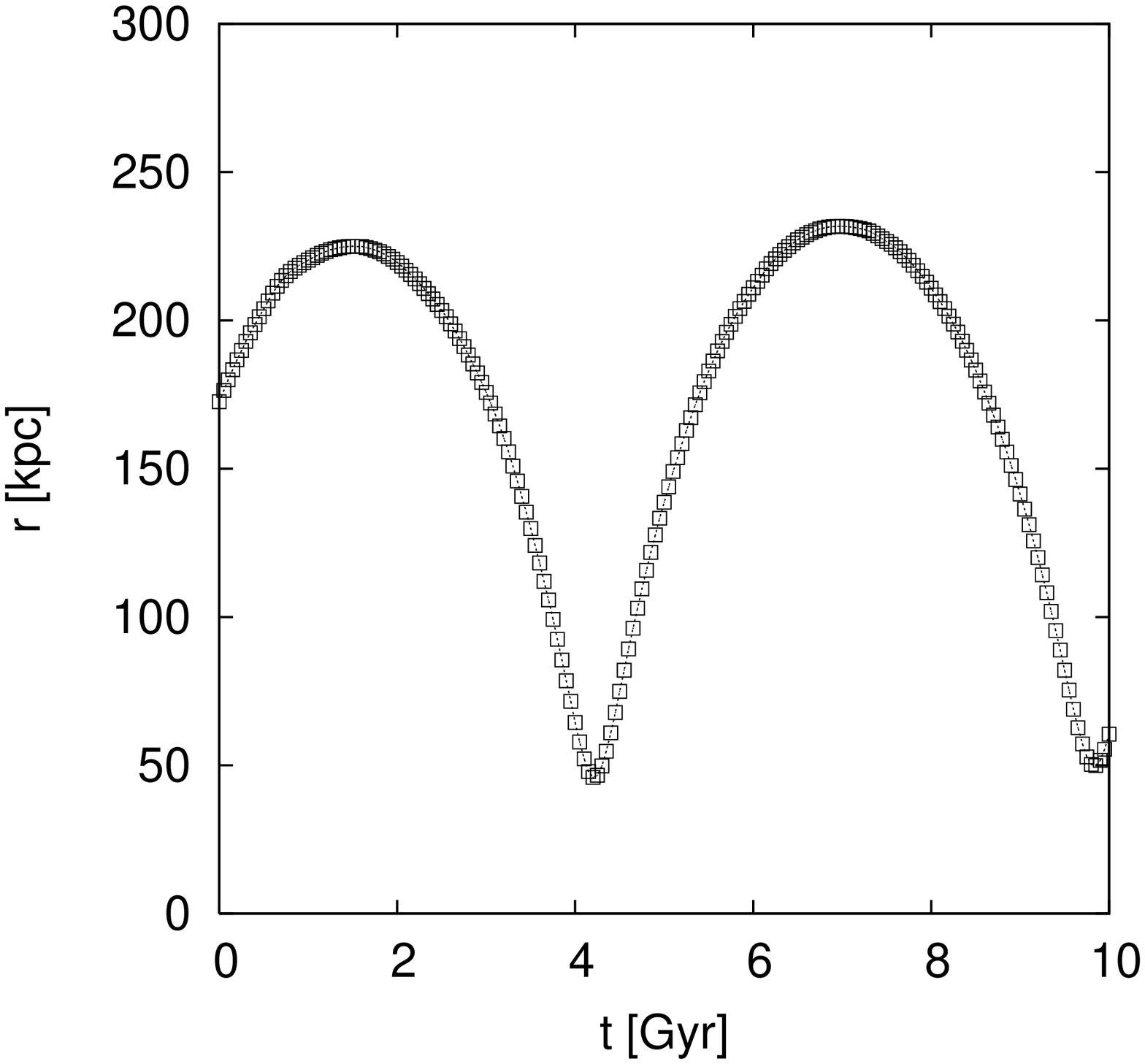}
    \hspace{0.1in}
\includegraphics[angle=-90,scale=.22]{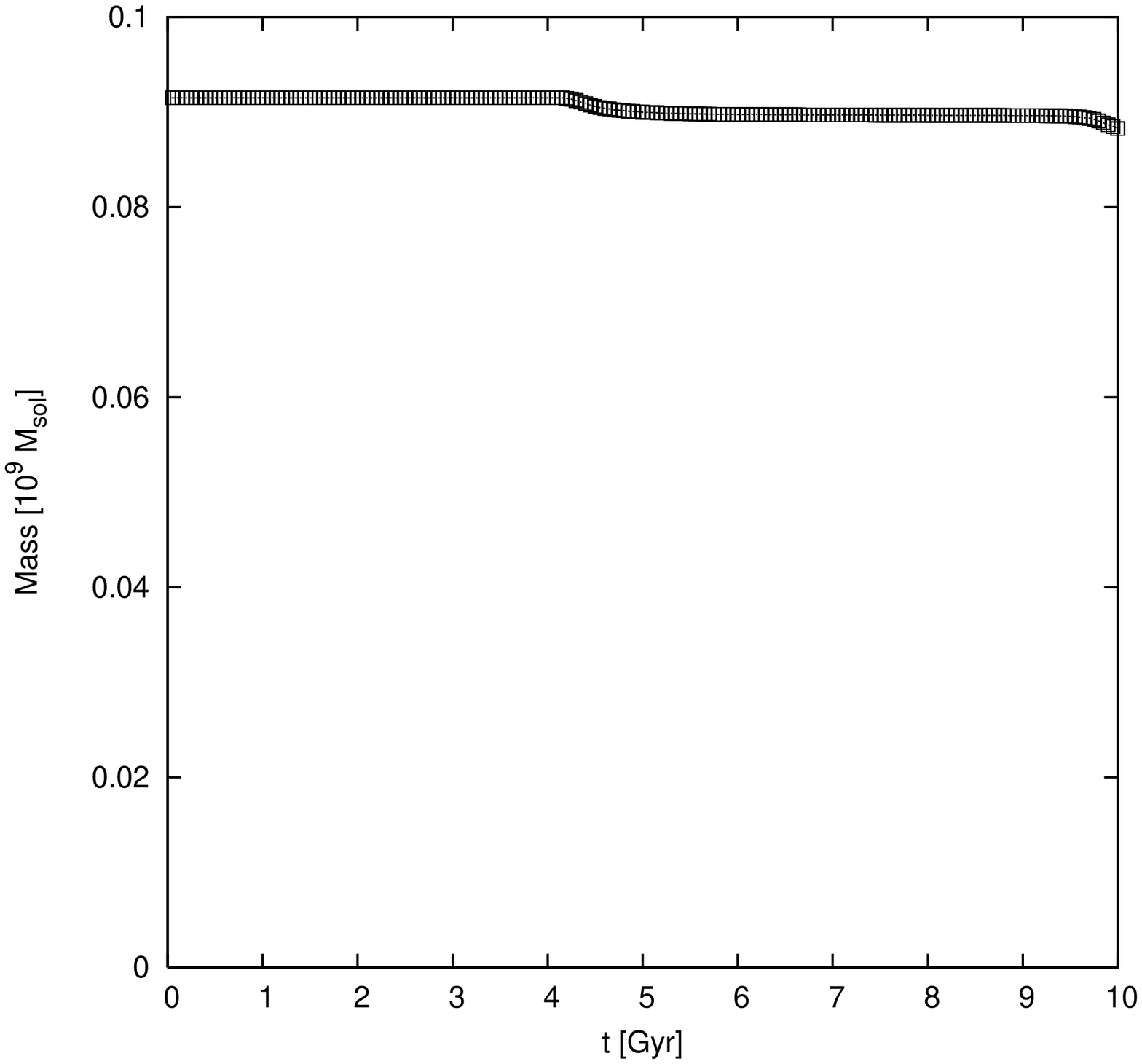}
    }
   }
  \vspace{7pt}
\centerline{\hbox{\hspace{0.0in}
\includegraphics[angle=-90,scale=.22]{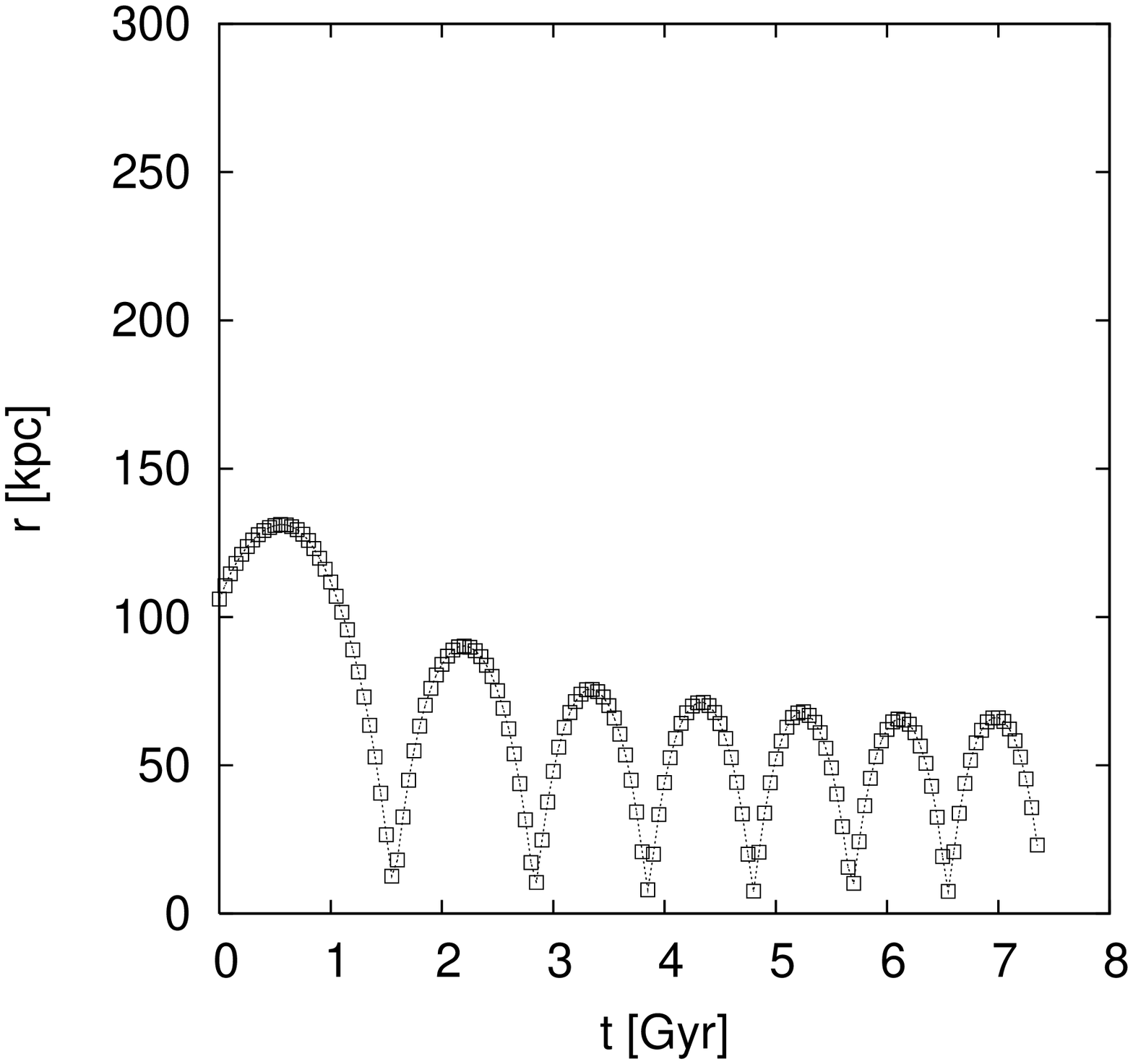}
     \hspace{0.1in}
\includegraphics[angle=-90,scale=.22]{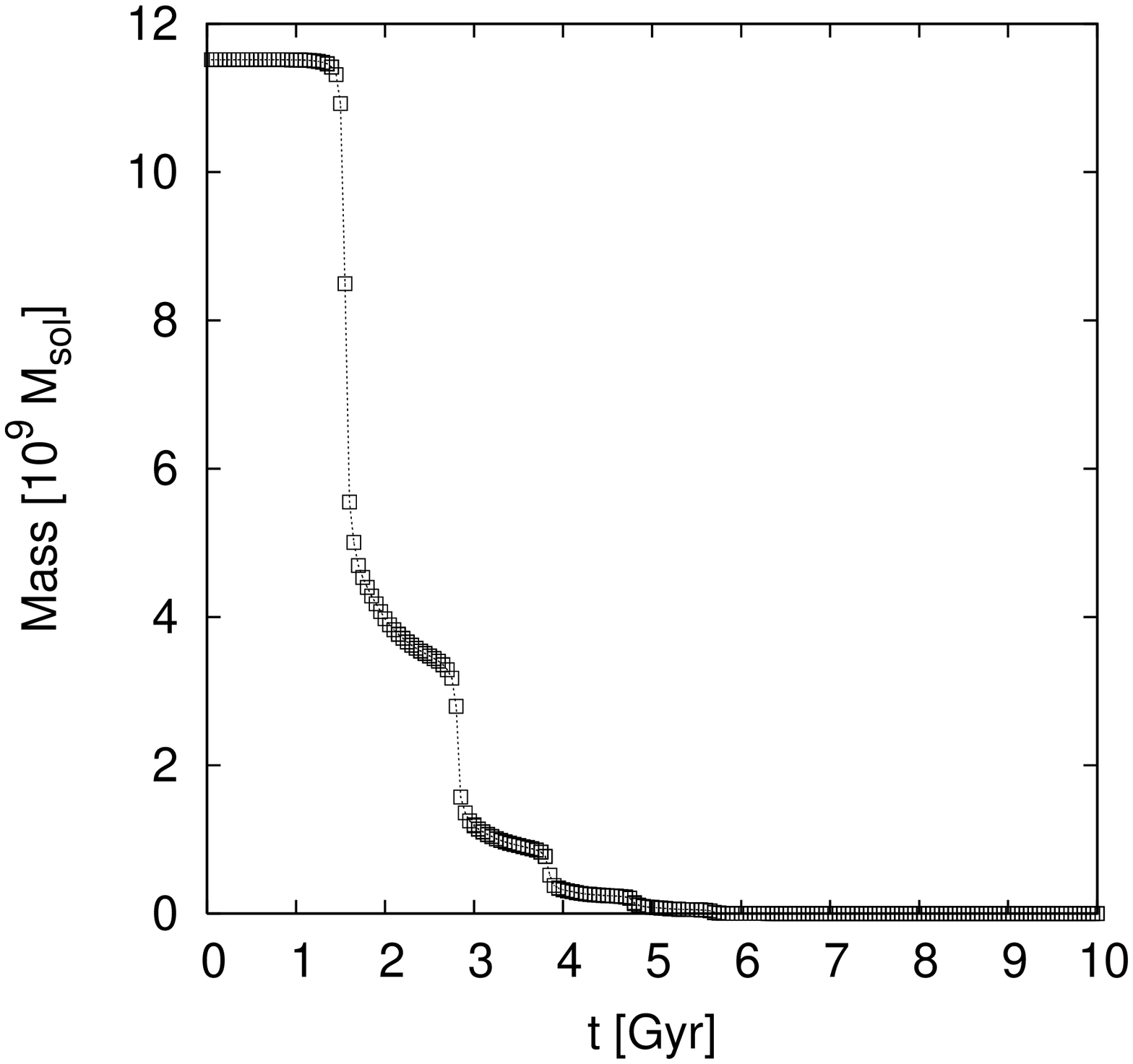}
     \hspace{0.1in}
\includegraphics[angle=-90,scale=.22]{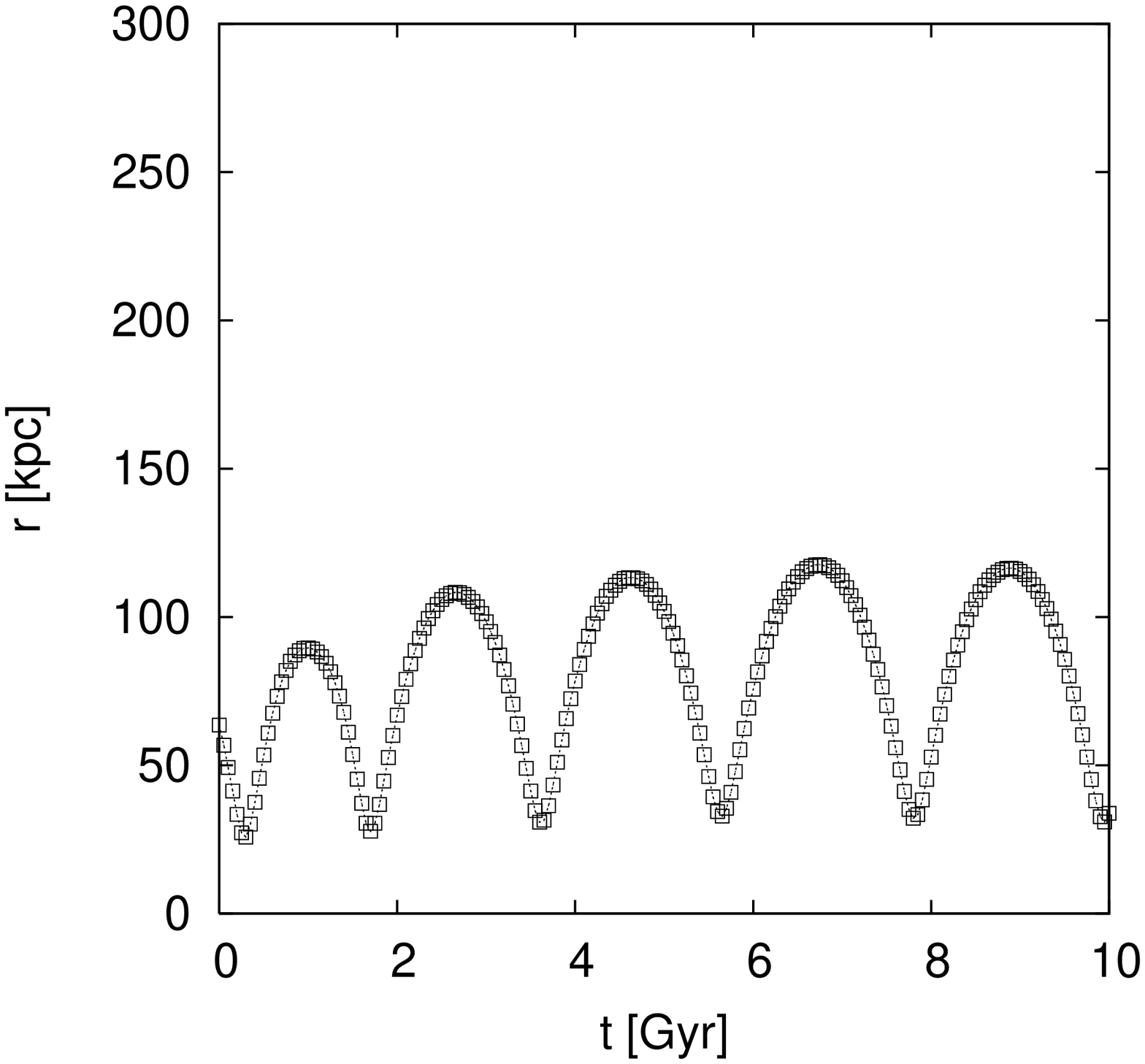}
     \hspace{0.1in}
\includegraphics[angle=-90,scale=.22]{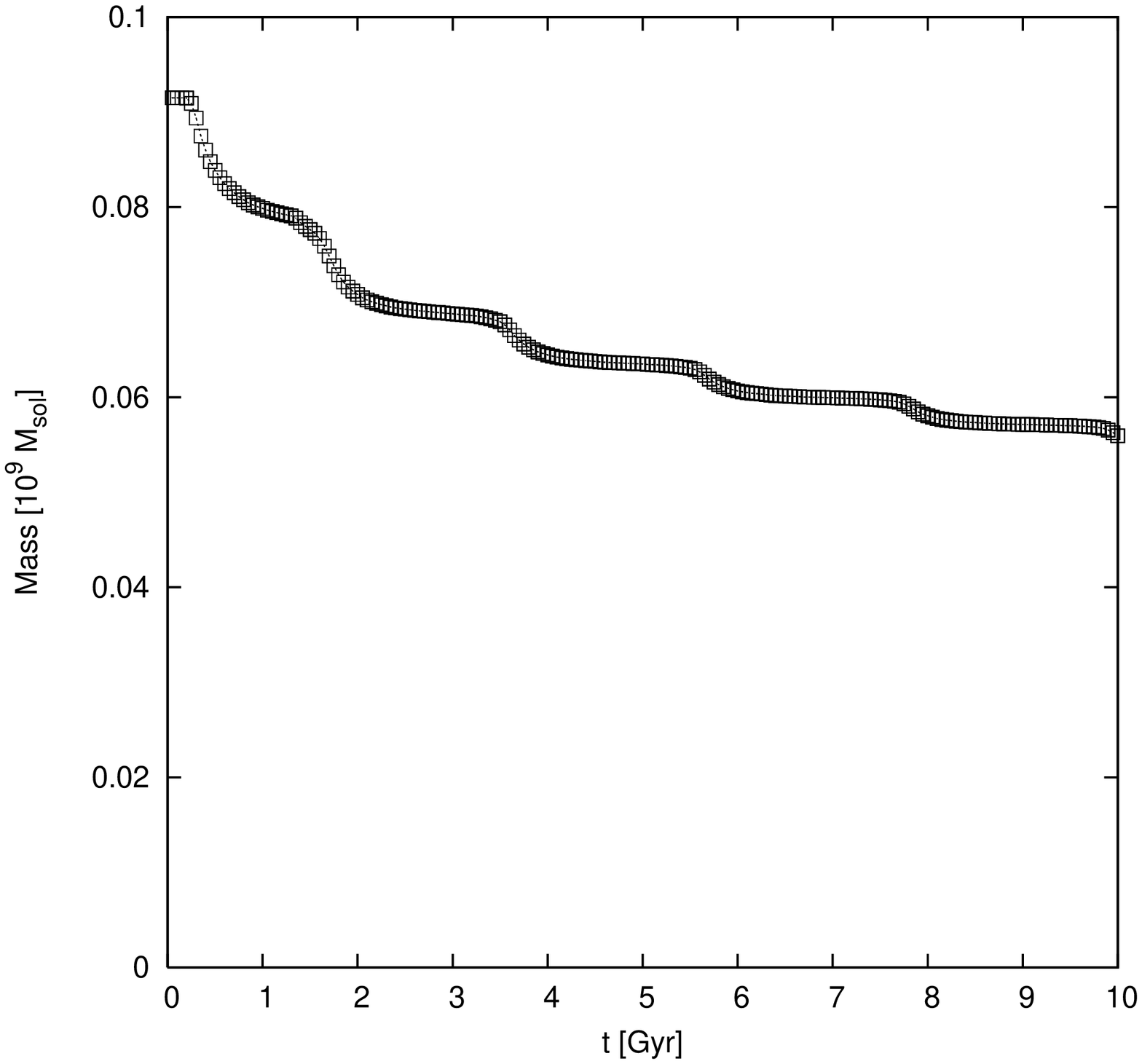}
    }
   }
  \vspace{7pt}
\centerline{\hbox{\hspace{0.0in}
\includegraphics[angle=-90,scale=.22]{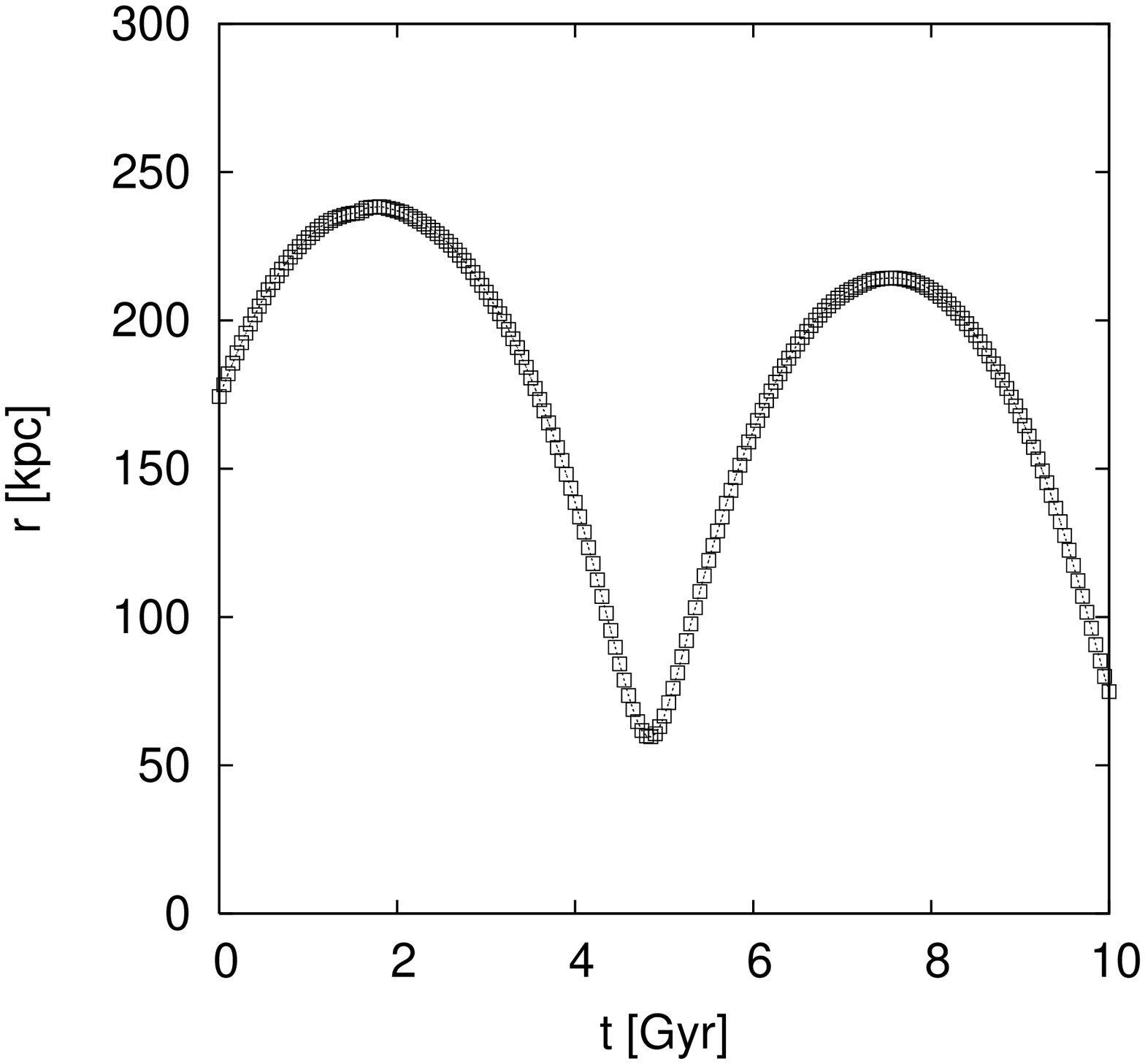}
     \hspace{0.1in}
\includegraphics[angle=-90,scale=.22]{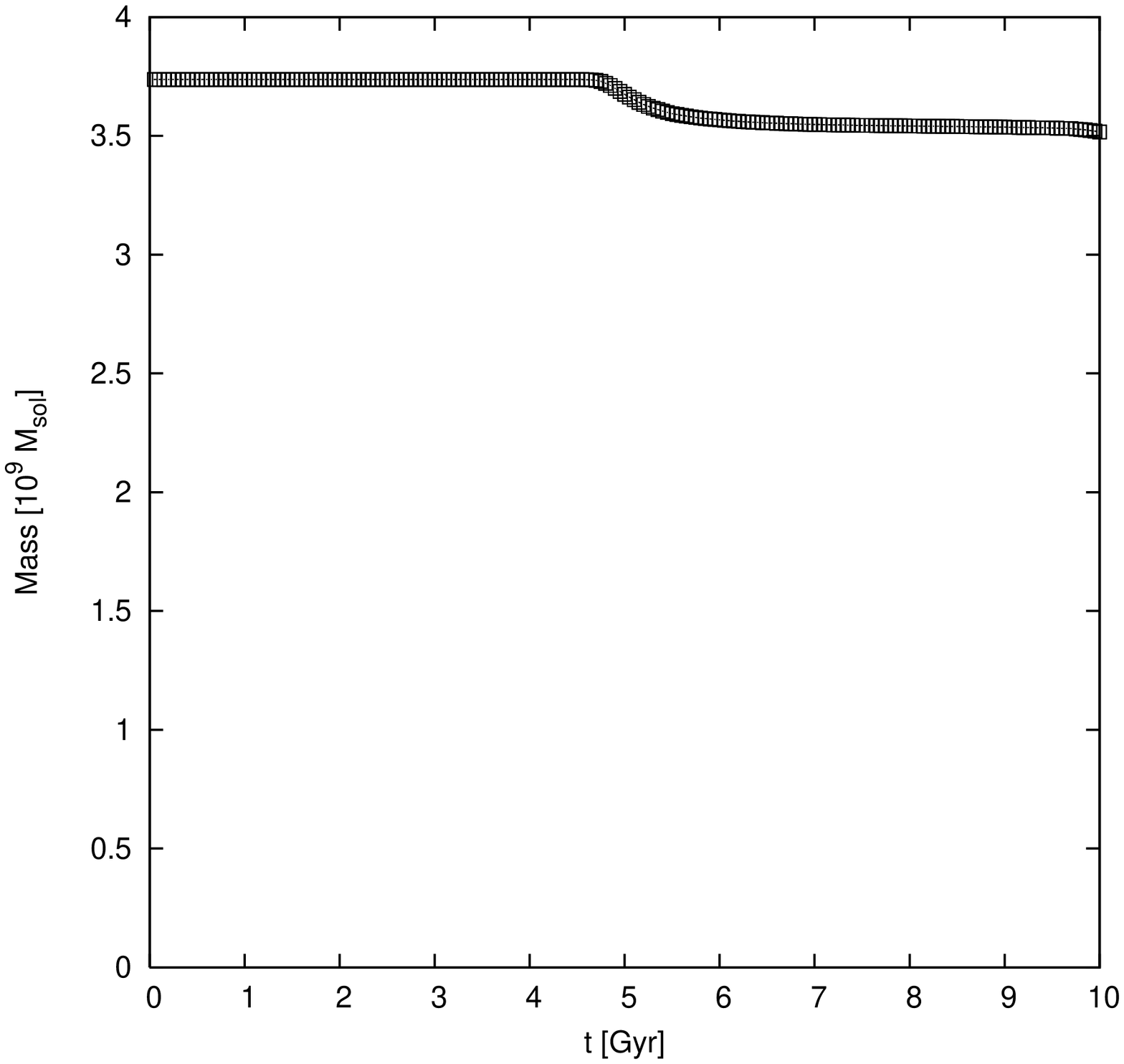}
     \hspace{0.1in}
\includegraphics[angle=-90,scale=.22]{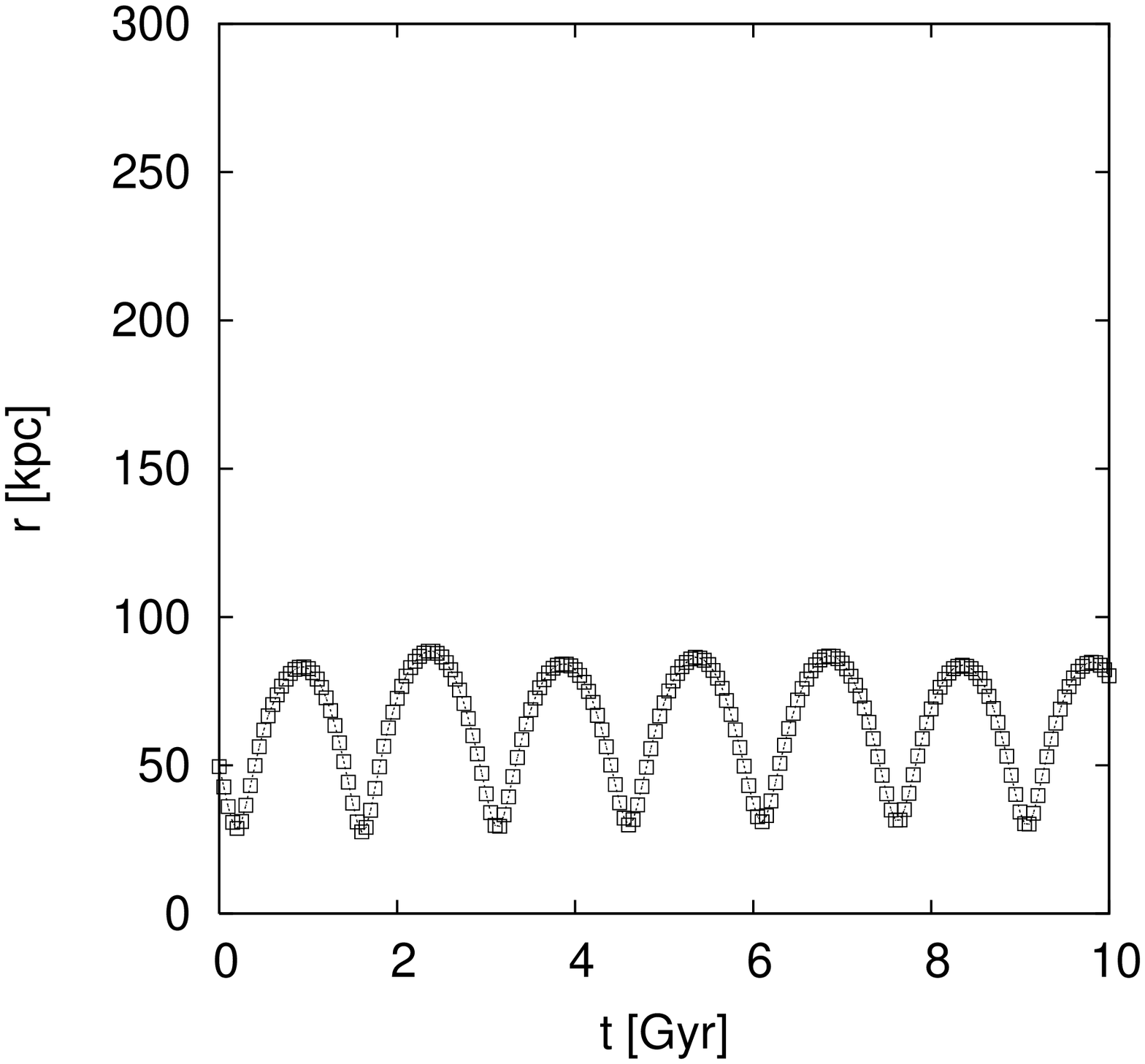}
     \hspace{0.1in}
\includegraphics[angle=-90,scale=.22]{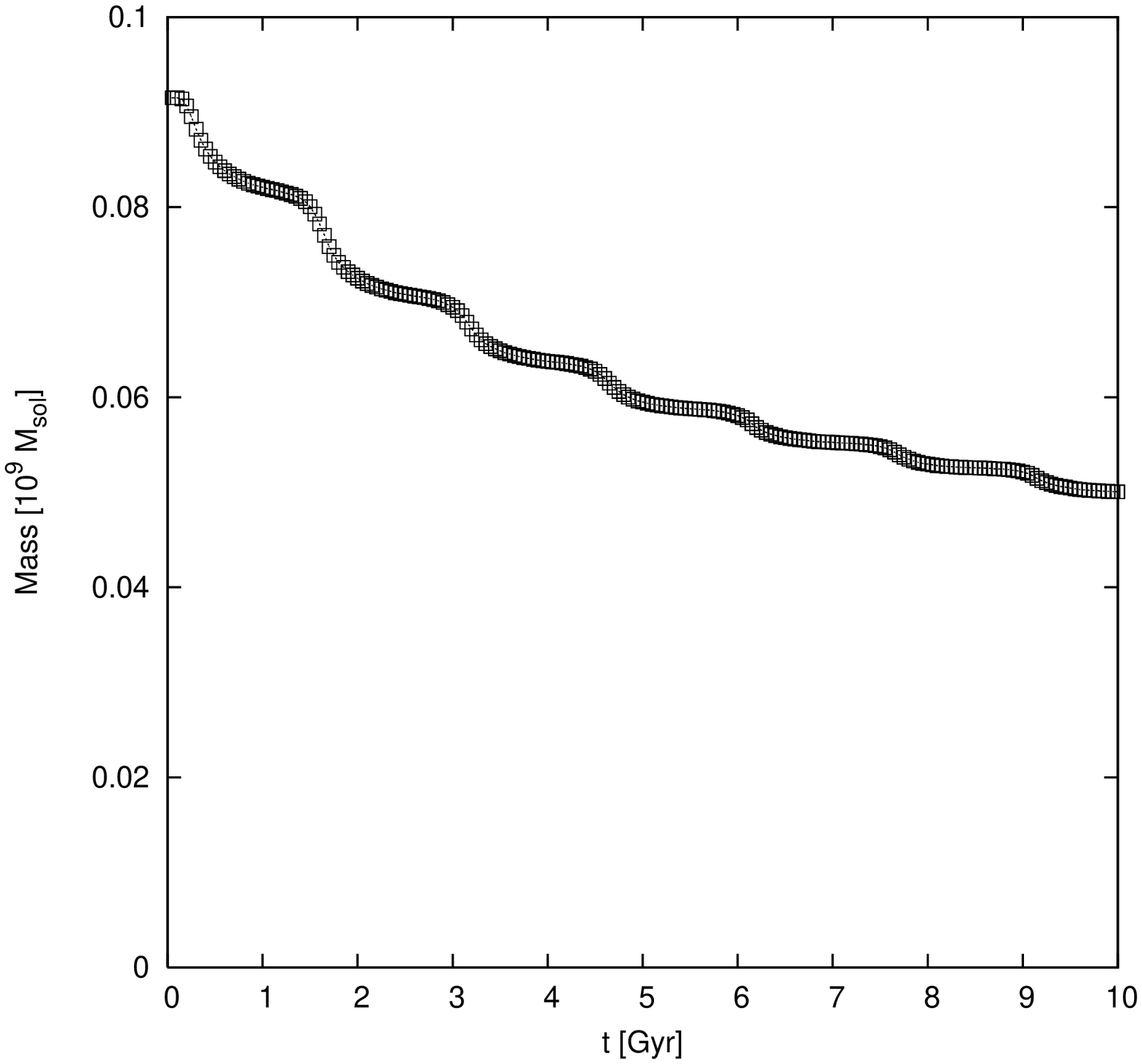}
    }
   }
   \vspace{7pt}
\centerline{\hbox{\hspace{0.0in}
\includegraphics[angle=-90,scale=.22]{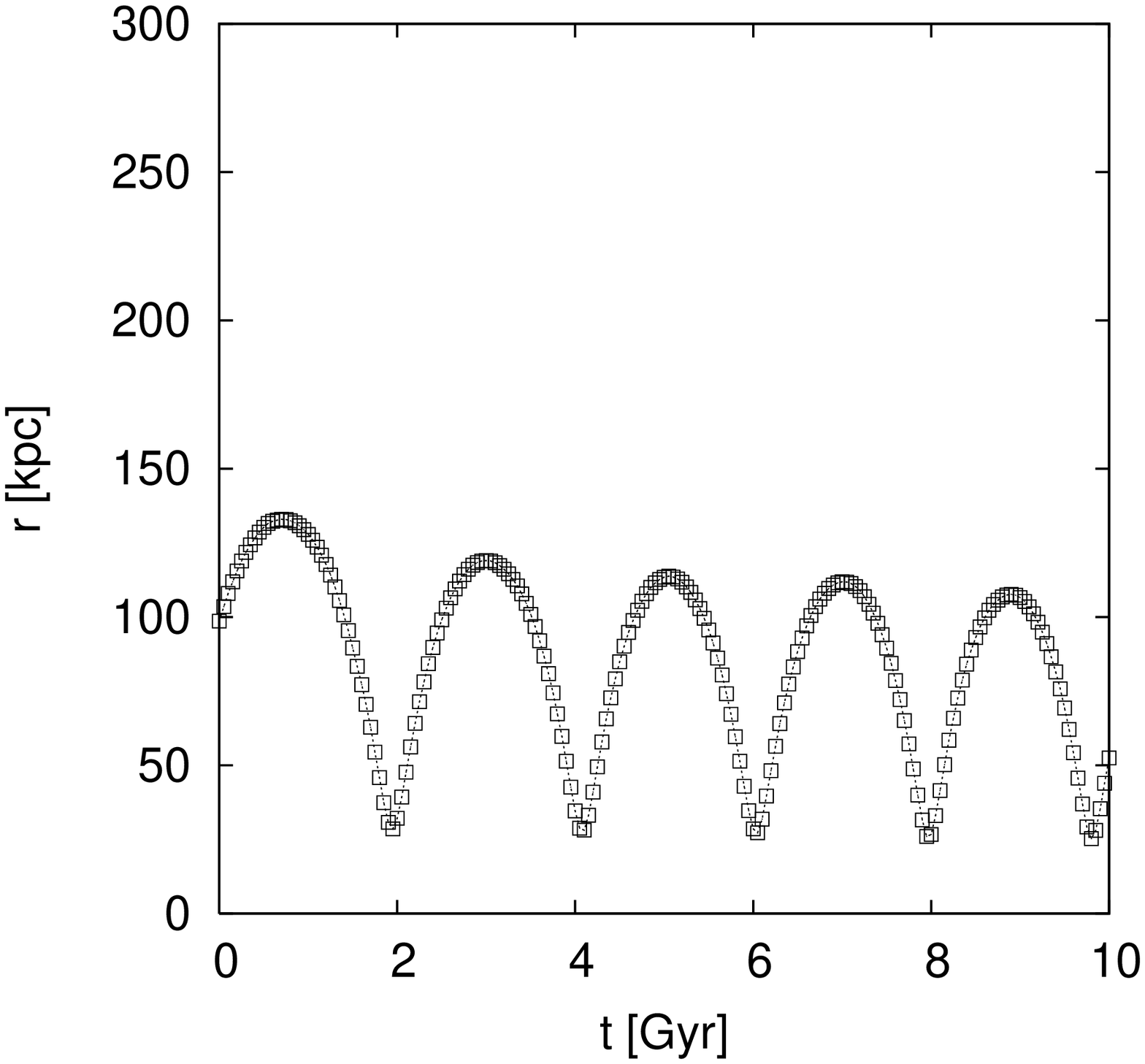}
     \hspace{0.1in}
\includegraphics[angle=-90,scale=.22]{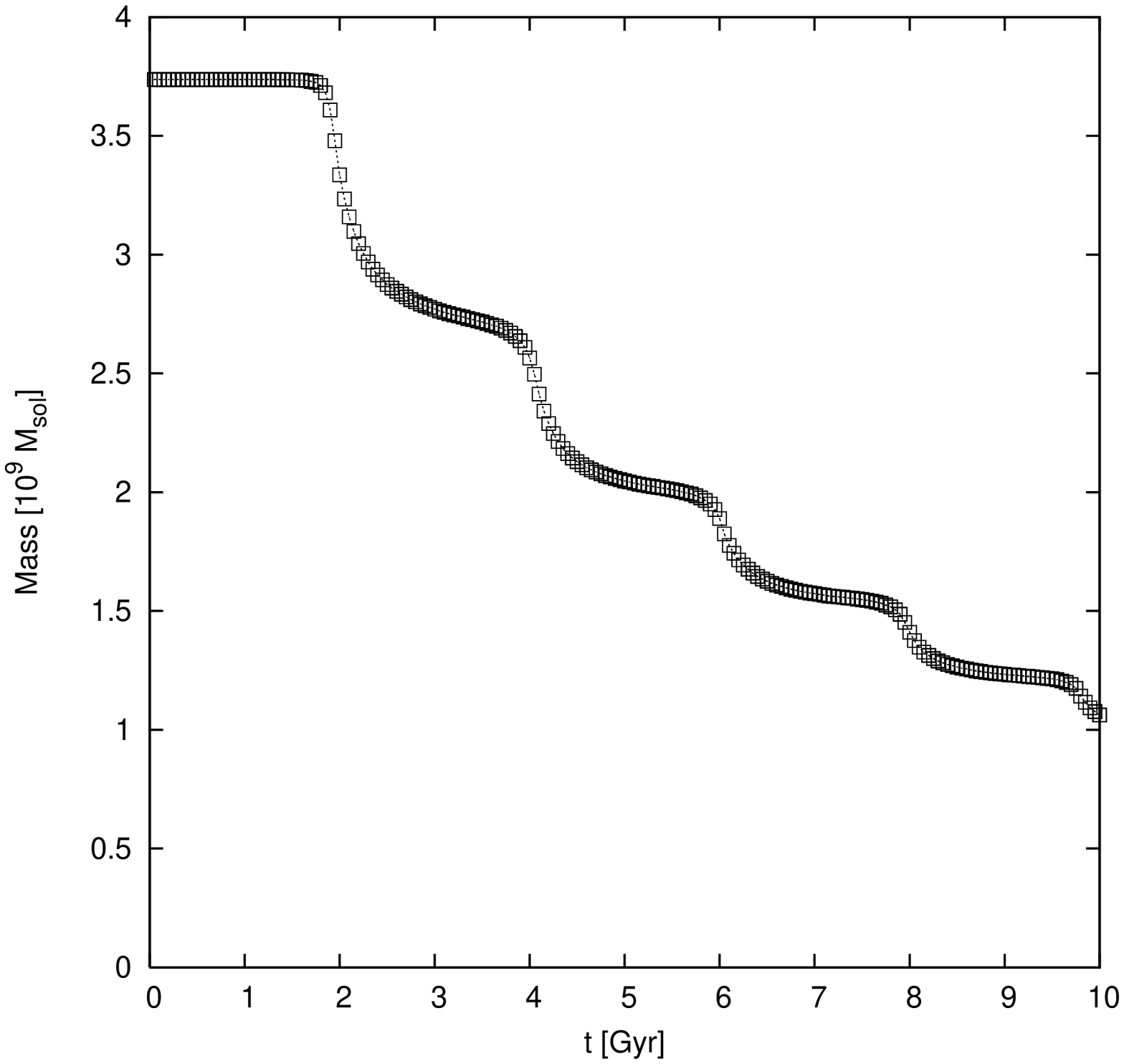}
     \hspace{0.1in}
\includegraphics[angle=-90,scale=.22]{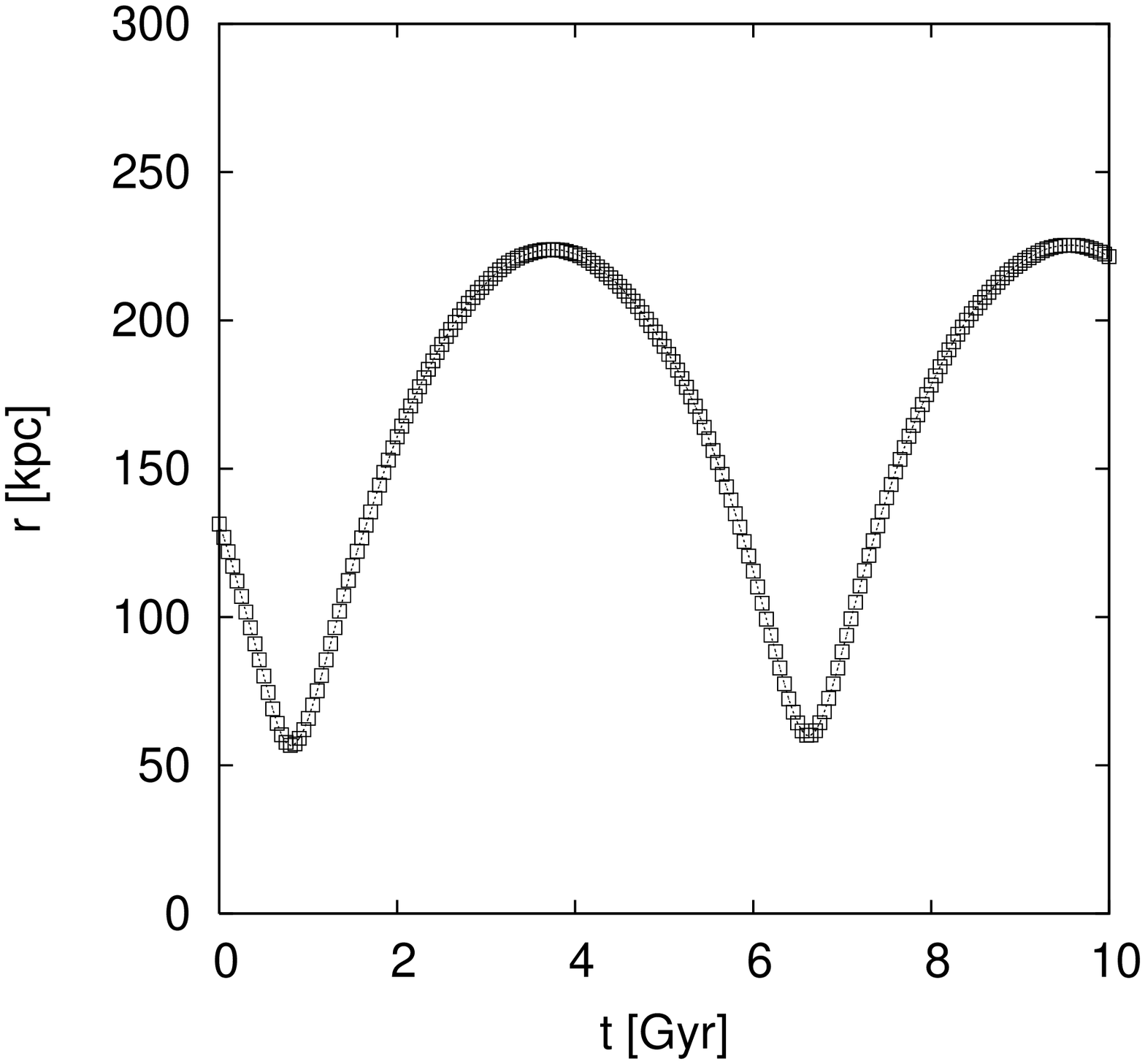}
     \hspace{0.1in}
\includegraphics[angle=-90,scale=.22]{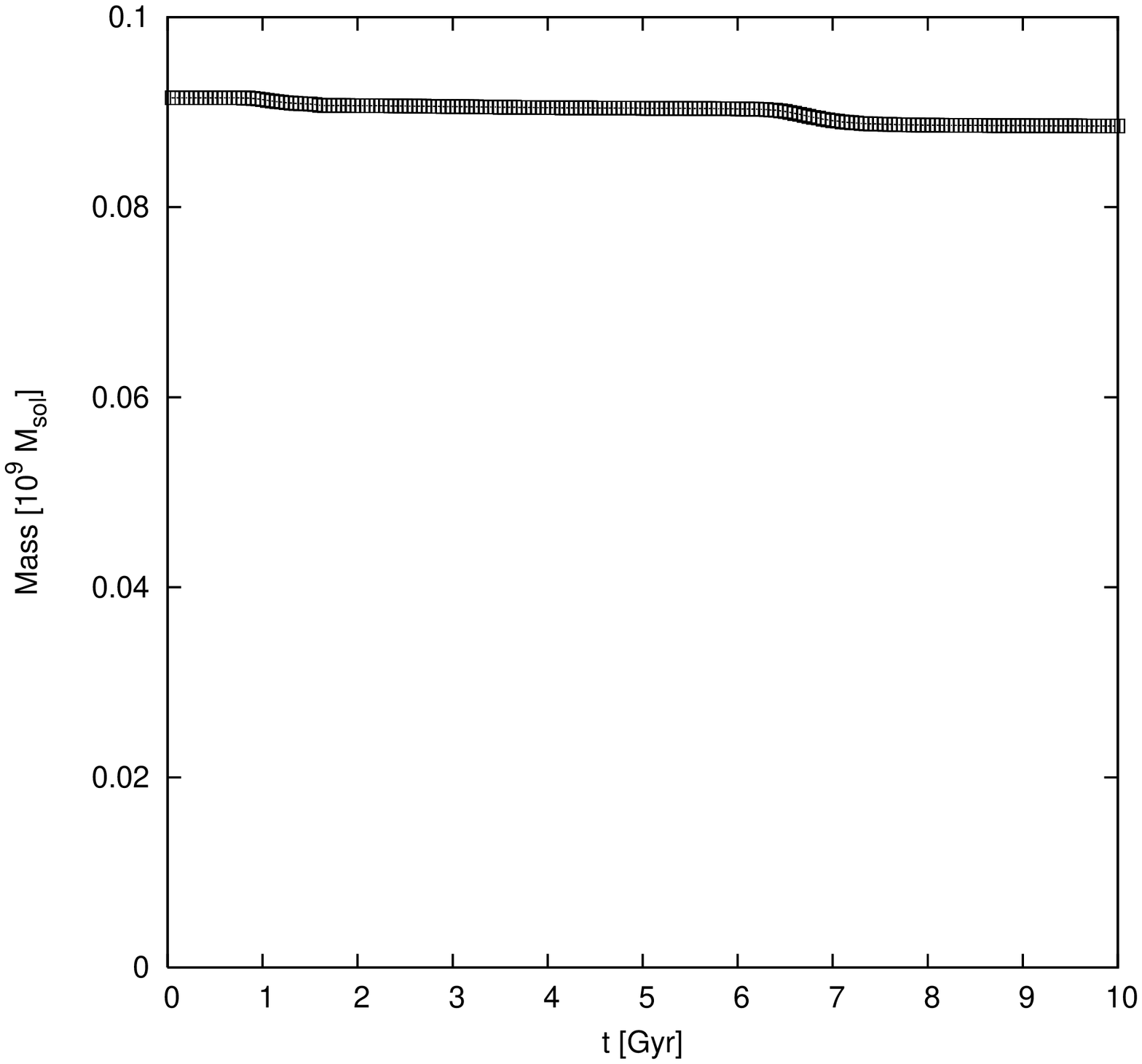}
    }
   }
    \vspace{7pt}
\centerline{\hbox{\hspace{0.0in}
\includegraphics[angle=-90,scale=.22]{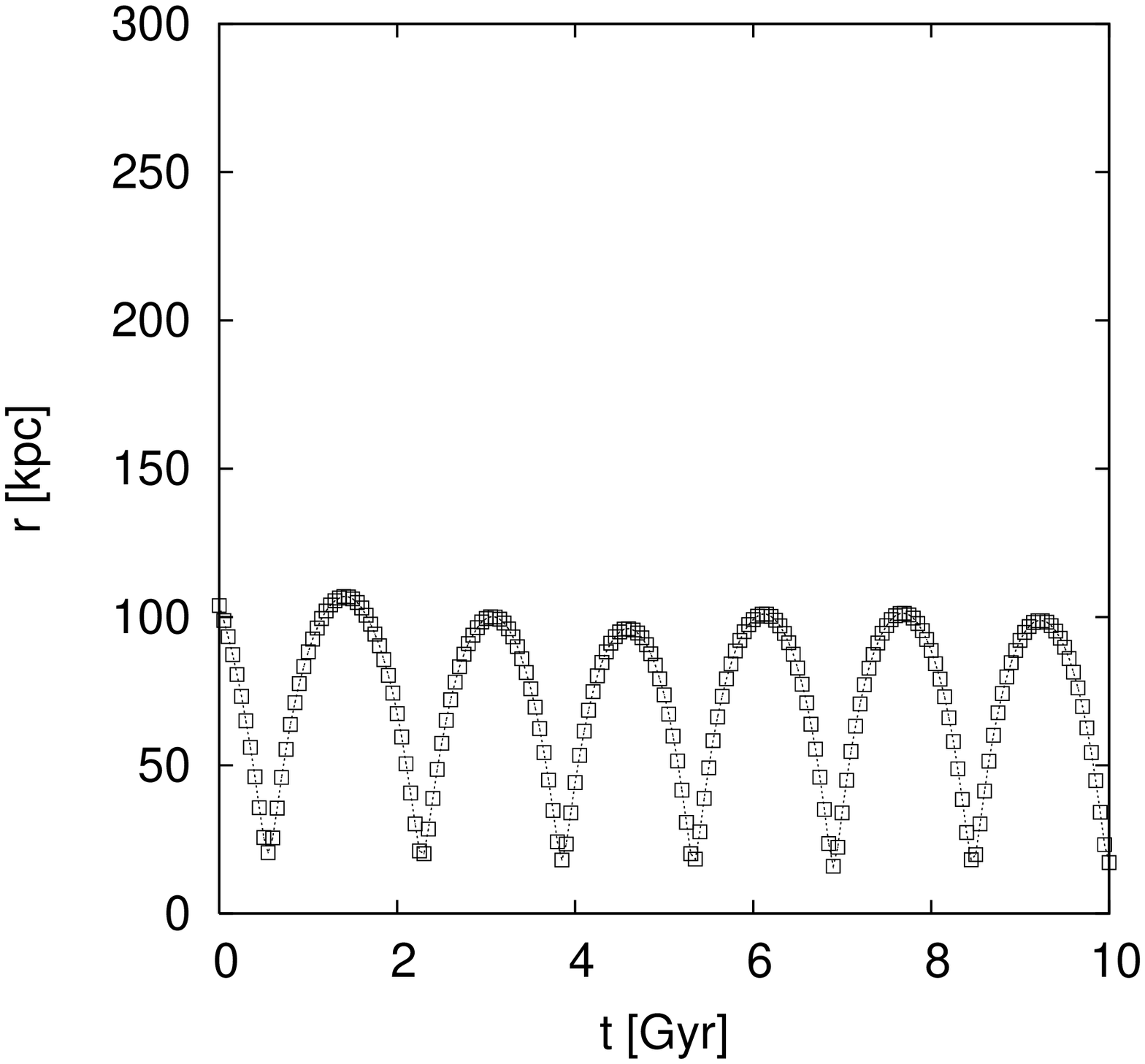}
      \hspace{0.1in}
\includegraphics[angle=-90,scale=.22]{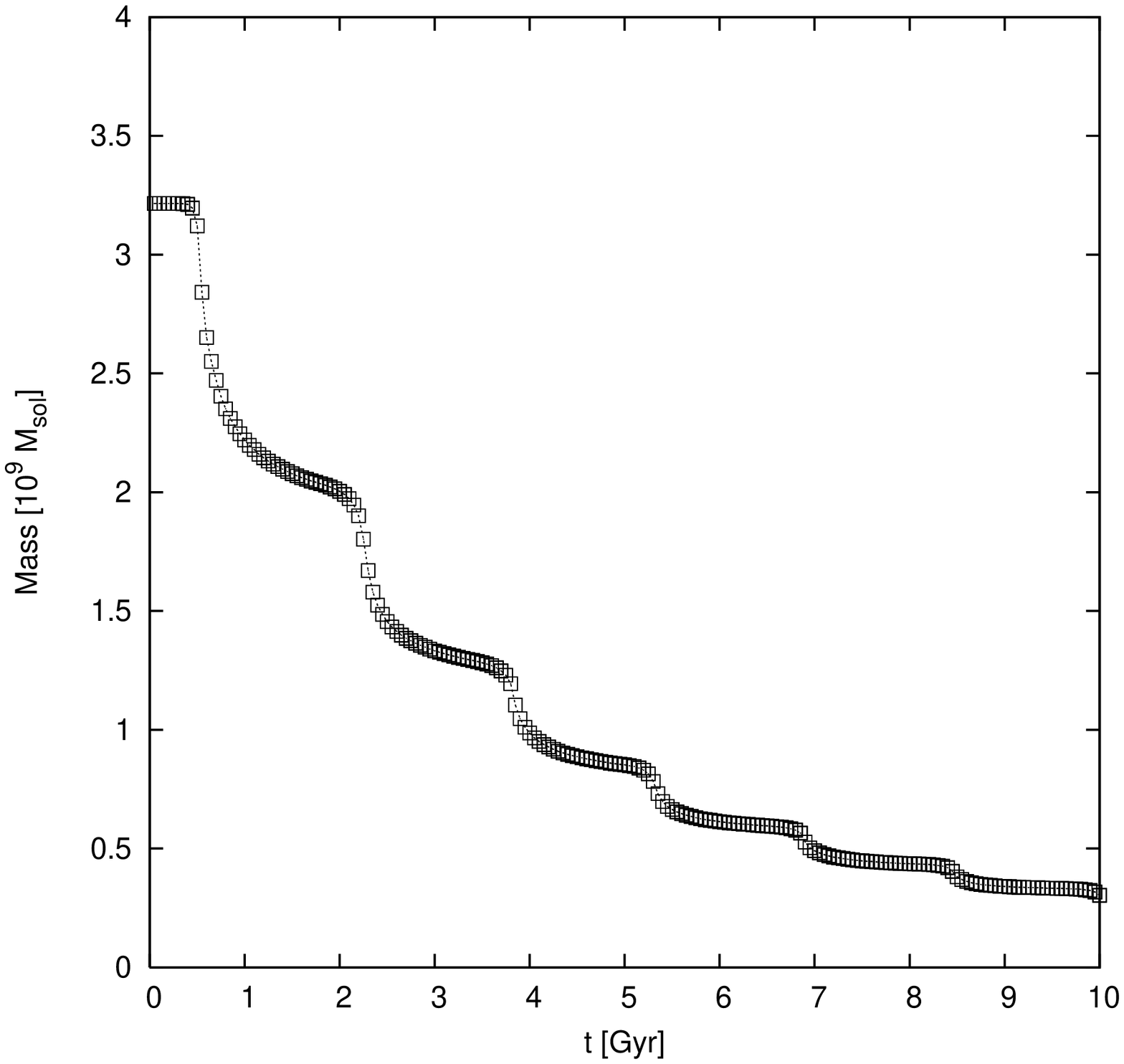}
      \hspace{0.1in}
\includegraphics[angle=-90,scale=.22]{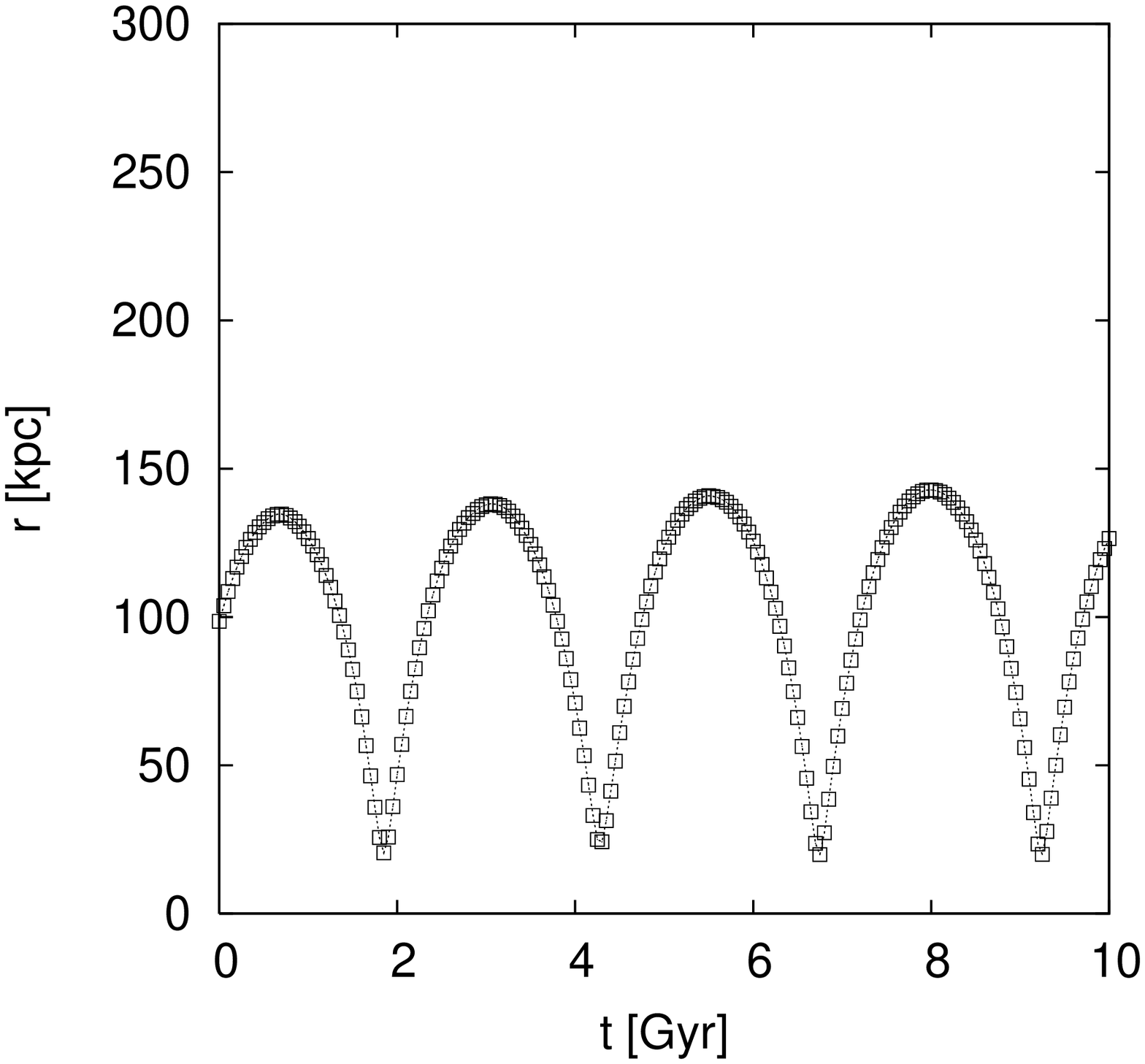}
      \hspace{0.1in}
\includegraphics[angle=-90,scale=.22]{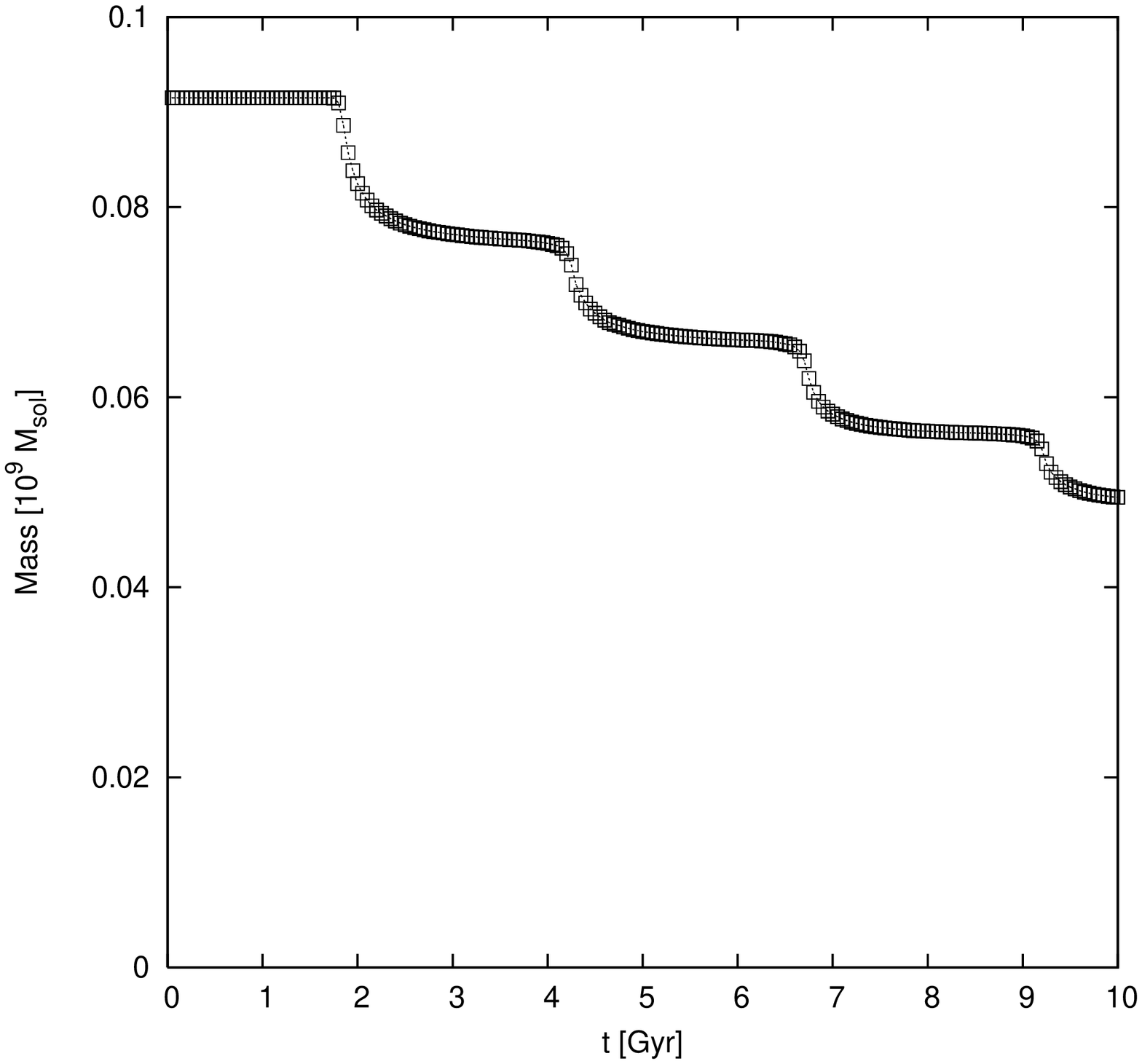}
     }
    }  
\vspace{7pt}
\caption{Decay and mass loss of individual satellites. The first 
two columns are for the five most massive satellites. The third and fourth ones are for five of the lightest satellites. 
First and third columns give the radius of the center of mass as a function of time. The figure demonstrates 
that dynamical friction does not play much of a role in evolution of the orbits.}
\label{ind_sats}
\end{figure}

\begin{figure}
\centering{\includegraphics[angle=-90,scale=.90]{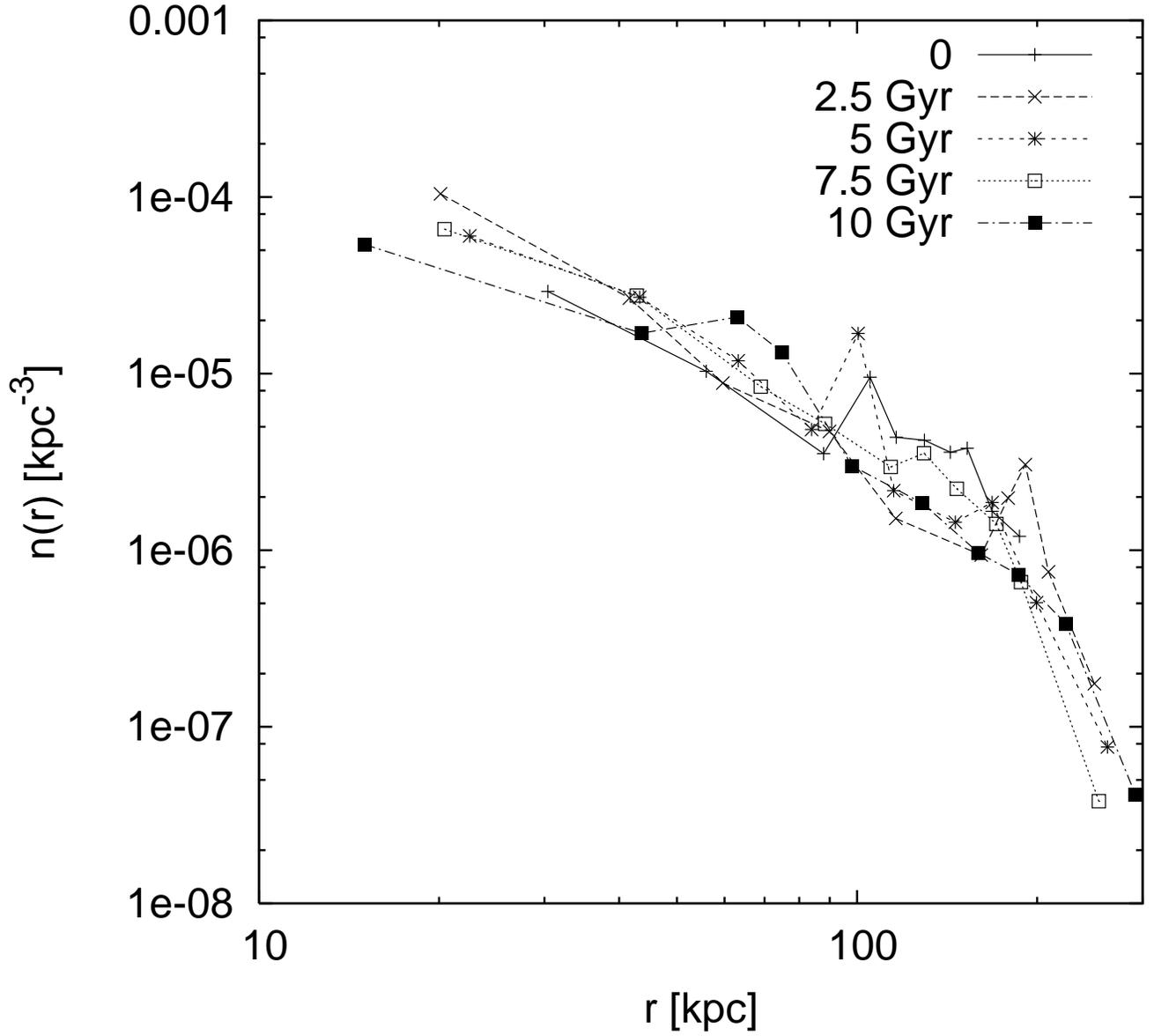}}
\caption{Evolution of the number density profile of satellites. The slope of the cusp remains constant over 10 Gyr.. }
\label{number_density_profile}
\end{figure}

\begin{figure}
 \vspace{5pt}
  \centerline{\hbox{ \hspace{0.0in}
    \includegraphics[angle=-90,scale=.35]{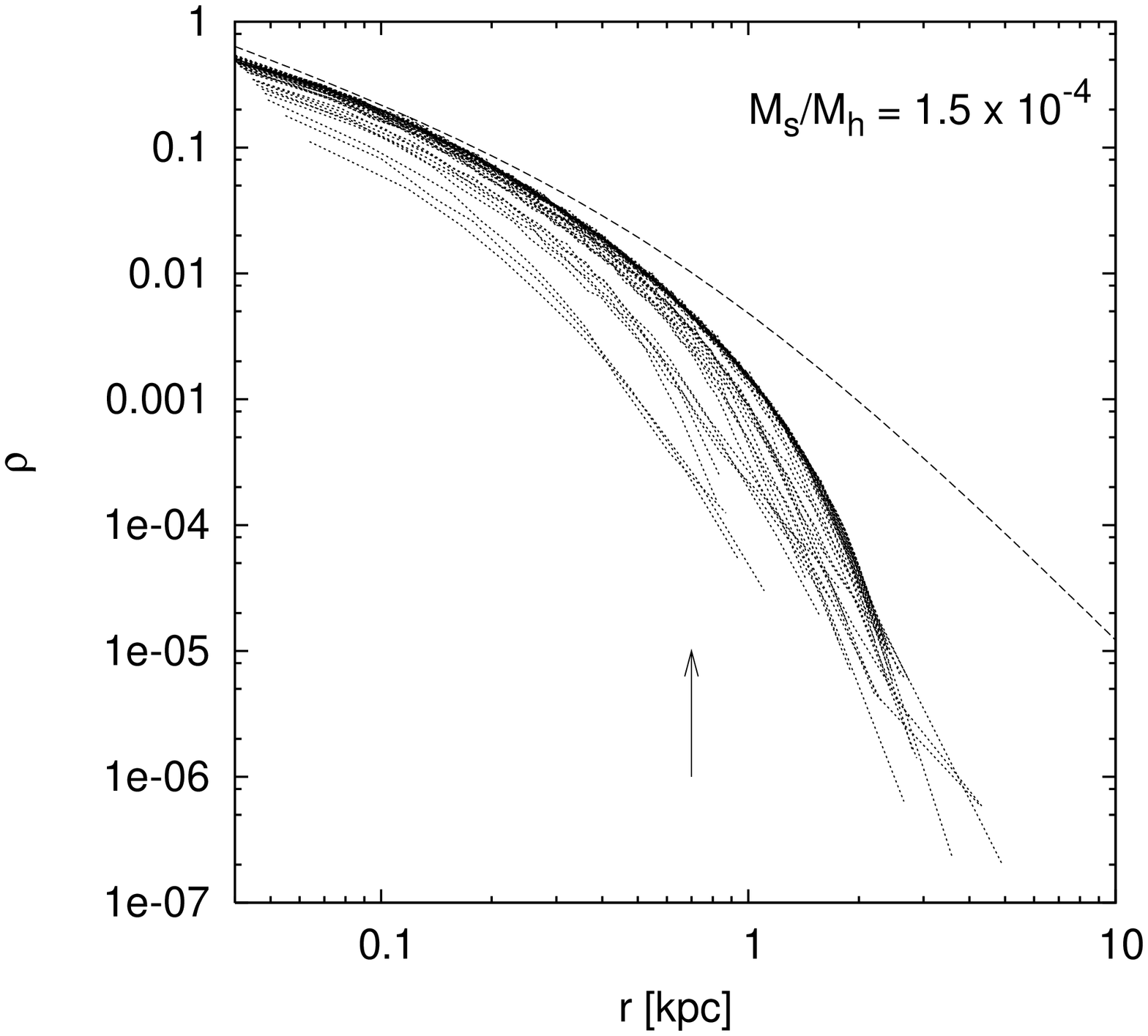}
    \hspace{0.1in}
    \includegraphics[angle=-90,scale=.35]{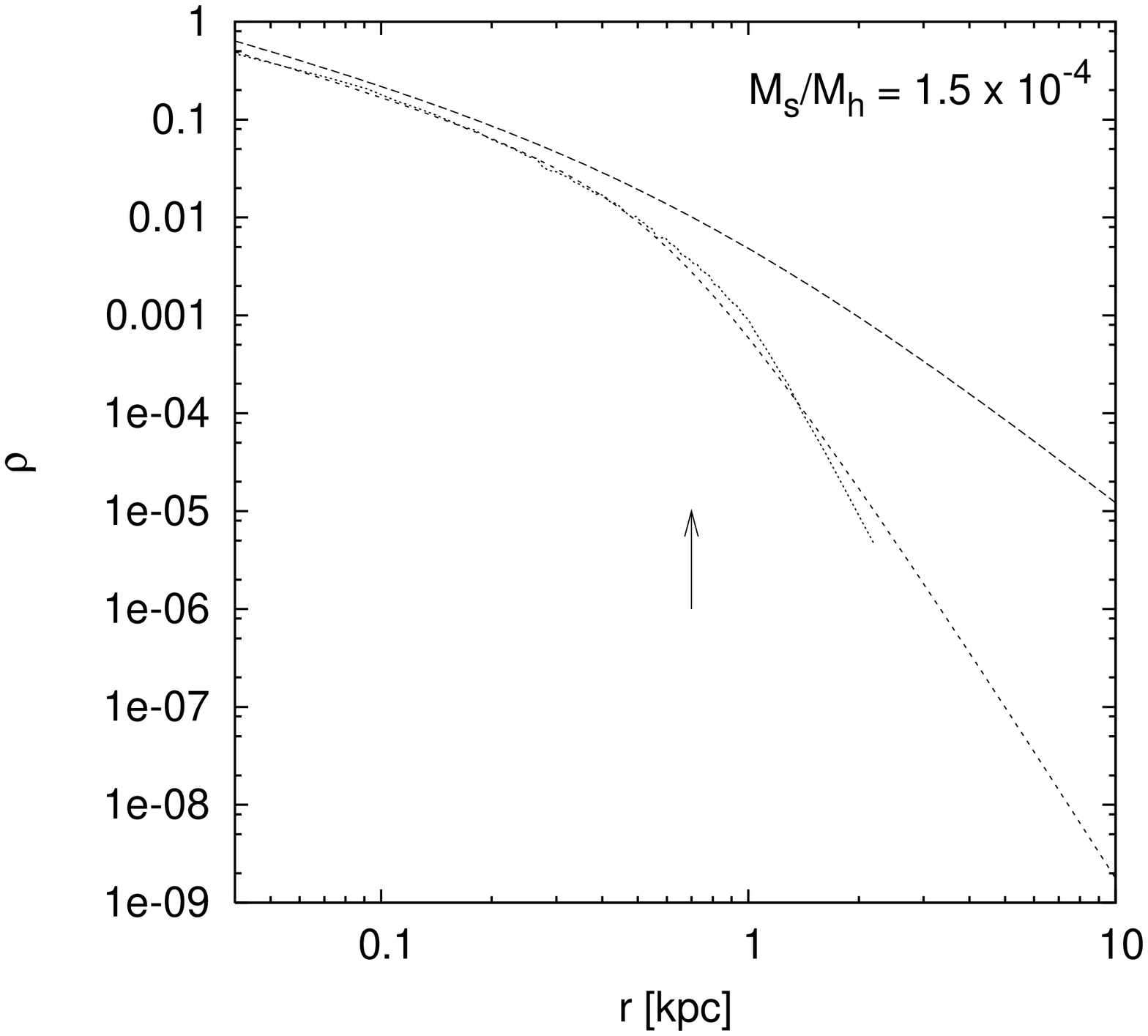}
    }
   }
  \vspace{5pt}
\centerline{\hbox{ \hspace{0.0in}
     \includegraphics[angle=-90,scale=.35]{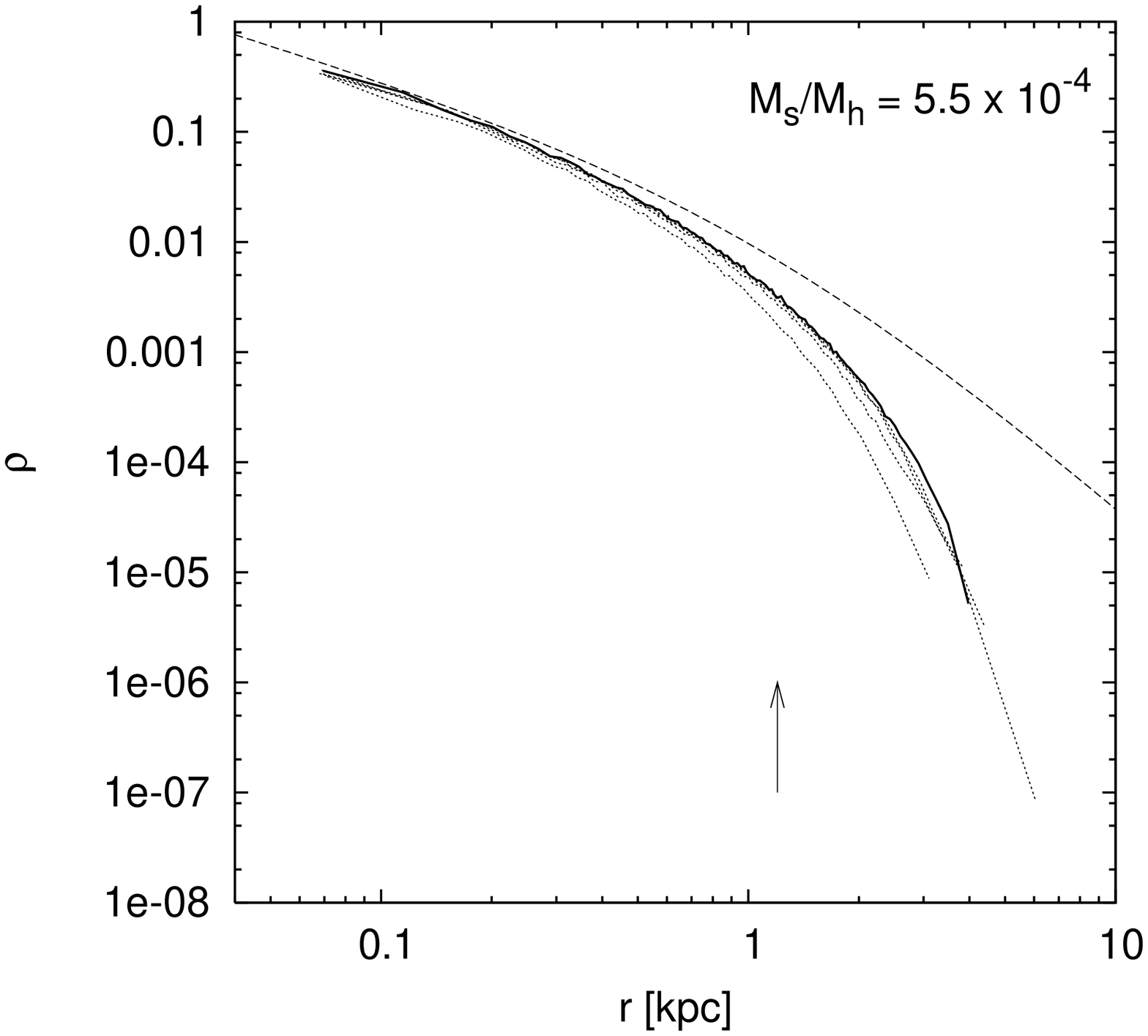}   
    \hspace{0.1in}
     \includegraphics[angle=-90,scale=.35]{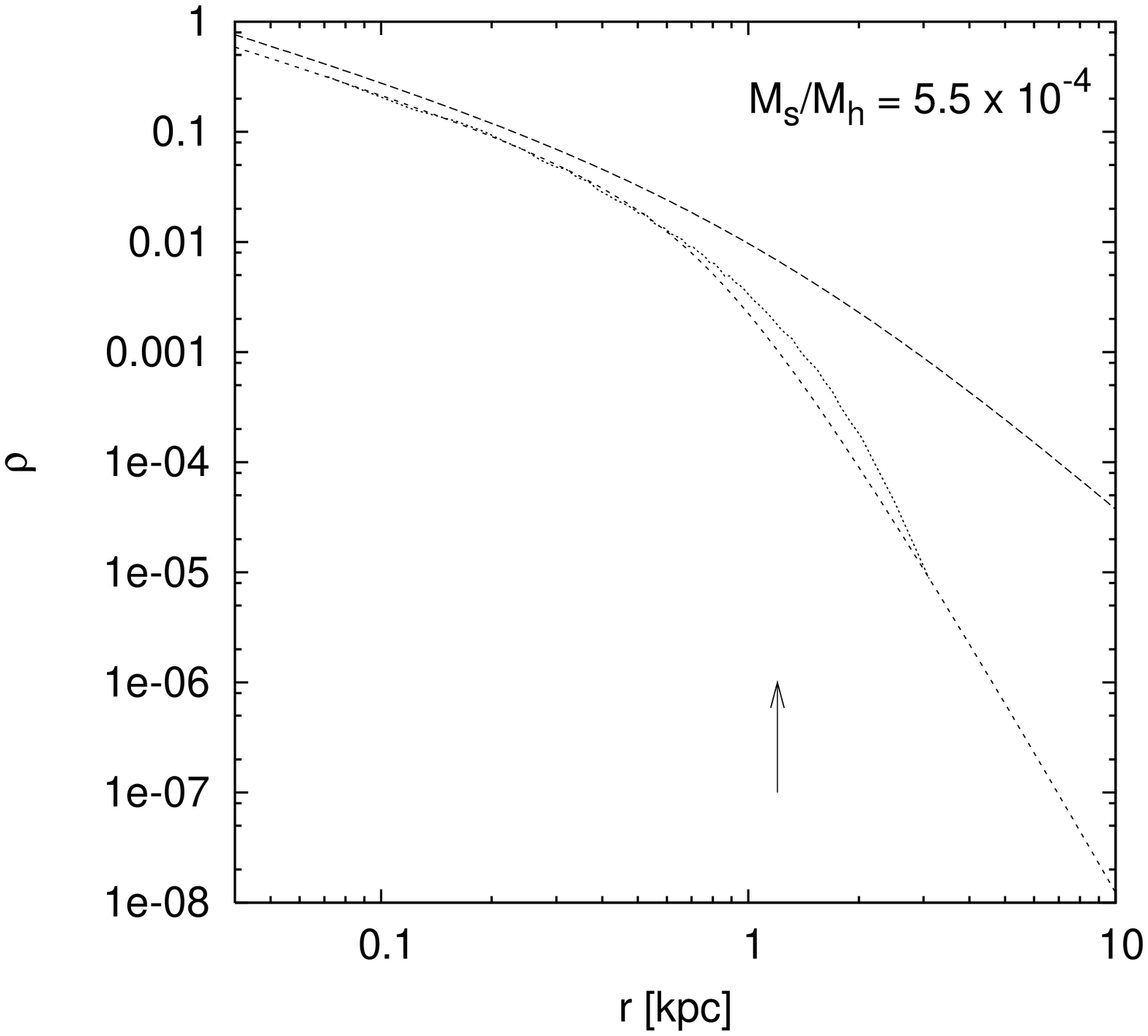}
    }
   }
  \vspace{5pt}
\centerline{\hbox{ \hspace{0.0in}
    \includegraphics[angle=-90,scale=.35]{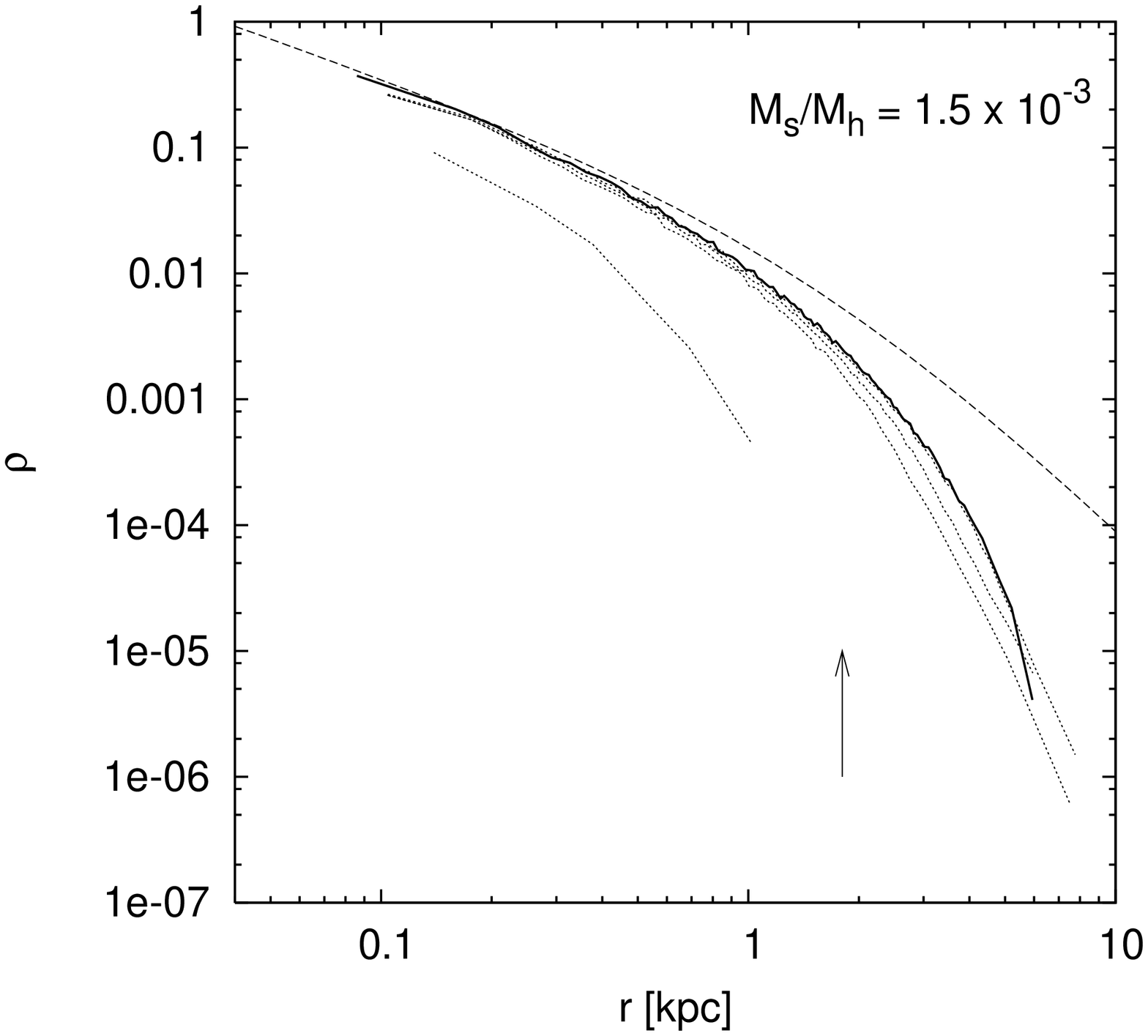}
    \hspace{0.1in}
    \includegraphics[angle=-90,scale=.35]{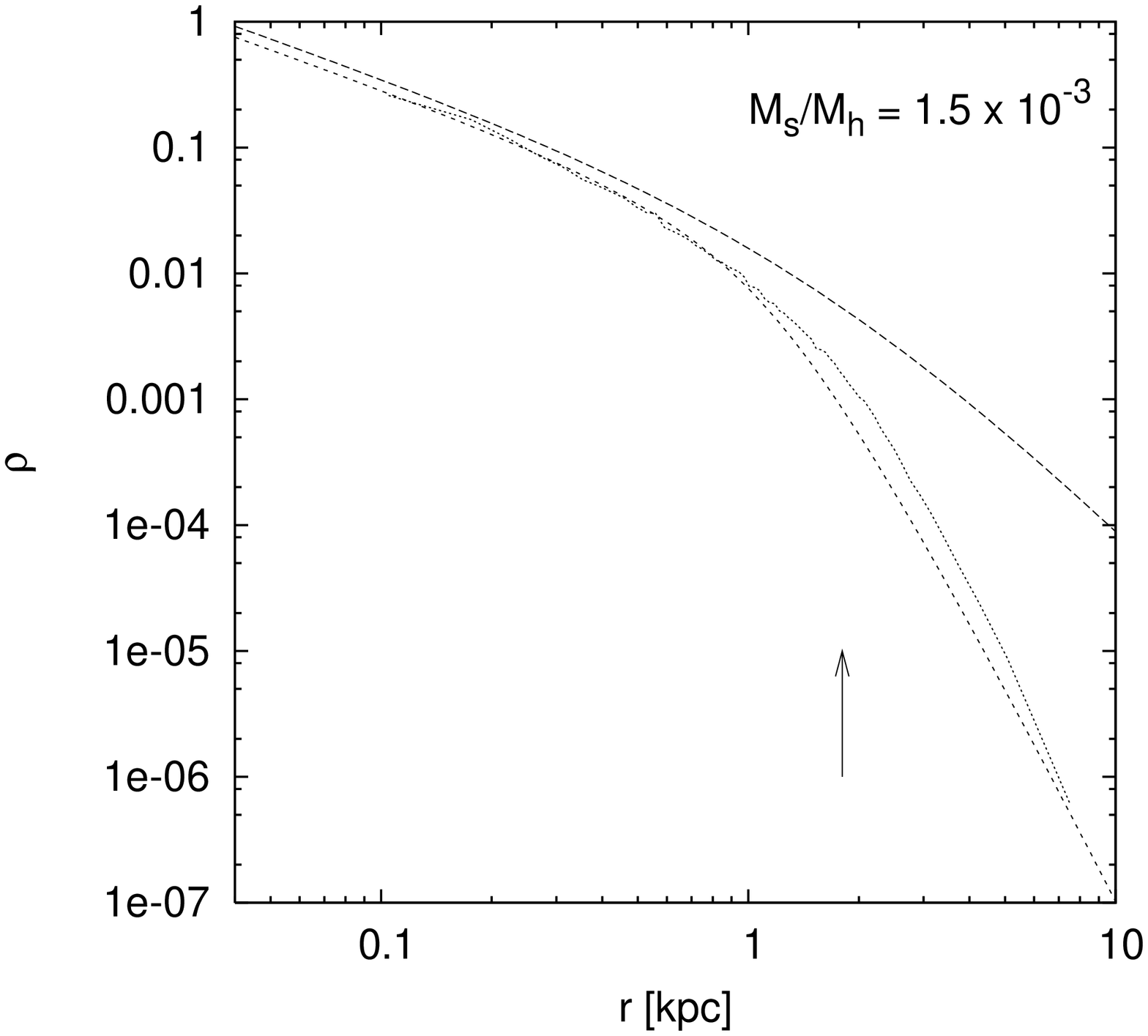}
    }
   }
 \vspace{5pt}
 \caption{Density profiles for three different satellite mass bins at 10 Gyr. Top : $M_{\rm sat}/M_{\rm halo} = 1.5 \times 10^{-4}$, middle :  $M_{\rm sat}/M_{\rm halo} = 5.5 \times 10^{-4}$ and bottom  $M_{\rm sat}/M_{\rm halo} = 1.5 \times 10^{-3}$. For the left plots, the thick line represents the unstripped density profile at the beginning of the simulation and the \emph{dashed} line 
is the pure NFW model used to model the satellite. Arrows indicate the position of the scale radius $r_s$. 
Satellites are initially truncated so that $r_{\rm{edge}}$ $\lesssim$ $r_{\rm{tidal}}$. The right plots show the fit given by equation (\ref{hayashi}) for a typical profile of one of our satellites in each mass bin. The \emph{short dashed} line is the best fit and the dotted line is the profile. As for the plots on the left, the \emph{dashed} line is the pure NFW model.    
 }\label{density_profiles.sats}
\end{figure}

\begin{figure}
\centering{\includegraphics[angle=-90,scale=.90]{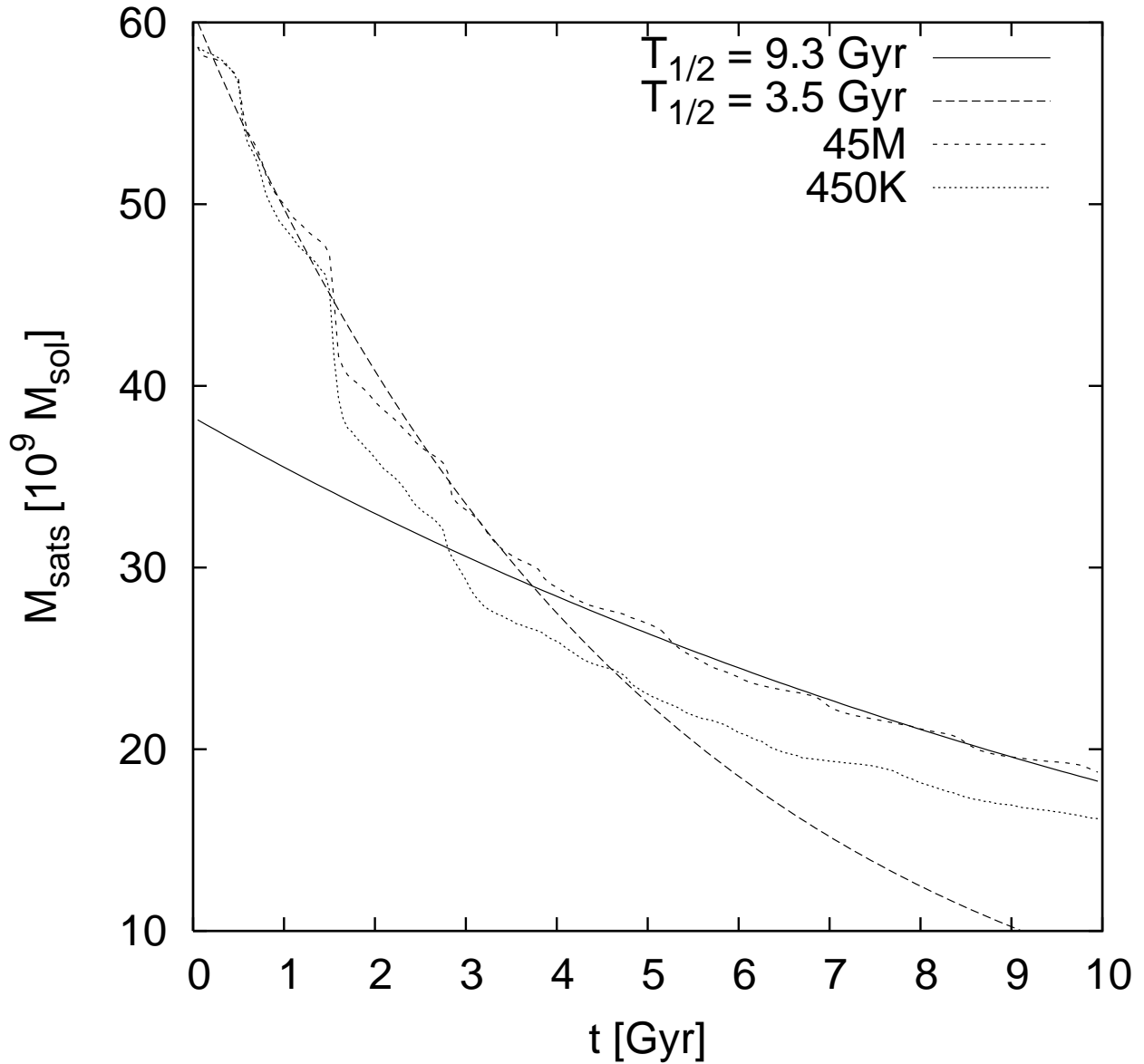}}
\caption{Evolution of the total mass bound in satellites as a function of time. 
The \emph{short-dashed} and \emph{dotted} lines are for the 45M and 450K 
simulation respectively. \emph{Long-dashed} and \emph{filled} lines are 
exponential decay fits to the 45M-particle run. The overall satellite decay is 
characterized by two distinct phases. The first one spans over the first 4 Gyr and corresponds to a very sharp mass loss with $t_{1/2}$ = 3.5 Gyr. For the second one, from 4 to 10 
Gyr, $t_{1/2}$ = 9.3 Gyr. 
}
\label{massbound}
\end{figure}

Though only seven satellites are completely destroyed by the end of the simulation most of the remaining satellites are 
stripped significantly. As shown in Figure \ref{massbound}, the time dependence of the total mass bound 
in satellites can be described by 
two distinct phases. During the first 4 Gyr, the satellites lose approximately half their mass. The time dependence
of the mass in the system is well-fit by an exponential decay : 
\begin{equation}
M_{\rm sat} \propto e^{-0.693t/t_{1/2}}
\end{equation}
where $t_{1/2} = 3.5$ Gyr. 
The second phase, from 4 to 10
Gyr, is quiet with $t_{1/2}$ = 9.3 Gyr. During their first complete orbit,
satellites are severely stripped and lose their outer mass layers as their
size becomes limited by the tidal radius at the pericentric passage. For the last 6 Gyr, 
the satellites radial extent is well constrained by the tidal field
of Andromeda and a smaller mass loss occurs at each pericenter passage. Clearly these numbers are 
affected by our initial conditions and especially by the constraints on the satellites size. 
The position at which we compute the tidal radius 
of our satellites (50 kpc) is somewhat arbitrary and a different value could lead to different results. 
As shown in Figure \ref{apoperi}, the distribution of the pericenter radii peaks at 50 kpc or so. Computing the tidal radius of the satellites at this position gives a good approximation of the radial extent of satellites who have completed several orbits around the galaxy (i.e. suppose to be in equilibrium with the host galaxy). Nevertheless, 
decreasing the radius to 30 - 40 kpc means increasing the concentration of the satellites to unrealistic values. 
It appears that the tidal field of the underlying halo potential is very strong and  satellites 
falling inside 30-40 kpc must have a very high concentration 
to survive multiple pericenter passages. On the other hand, computing the tidal radius at say, 100 kpc and constraining 
the satellites to have a size less than $r_{\rm{tidal}}$ means that the satellites will reach the disk galaxy with most of 
their mass stripped. We think that a value of 50 kpc (4 $r_s$) is a reasonable choice.

\subsection{Absence of Holmberg Effect}
\begin{figure}
\centering{\includegraphics[angle=-90,scale=0.8]{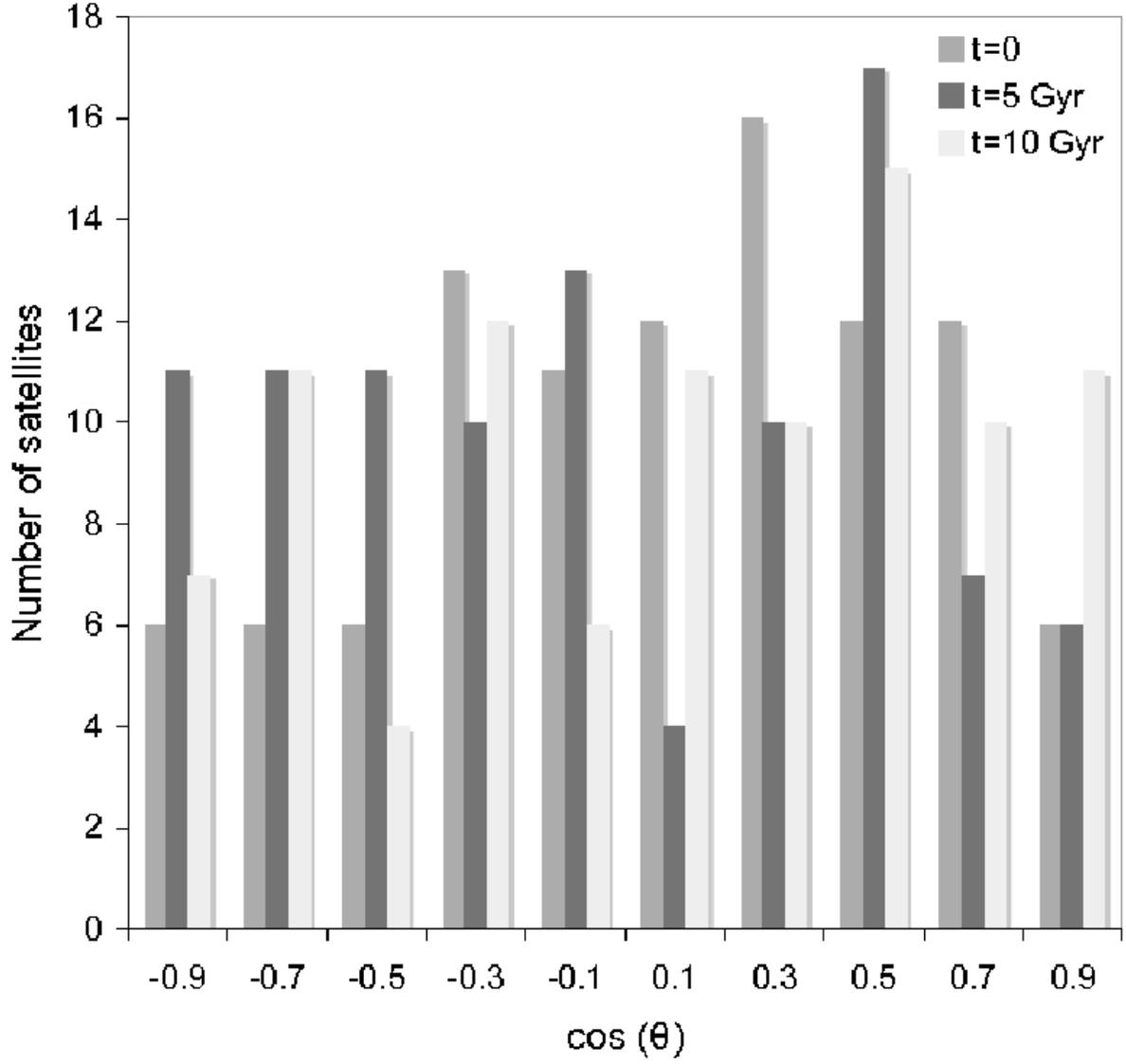}}
\caption{Number of satellites as a function of cos($\theta$). Large fluctuations are initially 
 present because of Poisson noise ($\sqrt(10) = 3$)}
\label{latitude}
\end{figure}

\citet{hol69} showed that there is a tendency for satellite galaxies to congregate near the poles of the spiral host
galaxy. His observations were then confirmed by \citet{zar97a} who showed that satellites located at large projected radii 
of isolated disk galaxies are aligned preferentially along the disk minor axis. 
However, recent observations by \citet{bra05} on a sample of SDSS galaxies showed that satellites 
are preferentially aligned with the major axis of the galaxy. This result contradicts Holmberg's previous observations. 
In this paper,  we examine if either of these conclusions are detected 
in our sample of evolved satellites and if a strong dynamical interaction between the disk and satellites 
at low latitudes could explain the Holmberg Effect. If dynamical friction is more 
important at lower latitudes, one would expect that satellites on almost 
coplanar orbits to sink on shorter timescales than 
satellites on polar orbits. That would lead to a deficit of satellites at low latitudes. 
In Figure \ref{latitude}, we show the distribution 
of satellites as a function of $\cos\theta$. The results are consistent with a uniform distribution in $\cos\theta$, 
that is a spatially isotropic distribution. In other words, we do not detect any Holmberg or anti-Holmberg effect in our simulation. We conclude 
that the anisotropic distribution of satellites observed around galaxies does not come from a dynamical interaction 
with the host galaxy but probably originates from the galaxy formation initial conditions. 

\subsection{Outer halo stellar density profile}

An important aspect of the satellite galaxy problem is their detectability. Our interest is in finding
the position of the stars coming from disrupted satellites and making predictions about their distribution and luminosity.
We assume that a good proxy for
the initial position of stars are particles deep in the potential well of the host satellite. For our purpose, we assume that
the 10\% most bound particles are kinematic tracers of the stellar population in each satellite. The rationale for it is simple :
as gas cools down, it loses energy and sinks in the potential well of the host satellite. Thus, we expect that most of the stars have large binding energy.  Once these particles
are identified, they are labeled and followed during the simulation. Each point particle represents a population of stars. 
To assign these particles a stellar mass, we assume a baryonic mass fraction and a star formation efficiency. 
We normalize the mass of these particles so that they correspond to a baryonic mass fraction of 0.171 \citep{ste03} and a star formation efficiency, the fraction
of baryonic mass turned into stars, between 
1 and 10\% \citep{ric05}. We assume a simple scenario in which 
the baryonic mass fraction and star formation efficiency are the same for all satellites. 

\begin{figure}
\centering{\includegraphics[angle=-90,scale=.90]{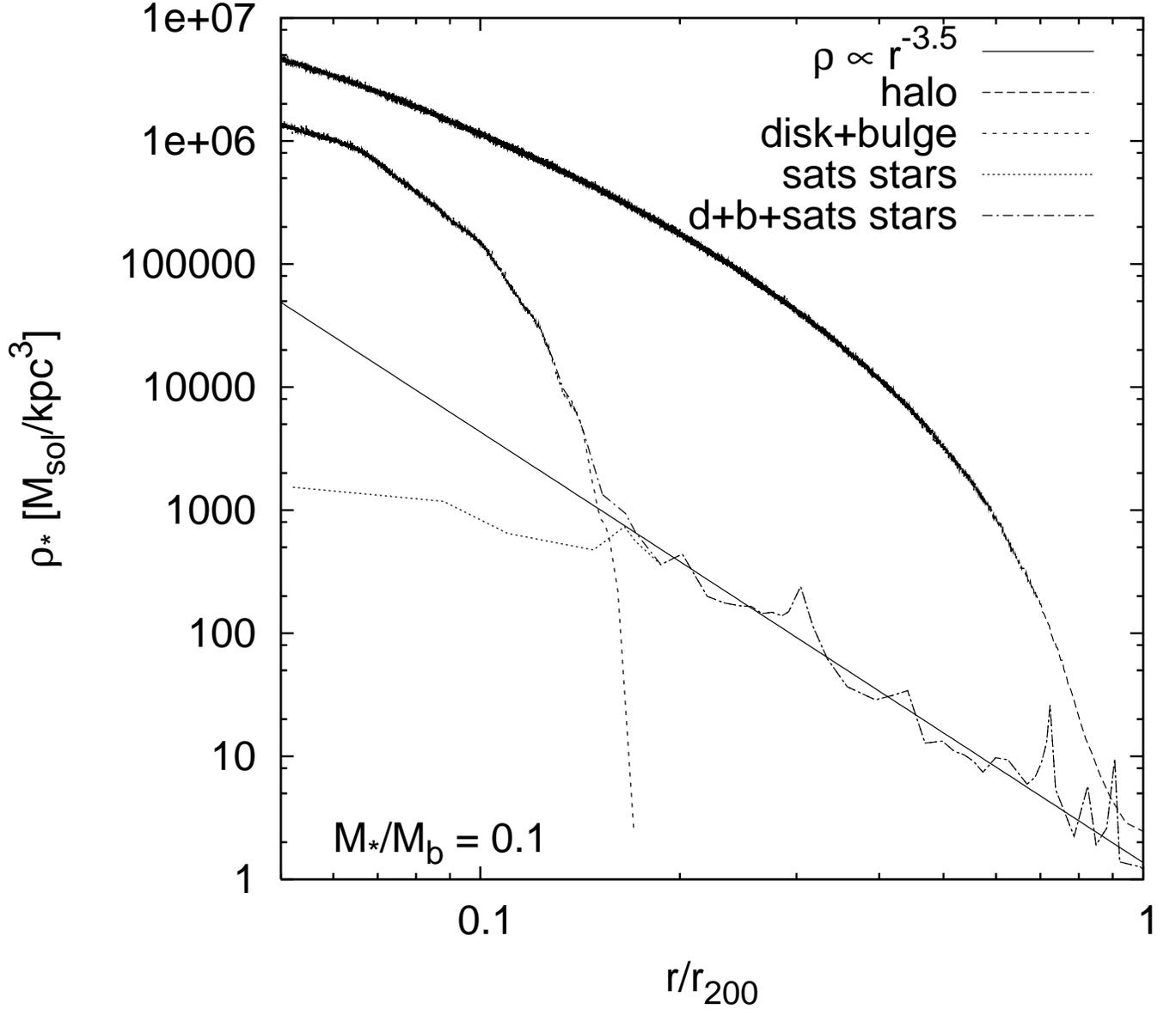}}
\caption{3D density profile of the stripped stars plus stars still in satellites after 10 Gyr. We assume a constant star formation efficiency $M_{*}/M_b$=$0.1$ and a baryon fraction of 0.171. The \emph{long-dashed} line represents the dark halo density profile; \emph{short-dashed} one the stellar component of the bulge+disk; the \emph{dotted} line is associated with satellites star only and the \emph{dash-dotted} line is the sum of the disk, bulge and satellites stellar component. For $r/r_{200} > 0.2$, the profile of satellites stars is well fitted by $\rho$ $\propto$ $r^{-3.5}$. The spikes in the satellite stellar component are associated with stars that 
are still bound to orbiting satellites.  }
\label{stars_radial_profile}
\end{figure}

We show in Figure \ref{stars_radial_profile} the spherically averaged density profile of ``stars". We include the stars that 
have been tidally disrupted from their host satellites and the ones that have not been displaced from the center 
of the potential well of individual satellites. The contribution from satellite stars dominate for $r/r_{200}
 \geq 0.2$ and the density profile of stellar material at larger radii can be well fit by an $r^{-3.5}$ power law that 
agrees with globular clusters system \citep{har76} of our Galaxy.     
This profile is also in good agreement with observations of the Milky Way's metal-poor stellar halo 
\citep{mor00,chi00,yan00} and with recent N-body simulations by \citet{aba05} and \citet{bul05} who show that the density profile of the accreted
stars goes as $\rho \propto r^{-\alpha}$, $\alpha$ $\simeq$ 3-4. In addition, recent analysis of 1047 SDSS edge-on disk galaxies by \citet{zib04} demonstrate the presence of 
stellar halos with spatial distribution that is well described by a power law $\rho \propto r^{-3}$. Our simulation predicts that, after 10 Gyr, the satellites contribute a total stellar mass of $1.2 \times 10^{9}$ $M_{\odot}$. About 20\% of this 
mass is found inside the edge radius of the disk and represents less than 1\% of the total stellar mass of the disk and bulge 
components. The overall contribution of the satellite debris to the total galactic stellar mass is about 1\%.  

\subsection{B-band photometric maps : Streams } 
In order to convert surface density (which one gets from the N-body distribution) to surface brightness, one requires the mass-to-light 
ratio of the stellar population.
This in turn requires a stellar evolution model and we use those  
by \citet{bru03}\footnote{See http://www.cida.ve/\~{}bruzual/bc2003}. 
The resulting mass-to-light ratios in the B-band at different epochs along with the baryon fraction
and the star formation history are listed in Table \ref{summary_maps}. 
We assume a constant $\Upsilon$ = 7.6 B-band mass-to-light ratio for M31 \citep{ber01,fab79}. 

We generate 2000 $\times$ 2000 pixels photometric maps showing a $10^{\circ}$  $\times$ $10^{\circ}$ field centered on M31 so that 
each pixel has a corresponding plate-scale of 18 arcsec.
To reduce shot noise, we smooth out our ``point mass stellar populations" with a 5 pixels (1.5 arcmin) gaussian window. 
The final results are presented in Figure \ref{photometric_maps3} and represent B-band photometric maps of the 
stellar population expected from disrupted satellites galaxies taken at different times. We rotate our N-body model to fit the orientation 
of M31 on the sky. For each panel, we vary the star formation efficiency and produce three different maps. 

We now examine the tidal streams in our simulations. Figure 
\ref{zoom-in} presents a close-up of photometric maps at 3.5 and 5.5 Gyr. At 3.5 Gyr, there is an obvious ``bridge" of stars connecting two subhalos and M31 in the upper map. This feature is similar both in morphology and brightness to the M31 giant stream detected in a number of surface density maps \citep{fer05,fer02,iba01} 
and constitutes the brightest stream detected in our photometric maps. 
The mean surface brightness of our giant stream proxy is about 28.5 mag $\rm{arcsec}^{-2}$ (in the B-band) 
and has comparable brightness to the actual measured value of 
$\mu_V$ $\approx$ 30 $\pm$ 0.5 mag $\rm{arcsec}^{-2}$ \citep{iba01}.
Even though we generate maps in the B-band, our photometric data in the B-band are very similar to those one would obtained in the V-band because, for a few billion year old simple stellar population, the mass-to-light ratio in the B-band is similar to the one in the V-band ($\mu_b$ $\approx$ $\mu_V$ in the simulations). 
Other streams have also been observed around M31 and have similar brightnesses. 
\citet{zuc04a} find a $3^{\circ}$ overdensity of luminous red giant stars (Andromeda NE) 
having a central g-band surface brightness of 29 mag $\rm{arcsec}^{-2}$. In addition, \citet{mcc04} 
observe an arc-like overdensity of blue, red giant stars in the west quadrant of M31. 
This tail has a $\mu_V$ = 28.5 $\pm$ 0.5 mag $\rm{arcsec}^{-2}$. Similarly, the G1 Clump, a stellar overdensity that is 
located 30 kpc along the southwestern major axis could 
be associated by an overdense clump locate at the lower right edge of 
Andromeda in bottom picture of Figure \ref{zoom-in}.

\subsection{Surface Brightness Profile}

In Figure \ref{SB} we show the surface brightness profiles of the stars no longer bound to the satellites and contributing to the metal-poor halo of M31. The plots are drawn assuming a star formation efficiency of 10\% and
a simple stellar population formed 1 Gyr prior to the beginning of the simulation. Comparisons with actual observations of the surface
brightness profile of M31 are quite challenging because the extremely low surface brightnesses involved pose a significant problem for observers (typically, 7-8 mag higher than the sky).
One of the deepest surveys of M31 halo surface brightness was carried out by \citet{irw05}. These authors have been able to go down to $\mu_V \approx 32$ mag $\rm{arcsec}^{-2}$
at a projected radius (along the minor axis) of $4^{\circ}$ (55 kpc). At that radius, they obtain a V-band surface brightness of about 30-31 mag $\rm{arcsec}^{-2}$ which is close to our B-band estimate of 29-30 mag $\rm{arcsec}^{-2}$. 
In Figure \ref{SB}, a star formation efficiency of 5\% would decrease the surface brightness at 
all radii by about +0.75 mag $\rm{arcsec}^{-2}$ and leads to a better agreement with the value measured at 55 kpc.

The value of the surface brightness for R $<$ 50 kpc flattens at about 28 mag $\rm{arcsec}^{-2}$.
Dynamical friction only weakly affects the trajectories of the satellites and thus, little mass is being deposited in the
center of the galaxy. Other studies (e.g. \citet{aba05}) show a constant increase of the surface brightness profile down to a radius of about 10 kpc. In many simulations, the satellites sink in the center of the galaxy and get heavily stripped within 30 kpc and deposit large amounts of mass in the center. It is not the case in this simulation.

\subsection{Satellites detected around M31}

A way of constraining the star formation history of the satellites is by counting the number
of satellites detected in the close vicinity of M31 as a function of the star formation 
efficiency and epoch of birth and compare their number and properties
with the current 16 M31 satellites claimed by \citet{mcc05}. 
The faintest of these satellites is Andromeda IX ($\mu_V$ = 26.8 mag $\rm{arcsec}^{-2}$) discovered
by \citet{zuc04b}. The survey is complete for satellites with $\mu_V$ $<$ 26.8 mag $\rm{arcsec}^{-2}$.
According to the results listed in Table \ref{number_detected10x10}, it appears
that a star formation rate of about 10\% combined with an epoch of formation 
between 3.5 and 1 Gyr prior to the beginning of the simulation can 
account for the actual number of satellites detected around M31. 
For the 5.5 Gyr snapshots, a star formation rate of 10\% results in exactly 15 detected satellites 
with $\mu_V$ $<$ 26.8 mag $\rm{arcsec}^{-2}$. On the other hand, a SFR of 1\% is insufficient since it reduces to 0 
the number of satellites detected below the 26.9 mag $\rm{arcsec}^{-2}$ limit.    

We expect the most massive satellites to have a light distribution
similar to M32 or NGC 205. The galaxies have, respectively, a central B-band surface
brightness of approximately 17.5 mag $\rm{arcsec}^{-2}$ and 20 mag $\rm{arcsec}^{-2}$ \citep{cho02} .
The brightest simulated satellites located in the vicinity of M31 have a central B-band surface
brightness of about 23 mag $\rm{arcsec}^{-2}$ assuming a simple stellar population
formed 1.5 Gyr before the beginning of the simulation and and a star formation
efficiency of 10\%. The fact that our simulations
are dissipationless is the cause of this difference. 
In a more realistic case, gas cools down and sink deep in the potential
well increasing the central density of stars. Our simulations provide lower limits to the value
of the central surface brightness one would expect. However, a central value of 22-23 mag $\rm{arcsec}^{-2}$ is consistent
with a recent study of dwarf galaxies in the Fornax cluster \citep{phi01} and the Andromeda (I-VII) dwarfs series 
\citep{mcc06b}. These authors find that
the majority of the dwarfs in their sample have a surface brightness spanning
20 - 24 mag $\rm{arcsec}^{-2}$ (see their Figure 1). Many of our satellites would fall in that range of central brightness.

The surface brightness at the half-light is a more representative measurement of the overall dwarf
light distribution and does not suffer much from dissipation effects. 
Our brightest satellites have $\mu_{eff,b} = 25.3$ mag $\rm{arcsec}^{-2}$.
By comparison, M32 has a half-light surface brightness $\mu_{eff,B} \approx 20$ mag $\rm{arcsec}^{-2}$ \citep{cho02}. A more recent episode of
star formation or a continuous star forming period could explain that difference since younger stars would increase the B-band flux emission. Our
simple model assume that all stars have the same age and in this case, formed 1.5 Gyr before the beginning of the simulation.
We conclude that our brightest satellites show photometric differences with the most massive companions of M31. However, our simulations are purely
collisionless and including dissipative processes would increase the central and half-light radius luminosities in such a way that the
brightest simulated satellites would be similar to M32 and NGC205.

Our choice of a satellite mass range of $1.5 \times 10^{-4}$ $<$ $M_{\rm sat}/M_{\rm halo}$ $<$ $0.02$ is consistent
with approximately 100 satellites according to measured numbers in cosmological simulations.
However, if we extend the mass range to lower and higher values there would be many more dark matter satellites. 
especially at the lower mass end. These satellites might potentially be detectable, add more stars to the tidal debris 
stellar halo, and thus change our conclusions. We therefore estimate the effects of the satellite mass range on the 
final total number of satellites that might be detected near M31. We sample the same mass function as 
equation (\ref{massfunction}) but this time with lower and upper limits 
$1.5 \times 10^{-5}$ $<$ $M_{\rm sat}/M_{\rm halo}$ $<$ $0.1$ that leads to about 1,000 satellites with a total 
combined mass of $0.25$ $M_{\rm halo}$ compared with a total satellite mass of $0.1$ $M_{\rm halo}$ for our 
100 satellite run. By extending the mass range, we add nearly 900 low mass halos and five high mass ones outside
the original range. The six most massive halos would account for $0.16$ $M_{\rm halo}$ and would be easily 
detected having a central surface brightness greater than 27 mag $\rm{arcsec}^{-2}$ even with significant 
stripping as seen in our simulation.
However, the 900 additional low mass halos would probably be fainter than 32 mag $\rm{arcsec}^{-2}$, the central 
surface brightness of the lowest mass satellites in our run. More than 60\% of them have a mass less than 
$2.5$ $\times$ $10^{-5}$ and have central brightnesses of 34 mag $\rm{arcsec}^{-2}$ according to a linear extrapolation 
of our results. None of this population would be detected in the current surveys of the M31 stellar halo region so our 
exclusion of the low mass tail of satellite population has little effect on the predictions for the number of satellites 
around M31. However, the low mass satellites contain a significant amount of mass and tidal stripping of this population 
could approximately double the amount of tidal debris in the stellar halo. Surface brightness maps might then be 
expected to be a magnitude higher than what our 45M particle simulation predicts. 

\section{Summary \& Conclusion}

This study presents the evolution of a self-consistent population of satellites in the presence of a parent galaxy. 
The internal structure, mass function and spatial
and orbital distribution of the satellite distribution is motivated by cosmological simulations.
The work improves upon previous self-consistent studies (e.g. \citealt{fon01}) by using a more realistic galaxy model and much improved numerical
resolution. It follows the lead of current work on satellite tidal disruption to make quantitative predictions of the
tidal debris field of galaxies (e.g. \citealt{joh01}). 

This work shows that members of a typical 
population of subhalos orbiting in a galactic will come close to the disk 
with a ten billion years timescale. In addition, it is possible to model 
these satellites and make direct predictions on the number one would expect to 
detect around a typical galaxy. We also agree that it is possible to 
maintain a disk in spite of substructure having a mass fraction of 
about 0.1$M_{\rm{halo}}$. 

Here we summarize the main results of this paper.
\begin{enumerate}

\item Dynamical friction plays only a minor role in the evolution of the satellite system and the number density profile
of the system is relatively unchanged over 10 Gyr. The orbits of the most massive satellites do show some decay but
since they suffer severe tidal stripping, dynamical friction quickly becomes unimportant.

\item The vast majority of the satellites survive in the galactic environment for more than a Hubble 
time; only 7 are completely disrupted. However, most of satellites lose 
a significant fraction of their mass due to tidal interactions.   

\item The satellite mass loss due to tidal stripping is described by two distinct phases. 
The first one is characterized by a sharp mass decline with $t_{1/2}$ = 3.5 Gyr. During that period, 
a small amount of mass loss, $5.3 \times 10^8$ $M_{\odot}$, is due to a transient state where satellites initially 
located inside 50 kpc are overfilling their tidal radius and start losing mass instantly.   
During the second phase, the satellites lose about 25\% of their initial mass with $t_{1/2} = 9.3$ Gyr. Over 10 Gyr, the satellite population mass diminishes by about 75\% and turns into tidal debris. These debris form long tidal tails around the galaxy and can be detected
in photometric maps of M31. The final density profile of the light satellites can be approximated by the fitting function given in
\citet{hay03}.

\item The spatial distribution of stars associated with infalling satellites follow a power law $\rho \propto r^{-3.5}$ at large radii. 
This result is in agreement with recent numerical studies and observations of the stellar halo in edge-on disk galaxies. 

\item No obvious Holmberg Effect is observed. 

\item The mock B-band photometric maps, computed under the assumption that the most tightly bound 
particles are kinematic tracers of the stars, show conspicuous features of bridges and 
tails comparable to what is actually observed. A star formation efficiency of 10\% is necessary to match the morphological 
and photometric properties of our Giant Stream proxy with the real one.

\item The number of satellites detected around M31 below $\mu_V$ $\approx$ 26.8 mag $\rm{arcsec}^{-2}$ can match our results 
with a star formation efficiency of 10\%. For a value of 1\%, the number of satellites detected goes down to 0. A value of 5\% 
can not be discarded since dissipation is not taken into account. 
Many of the satellites between 26.8 and 29.4 mag $\rm{arcsec}^{-2}$ in a 5\% star 
formation efficiency scenario would probably be detected under the limit of 26.8 mag $\rm{arcsec}^{-2}$ if dissipation was included in the simulations. 
 
\item A star formation efficiency of about 5\% is necessary to account 
for the actual M31 value of the surface brightness of the stellar halo measured at 55 kpc. 
Comparisons for radii larger than 55 kpc are observationally challenging
because of the very low surface brightness in the outer parts of the galaxy. The simulated maps show a 
flattening in the inner part of the surface brightness profile 
(for R $<$ 55 kpc, $\mu_B \approx$ 28 mag $\rm{arcsec}^{-2}$) showing
that dynamical friction is unable to bring
the satellites close to the center. Most of the mass is lost at larger radii.   
  
\item The formation of a bar around 5 Gyr is the result of interactions of satellites with the disk. 
This intriguing result will be discussed in a forthcoming paper.  

\end{enumerate}

\acknowledgments

JRG would like to acknowledge support from NSERC postgraduate scholarship. 
JD and LMW acknowledge research support from NSERC.
We would like to thank Amr A. El-Zant for helpful discussions and comments. All simulations 
were run on the McKenzie Beowulf Cluster at the Canadian Institute for Theoretical Astrophysics funded by the Canadian Foundation for
Innovation and the Ontario Innovation Trust.

\clearpage

\begin{deluxetable}{lll}
\tabletypesize{\scriptsize}
\tabletypesize{\scriptsize}
\tablecaption{Andromeda (M31) model}
\tablewidth{0pt}
\tablehead{
\colhead{Component} & \colhead{Parameter} & \colhead{Value}
}
\startdata
Disk & &  \\
& Mass & $7.77 \times 10^{10}$ $M_{\odot}$ \\
& $R_D$ & 5.57 kpc \\
& $z_D$ & 0.3 kpc \\
& $R_{out}$ & 30 kpc \\
Bulge & & \\
& Mass & $2,88 \times 10^{10}$ $M_{\odot}$\\
& $r_b$ & 1.82 kpc\\
& $v_b^a $ & 460 km s$^{-1}$\\
& cut-off & 0.929 \\
Halo & & \\
& Mass & $5.83 \times 10^{11}$ $M_{\odot}$\\
& $r_s$ & 12.93 kpc\\
& $v_s$ & 337 km s$^{-1}$\\
& $c$ & 13.37 \\
& edge radius & 232.3 kpc\\
& cut-off & 0.75
\enddata
\label{m31}
\end{deluxetable}
\clearpage

\begin{deluxetable}{lcccccc}
\tabletypesize{\scriptsize}
\tabletypesize{\scriptsize}
\tablecaption{Physical parameters of the satellite models.}
\tablewidth{0pt}
\tablehead{
\colhead{Name} & \colhead{$M_{sat}/M_{halo}$} & \colhead{$r_s$ [kpc]}  & \colhead{$v_s$ [km s$^{-1}$]}
& \colhead{cutoff} & \colhead{$M(<r_{\rm tid}/M_{\rm sat})$} & \colhead{c}
}
\startdata
NFW1 & $1.5 \times 10^{-4}$ & 0.7 & 50 & 0.4 & 0.97 & 29.4 \\
NFW2 & $2.5 \times 10^{-4}$ & 0.85 & 56 & 0.4 & 0.96 & 27.7 \\
NFW3 & $3.5 \times 10^{-4}$ & 0.98 & 62 & 0.4 & 0.92 & 26.8 \\
NFW4 & $4.5 \times 10^{-4}$ & 1.1 & 65 & 0.4 & 0.91 & 25.4 \\
NFW5 & $5.5 \times 10^{-4}$ & 1.2 & 70 & 0.4 & 0.95 & 25.2 \\
NFW6 & $6.5 \times 10^{-4}$ & 1.33 & 75 & 0.4 & 0.94 & 24.5 \\
NFW7 & $7.5 \times 10^{-4}$ & 1.39 & 77 & 0.4 & 0.94 & 24.2 \\
NFW8 & $8.5 \times 10^{-4}$ & 1.46 & 80 & 0.4 & 0.95 & 24.0 \\
NFW9 & $9.5 \times 10^{-4}$ & 1.54 & 82 & 0.4 & 0.94 & 23.5 \\
NFW10 & $1.5 \times 10^{-3}$ & 1.79 & 93 & 0.4 & 0.93 & 23.0 \\
NFW11 & $2.5 \times 10^{-3}$ & 2.2 & 112 & 0.4 & 0.91 & 22.7 \\
NFW12 & $3.5 \times 10^{-3}$ & 2.5 & 122 & 0.4 & 0.89 & 21.9 \\
NFW13 & $5.5 \times 10^{-3}$ & 3.09 & 141 & 0.4 & 0.88 & 20.8 \\
NFW14 & $6.5 \times 10^{-3}$ & 3.35 & 146 & 0.4 & 0.87 & 20.1 \\
NFW15 & $2 \times 10^{-2}$ & 5.5 & 200 & 0.4 & 0.77 & 17.5
\enddata
\label{15models}
\end{deluxetable}

\clearpage

\begin{deluxetable}{lr}
\tabletypesize{\scriptsize}
\tabletypesize{\scriptsize}
\tablecaption{B-band photometric maps parameters}
\tablewidth{0pt}
\tablehead{
\colhead{} & \colhead{}
}
\startdata
Kinematic tracers : & 10 \% most bound particles  \\
Baryonic mass fraction ($M_{\rm b}/M_{\rm DM}$) : & $0.171^a$ \\
Star formation $\rm{efficiency}^b$ ($M_{\star}/M_{\rm b}$) : & 0.01, 0.05 and 0.1  \\
M31 $M/L_{\rm{B}}$ $\rm{(constant)}^c$ : 7.6 \\
For a simple stellar population formed at t = -1$^e$ Gyr : & \\
$M/L_{\rm{B}}$ ratio at t=3.5 Gyr : & $1.5476^d$  \\ 
$M/L_{\rm{B}}$ ratio at t=5.5 Gyr : & 2.0771 \\
$M/L_{\rm{B}}$ ratio at t=9.5 Gyr : & 2.9688 \\
For a simple stellar population formed at t = -2.5 Gyr : & \\
$M/L_{\rm{B}}$ ratio at t=3.5 Gyr : & 1.9485  \\
$M/L_{\rm{B}}$ ratio at t=5.5 Gyr : & 2.4115 \\
$M/L_{\rm{B}}$ ratio at t=9.5 Gyr : & 3.2855 \\
For a simple stellar population formed at t = -3.5 Gyr : & \\
$M/L_{\rm{B}}$ ratio at t=3.5 Gyr : & 2.1877  \\ 
$M/L_{\rm{B}}$ ratio at t=5.5 Gyr : & 2.6631 \\
$M/L_{\rm{B}}$ ratio at t=9.5 Gyr : & 3.4411 
\enddata
\label{summary_maps}
\tablenotetext{a}{\citet{ste03}}
\tablenotetext{b}{\citet{ric05}}
\tablenotetext{c}{\citet{fab79}}
\tablenotetext{d}{GALAXEV code, \citet{bru03}}
\tablenotetext{e}{Note that the simulation starts at t=0. We imply here that the stars formed prior to the beginning 
of the simulation.}
\end{deluxetable}

\clearpage

\begin{deluxetable}{lllccc}
\tabletypesize{\scriptsize}
\tabletypesize{\scriptsize}
\tablecaption{Number of satellite galaxies detected in a $10^{\circ}$ $\times$ $10^{\circ}$ 
field centered on M31 for different stellar population ages, star formation efficiency and isophotal thresholds}
\tablewidth{0pt}
\tablehead{
\colhead{$t_{\rm form}$ (Gyr) } & \colhead{$t_{\rm meas}^a$ (Gyr)} & \colhead{$M_{\star}/M_{\rm bar}$} & \colhead{$N\leqslant 26.89$ mag $\rm{arcsec}^{-2}$} &\colhead{$N\leqslant 29.4$ mag $\rm{arcsec}^{-2}$} & \colhead{$N\leqslant 31.9$ mag $\rm{arcsec}^{-2}$}
}
\startdata
-1 & 3.5 & 0.01 &1 &23 &26 \\
-1 & 3.5 & 0.05 &11 &26 &26 \\
-1 & 3.5 & 0.1 &25 &26 &26 \\
-1 & 5.5 & 0.01 &0 &15 &30 \\
-1 & 5.5 & 0.05 &6 &30 &30 \\
-1 & 5.5 & 0.1 &15 &30 &30 \\
-1 & 9.5 & 0.01 &0 &8 &29 \\
-1 & 9.5 & 0.05 &4 &30 &30 \\
-1 & 9.5 & 0.1 &8 &30 &30 \\
-2.5 & 3.5 & 0.01 &0 &23 & 26 \\
-2.5 & 3.5 & 0.05 &10 &26 &26 \\
-2.5 & 3.5 & 0.1 &23 &26 &26 \\
-2.5 & 5.5 & 0.01 &0 &13 &30 \\
-2.5 & 5.5 & 0.05 &6 &30 &30 \\
-2.5 & 5.5 & 0.1 &15 &30 &30 \\
-2.5 & 9.5 & 0.01 &0 &8 &29 \\
-2.5 & 9.5 & 0.05 &3 &30 &30 \\
-2.5 & 9.5 & 0.1 &7 &30 &30 \\
-3.5 & 3.5 & 0.01 &1 &20 &26 \\
-3.5 & 3.5 & 0.05 &9 &26 &26 \\
-3.5 & 3.5 & 0.1 &22 &26 &26 \\
-3.5 & 5.5 & 0.01 &0 &12 &30 \\
-3.5 & 5.5 & 0.05 &6 &30 &30 \\
-3.5 & 5.5 & 0.1 &15 &30 &30 \\
-3.5 & 9.5 & 0.01 &0 &7 &29 \\
-3.5 & 9.5 & 0.05 &3 &30 &30 \\
-3.5 & 9.5 & 0.1 &7 &30 &30 \\
\enddata
\label{number_detected10x10}
\tablenotetext{a}{The time at which the measurements are made after the beginning of the simulation. }
\end{deluxetable}

\clearpage

\clearpage 

\clearpage

\clearpage

\clearpage

\clearpage

\clearpage 

\clearpage

\clearpage

\clearpage

\newpage

\newpage
\begin{figure}
\centering{\includegraphics[angle=-90,scale=.90]{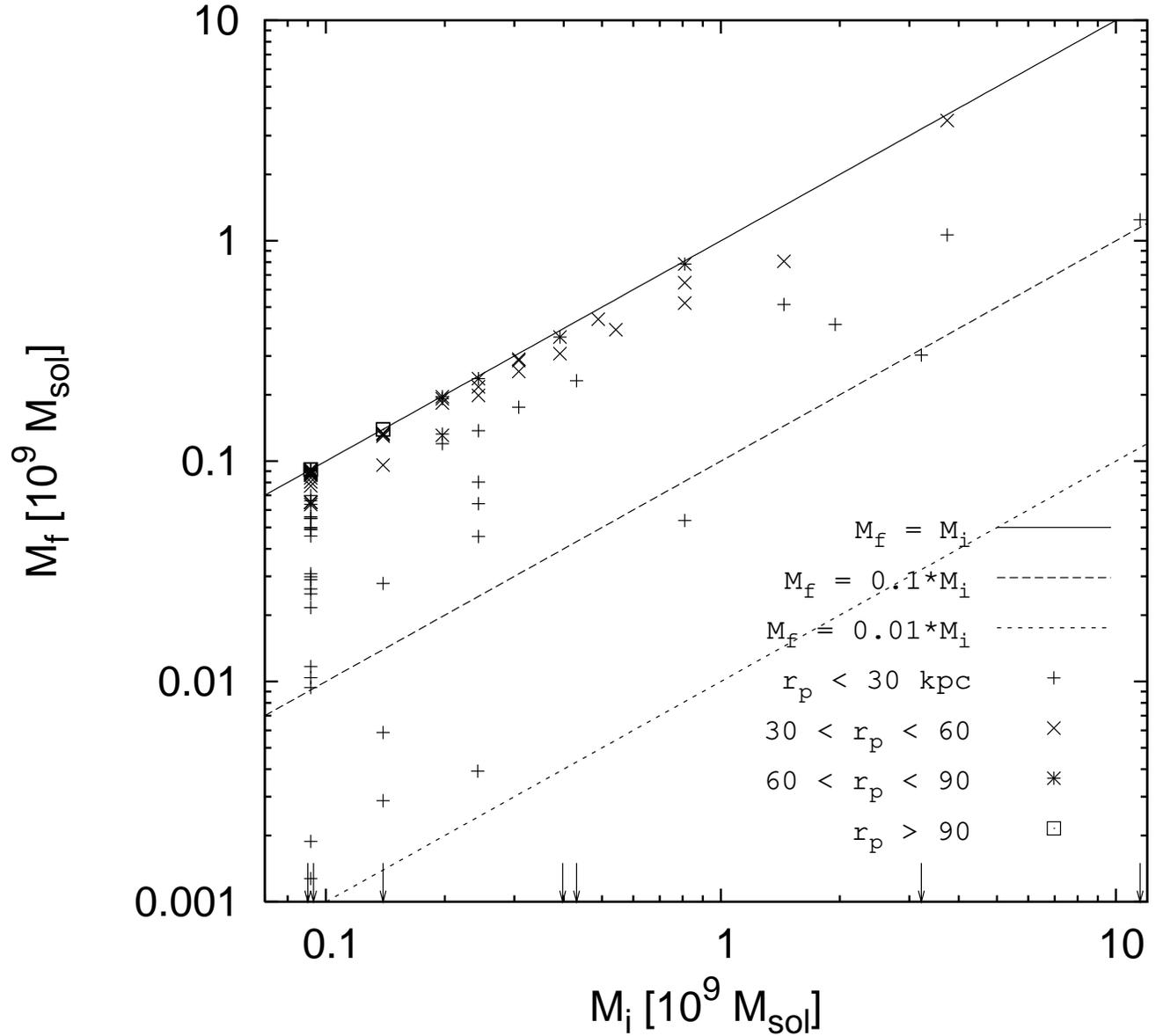}}
\caption{
Scatter plot of the final mass $m_f$ vs initial mass $m_i$ fo the 100 satellites in the experiment. Different symbols 
distinguish satellites according to pericenter. The arrows represent the seven satellites that are completely destroyed
by the end of the simulation.
}
\label{mass_evolution}
\end{figure}

\newpage

\begin{figure}
 \vspace{5pt}
  \centerline{\hbox{ \hspace{0.0in}
    \includegraphics[angle=-90,scale=.5]{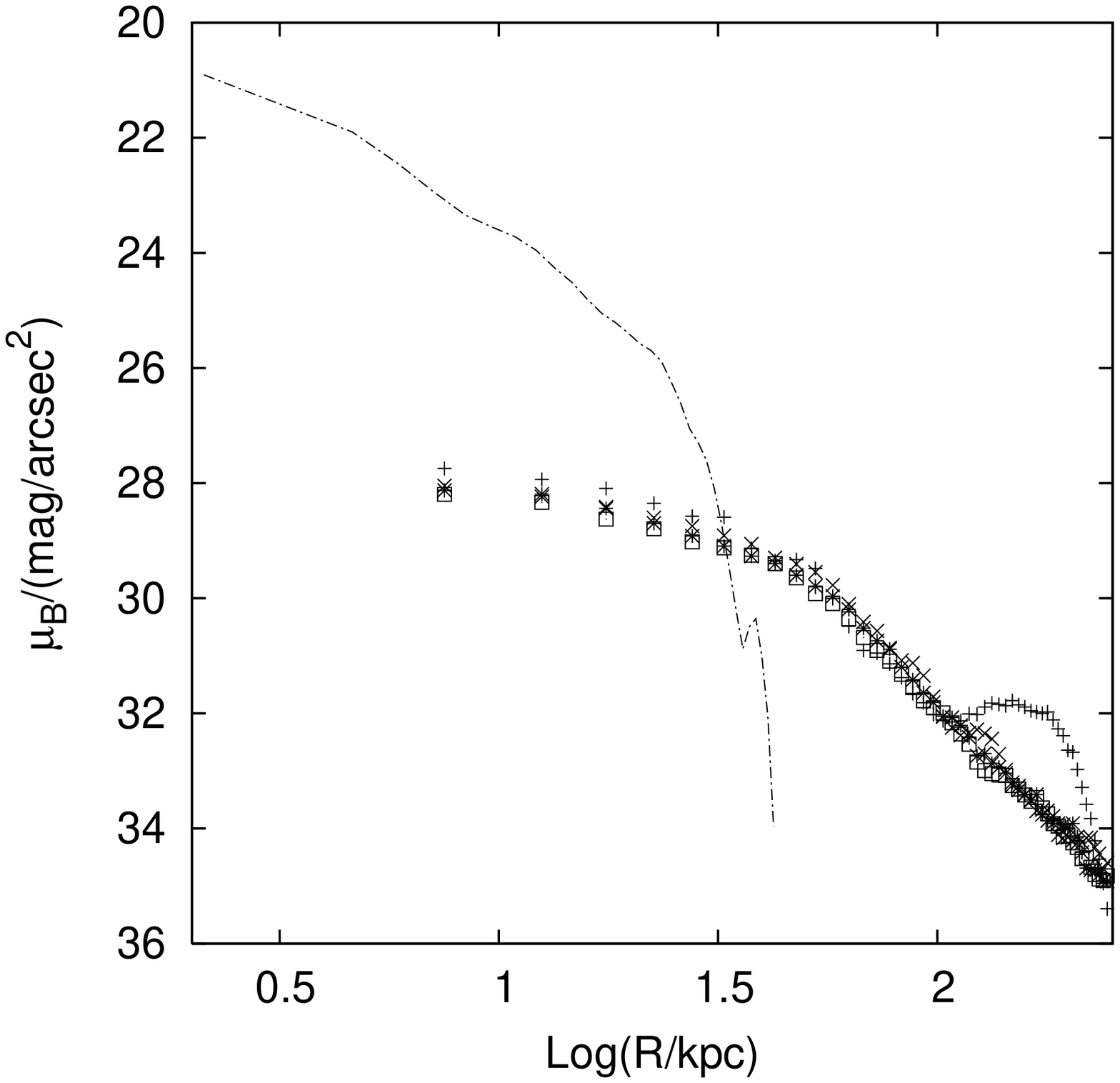}
    \includegraphics[angle=-90,scale=.5]{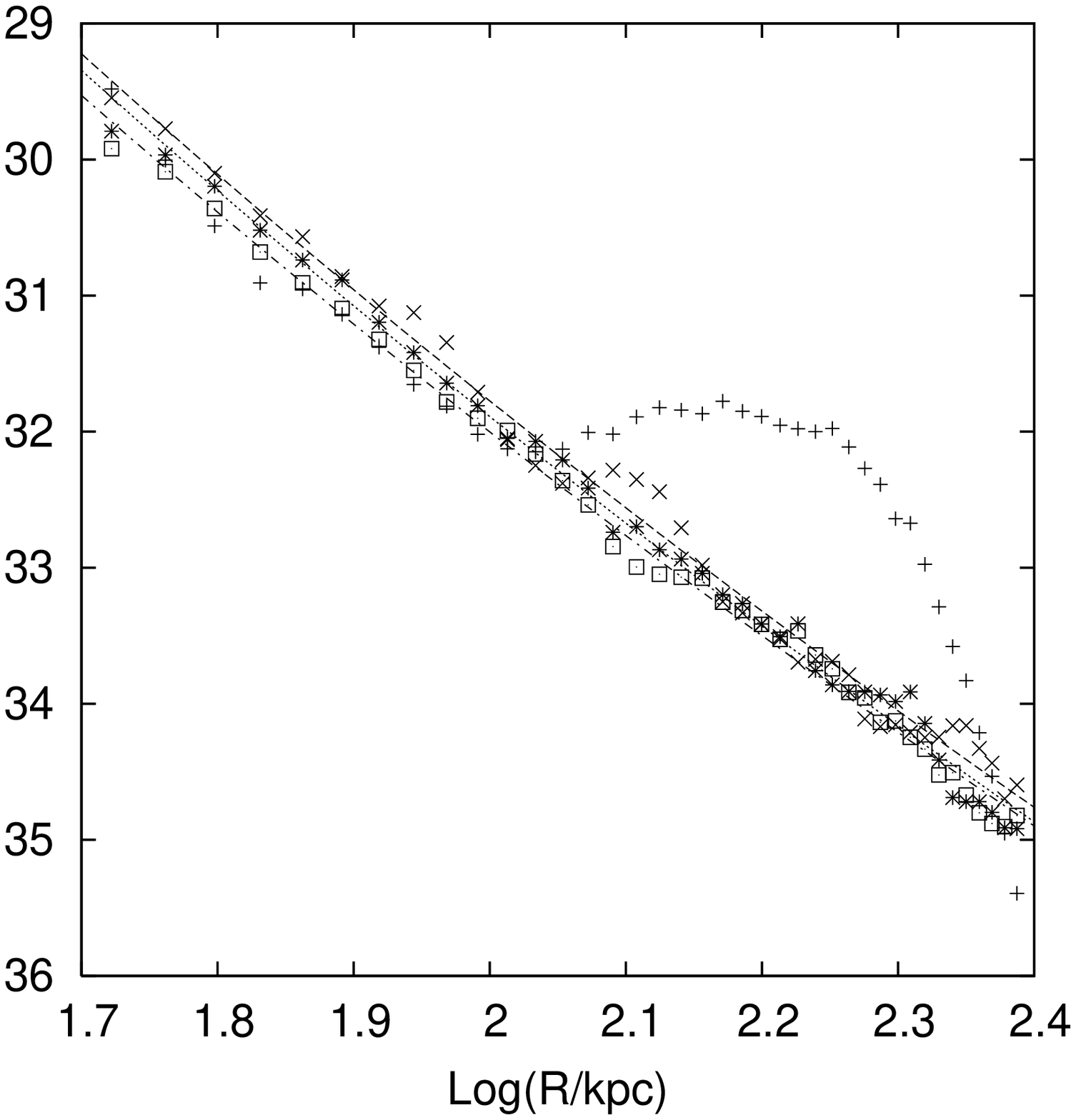}
    }
   }
 \vspace{5pt}
\caption{B-band surface brightness profiles of the stars tidally stripped from the satellites. The profiles are drawn assuming a star formation efficiency of 10\% and a simple stellar population formed
1 Gyr prior to the beginning of the simulation. On both plots, the symbols are for four different snapshots : \emph{plus} 2.5 Gyr, \emph{X} 5.5 Gyr, \emph{stars} 7.5 Gyr and \emph{squares} 10 Gyr. On the left-hand side, the \emph{dot-dashed} line shows the contribution of disk+bulge of M31 to the total surface brightness. On the right-hand side, a closeup shows straight lines corresponding to the best fit de Vaucouleurs profiles at 5, 7.5, and 10 Gyr. 
The feature around log R = 2.2 for the 2.5-Gyr snapshot is a consequence of the transient start and tends to disappear before 5 Gyr.
}
\label{SB}
\end{figure}

\clearpage

\begin{figure}
 \vspace{7pt}
  \centerline{\hbox{ \hspace{0.0in}
    \includegraphics[angle=0,scale=.25]{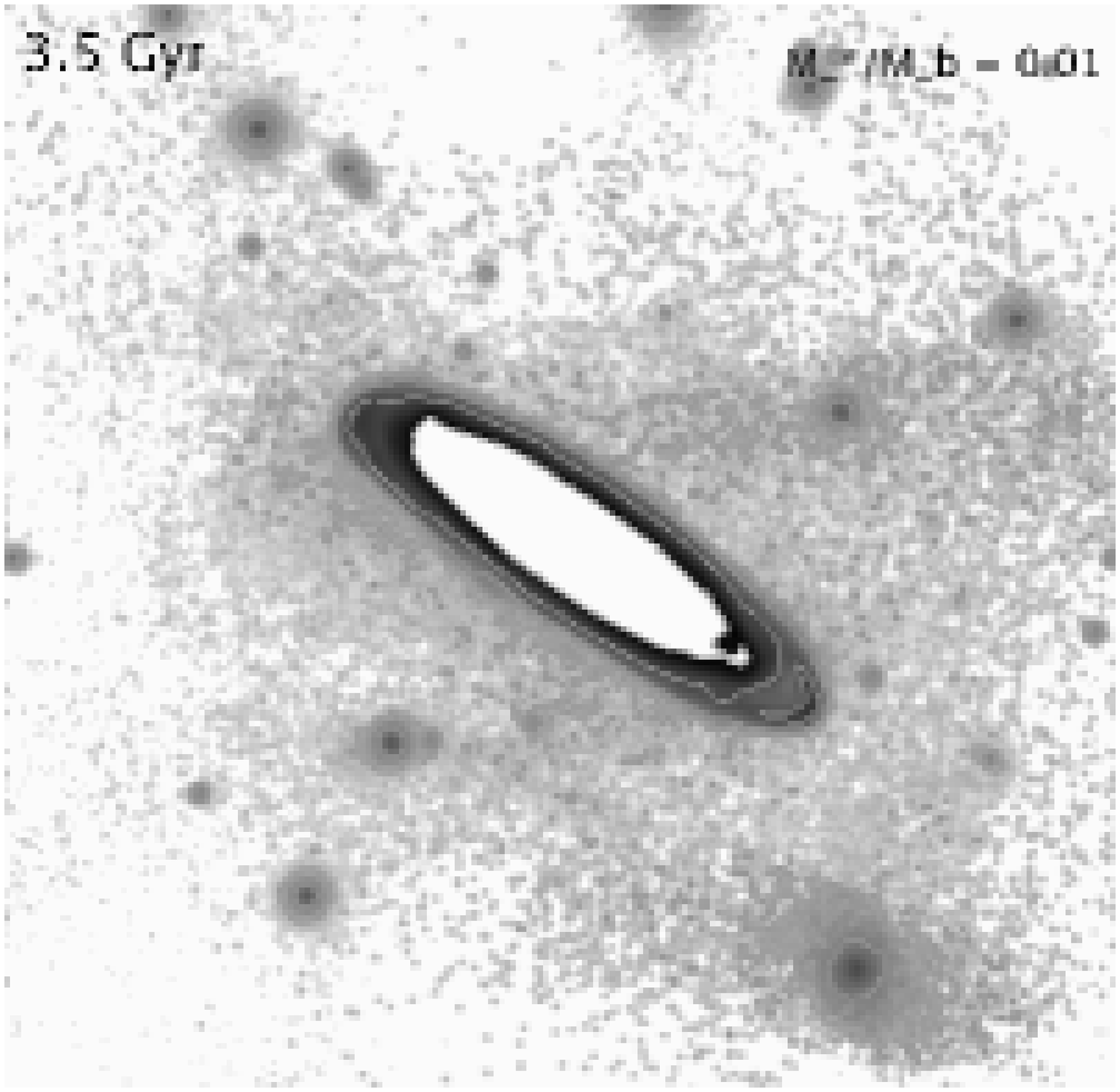}
    \hspace{0.1in}
    \includegraphics[angle=0,scale=.25]{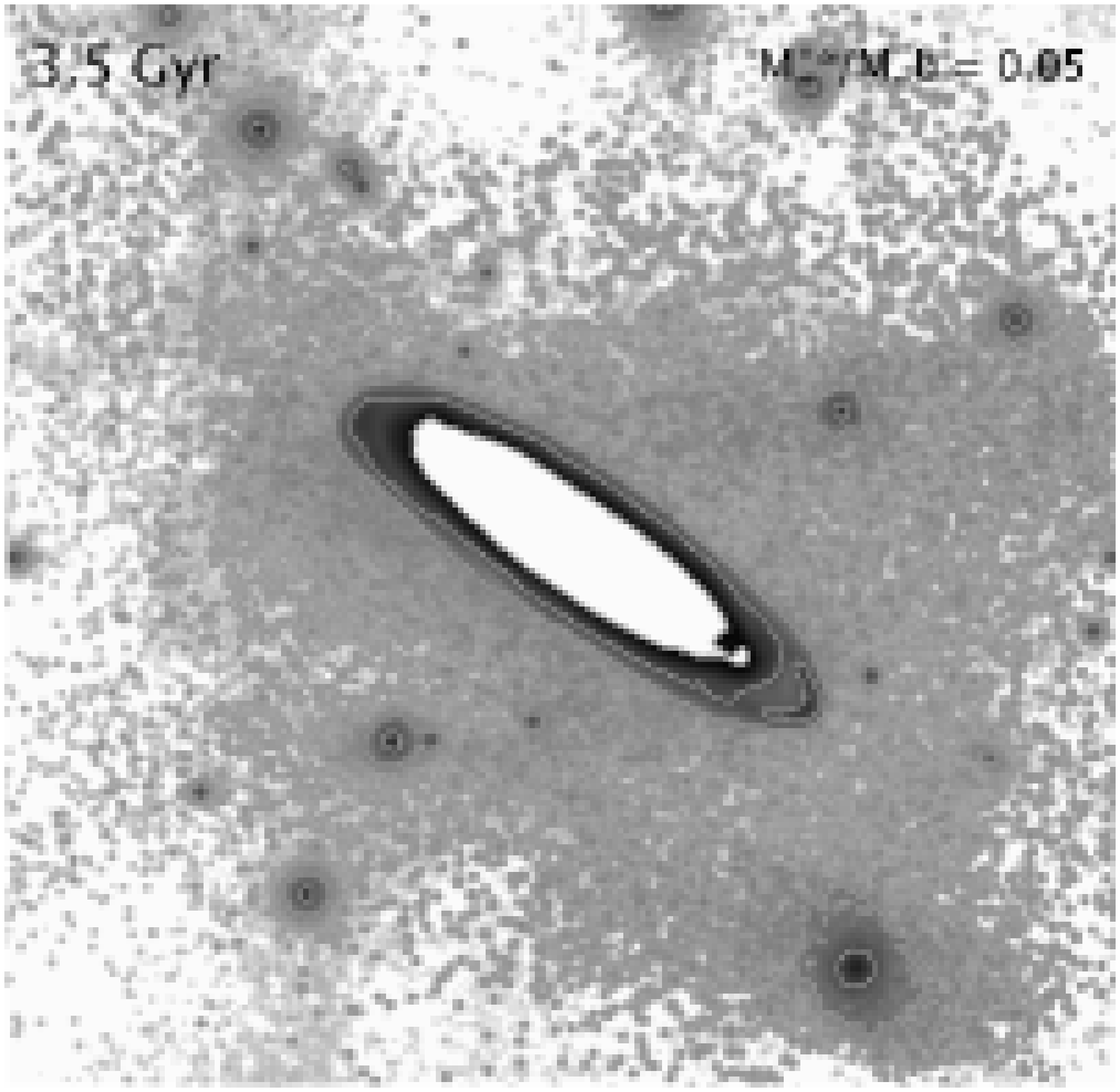}
    \hspace{0.1in}
    \includegraphics[angle=0,scale=.25]{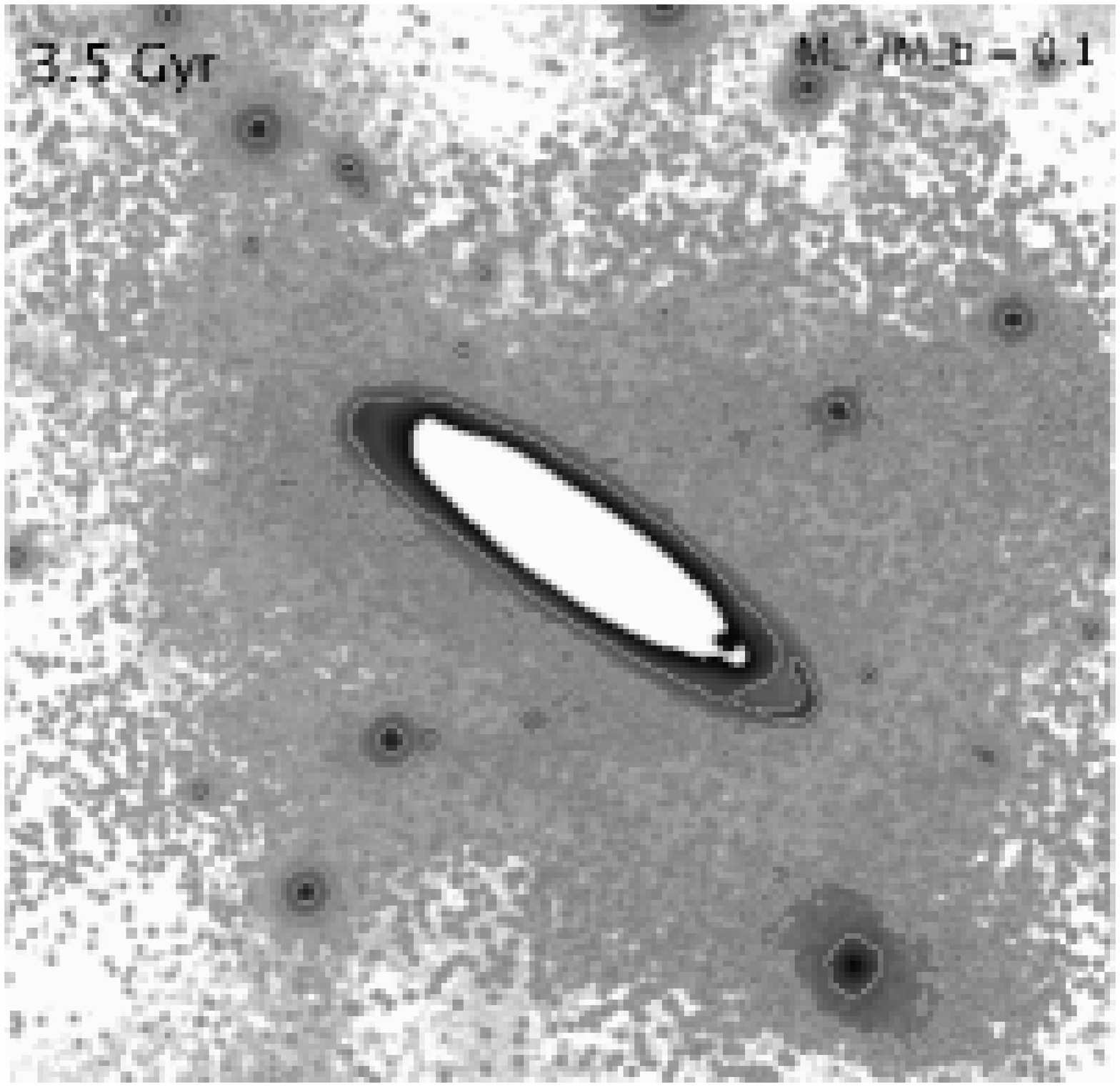}
    }
   }
  \vspace{7pt}
\centerline{\hbox{ \hspace{0.0in}
    \includegraphics[angle=0,scale=.25]{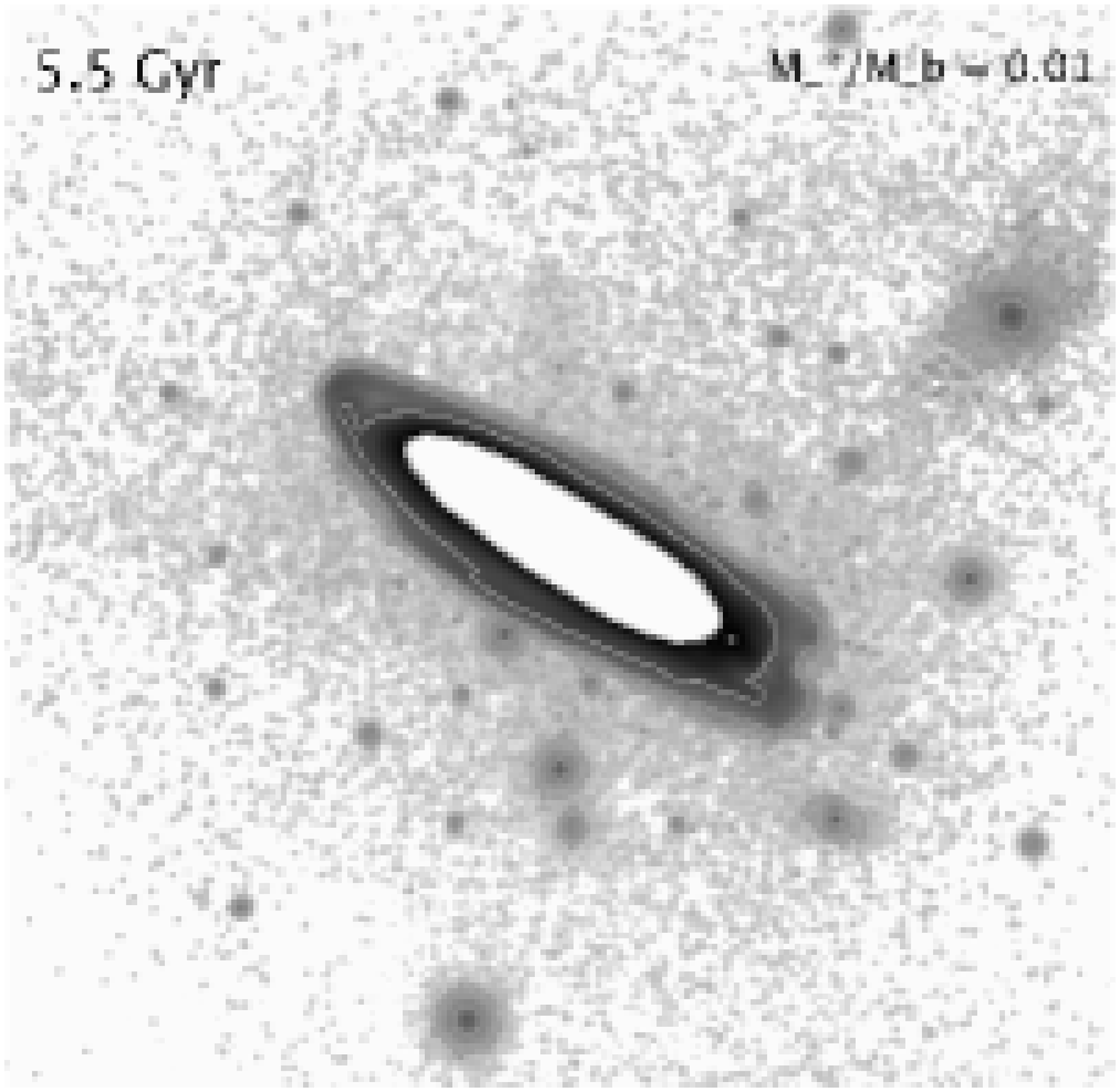}
    \hspace{0.1in}
    \includegraphics[angle=0,scale=.25]{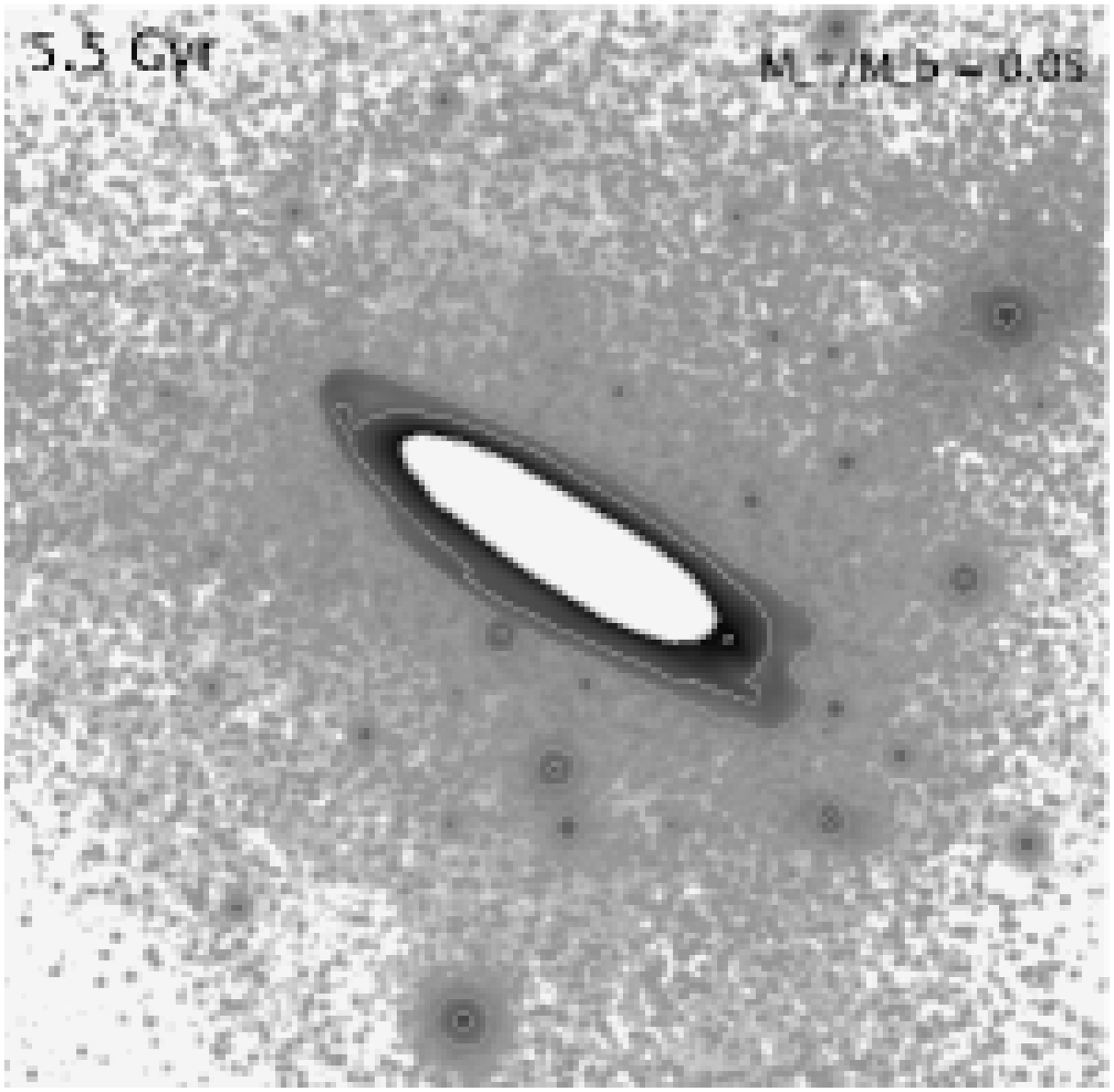}
    \hspace{0.1in}
    \includegraphics[angle=0,scale=.25]{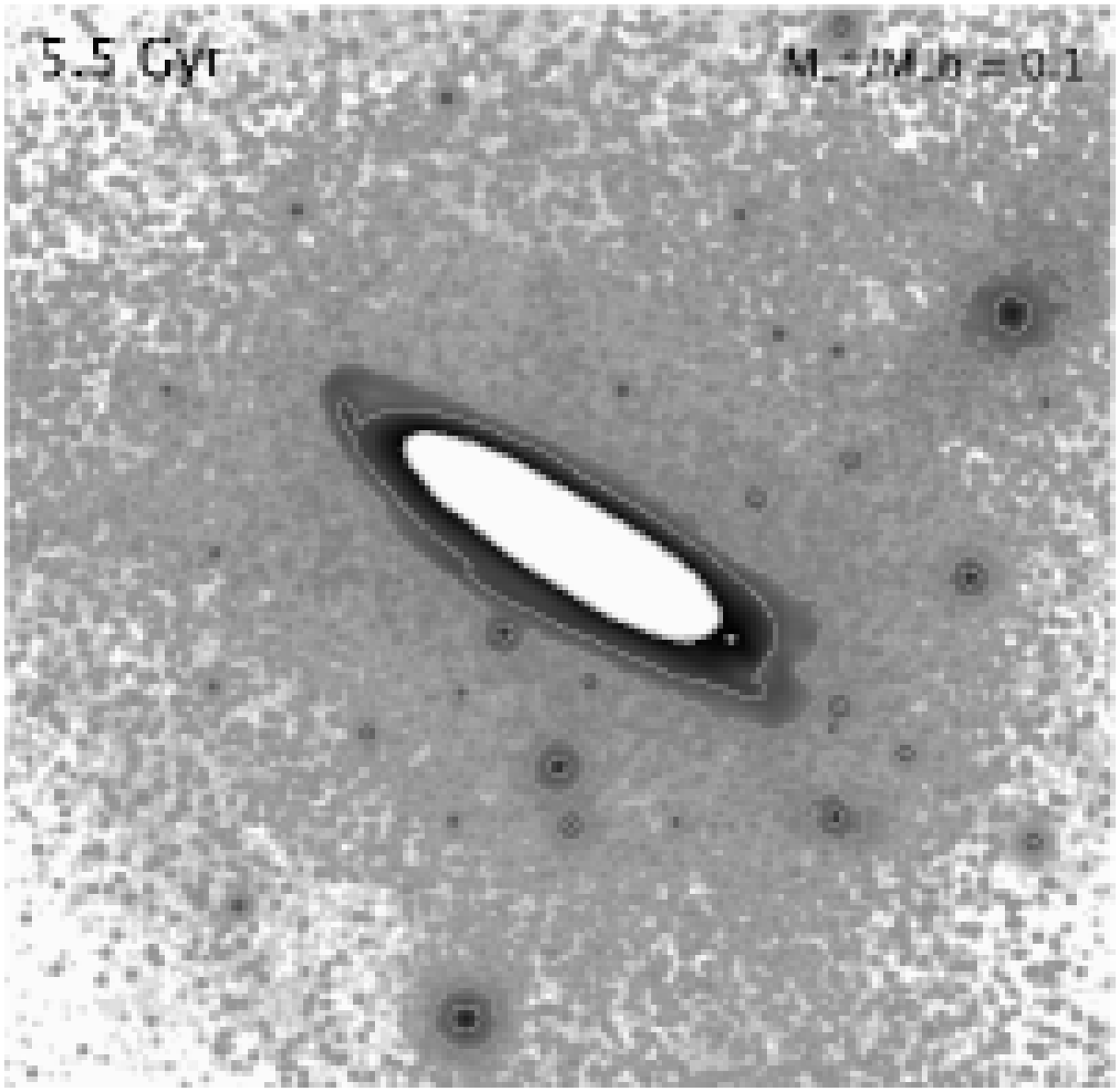}
    }
   }
  \vspace{7pt}
\centerline{\hbox{\hspace{0.0in}
    \includegraphics[angle=0,scale=.25]{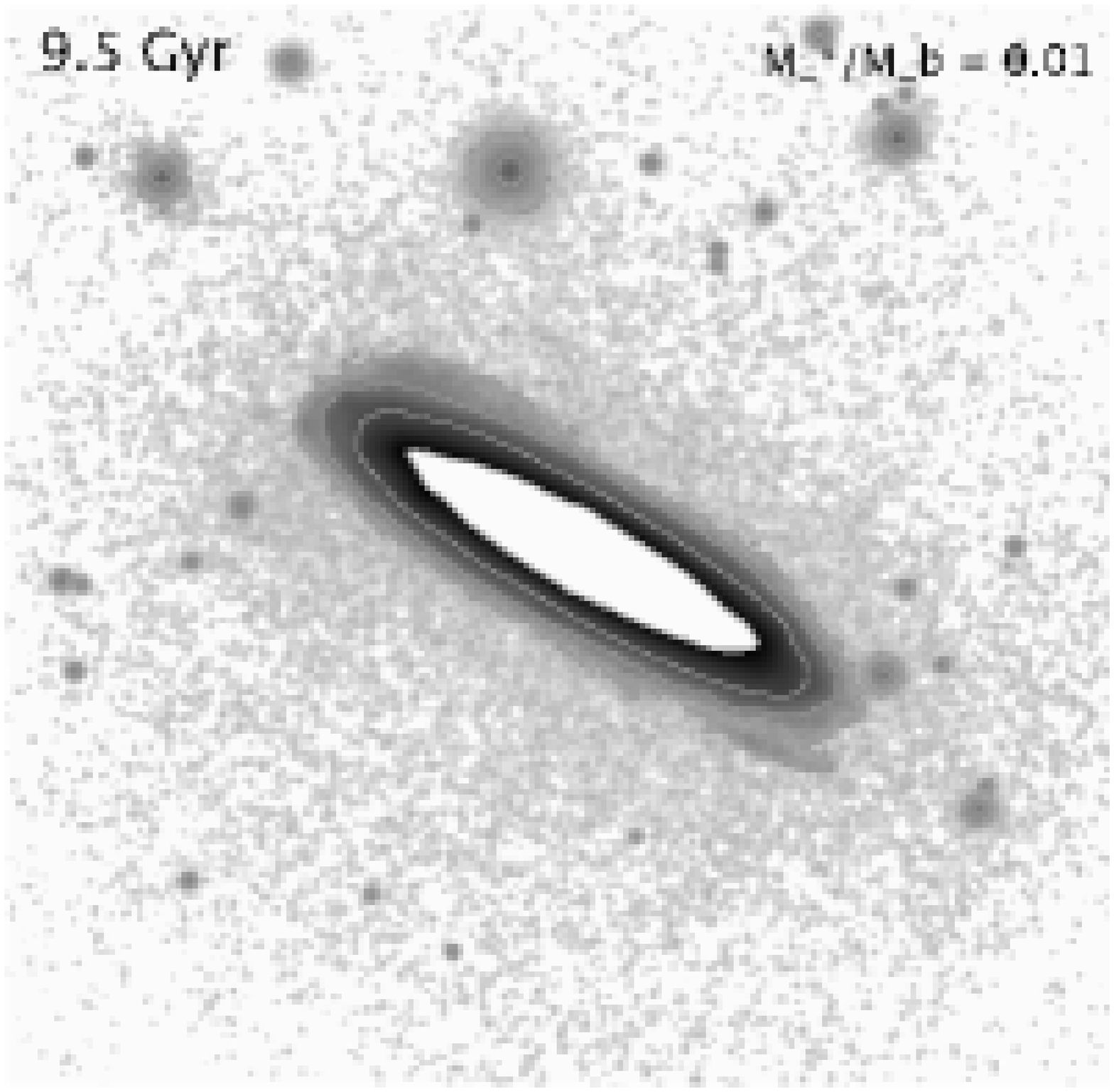}
    \hspace{0.1in}
    \includegraphics[angle=0,scale=.25]{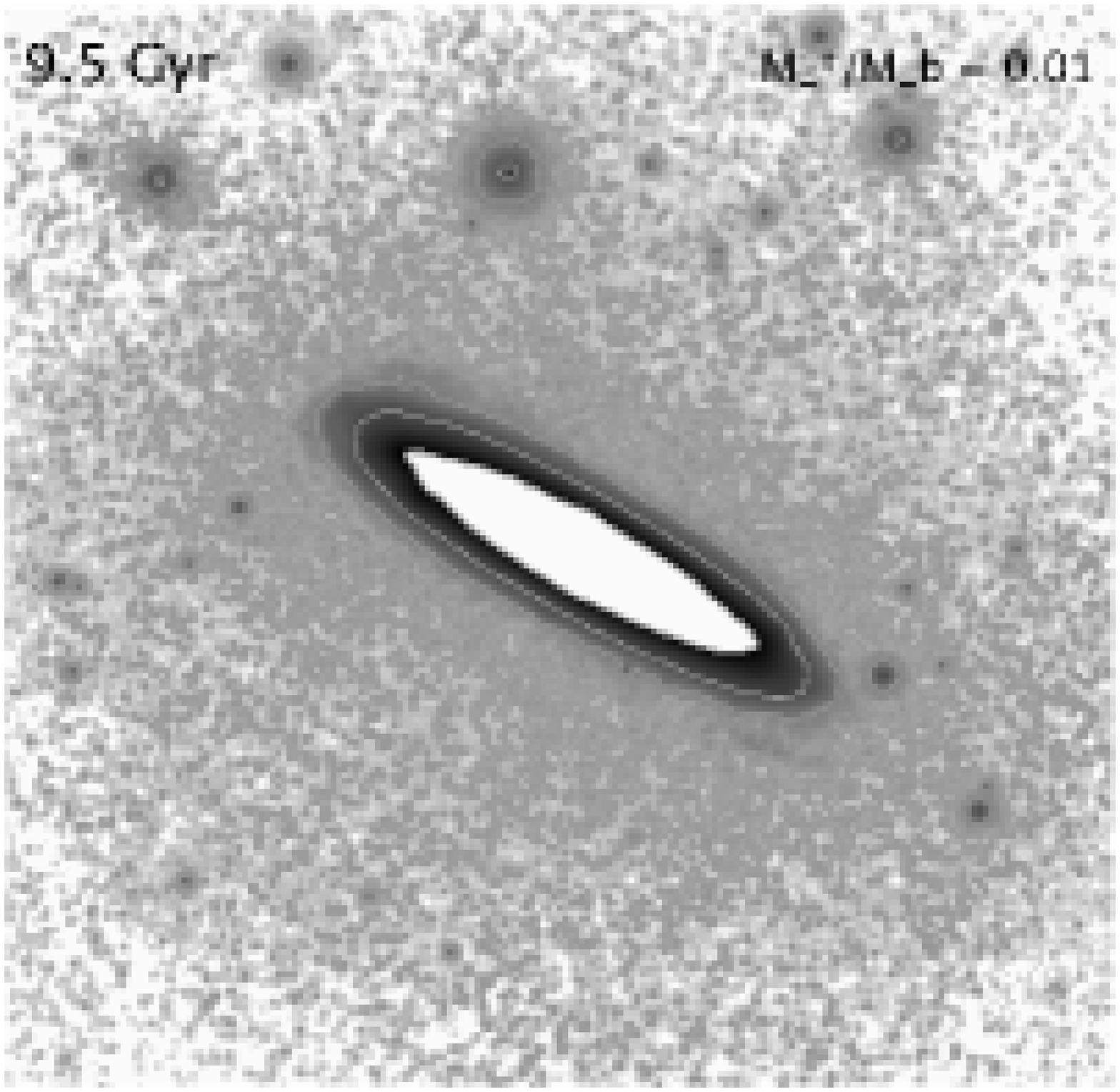}
    \hspace{0.1in}    
    \includegraphics[angle=0,scale=.25]{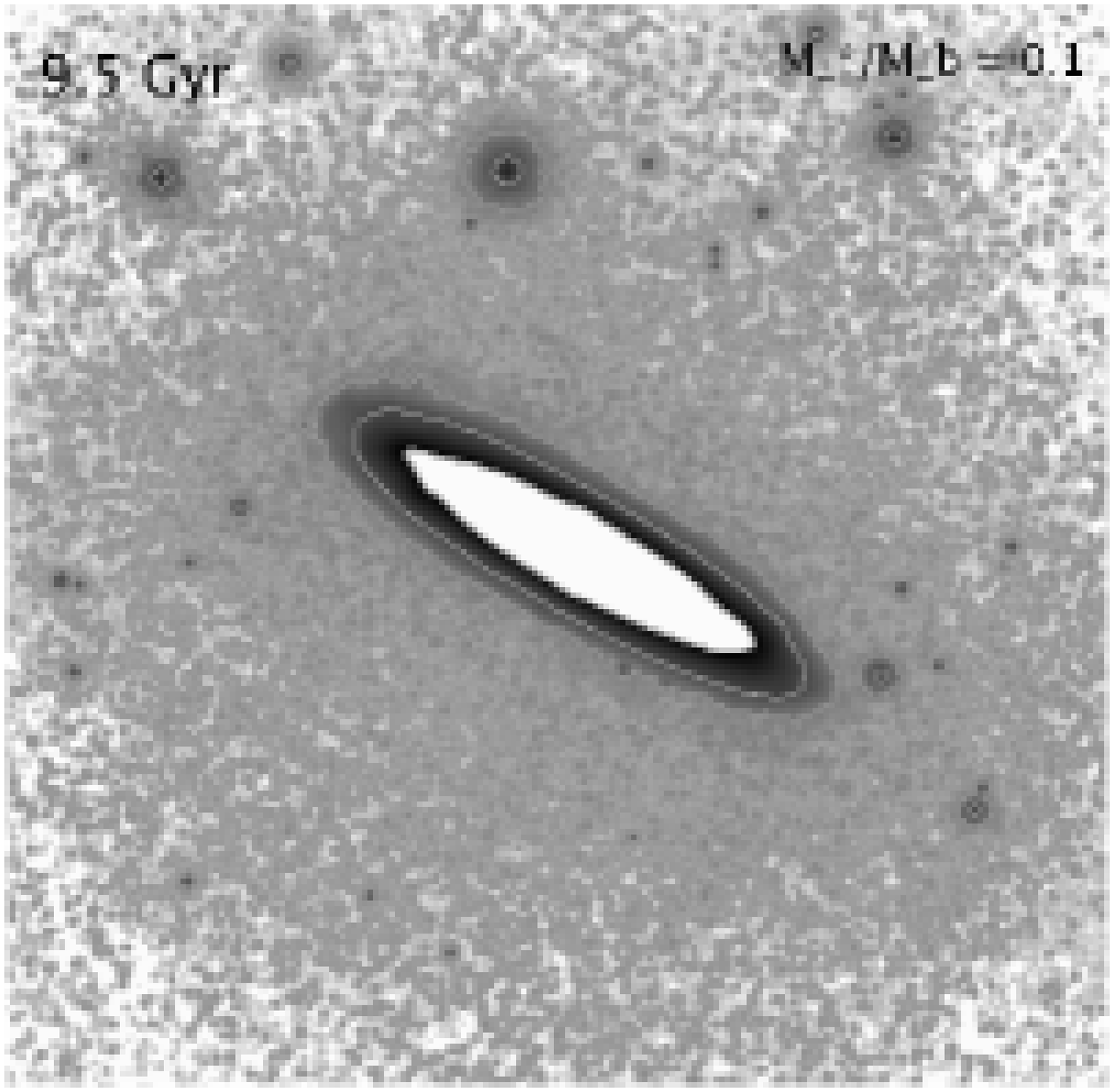}
    }
   }
 \vspace{7pt}
 \caption{B-band CCD Photometric maps at 2000 $\times$ 2000 pixel resolution showing a $10^{\circ}$ $\times$ $10^{\circ}$ field of the satellite stellar population taken at from top to bottom 3.5, 5.5 and 9.5 Gyr. For each map, we assume a simple stellar population formed at t=-2.5 Gyr. We show the contours corresponding to a surface brightness of 34.4, 31.9, 26.89, and 24.59 mag $\rm{arcsec}^{-2}$. From one figure to another, we vary the star formation efficiency i.e. how much gas is converted into stars. \emph{left column} : $M_{\rm stars}/M_{\rm baryons}$ = 0.01, \emph{middle} : $M_{\rm stars}/M_{\rm baryons}$ = 0.05 and \emph{right column} : $M_{\rm stars}/M_{\rm baryons}$ = 0.1. We assume a constant $M_{\rm baryons}/M_{\rm sat} = 0.171$ for all satellites. The white patch in the center of M31 is to avoid saturation of the chip and corresponds to a limiting isophote of 24.59 mag $\rm{arcsec}^{-2}$. (\emph{See high-resolution color version of these maps at http://www.cita.utoronto.ca/\~{}jgauthier/m31}) }
 \label{photometric_maps3}
\end{figure}

\clearpage

\begin{figure}

 \vspace{7pt}
  \centerline{\hbox{ \hspace{0.0in}
    \includegraphics[angle=0,scale=.4]{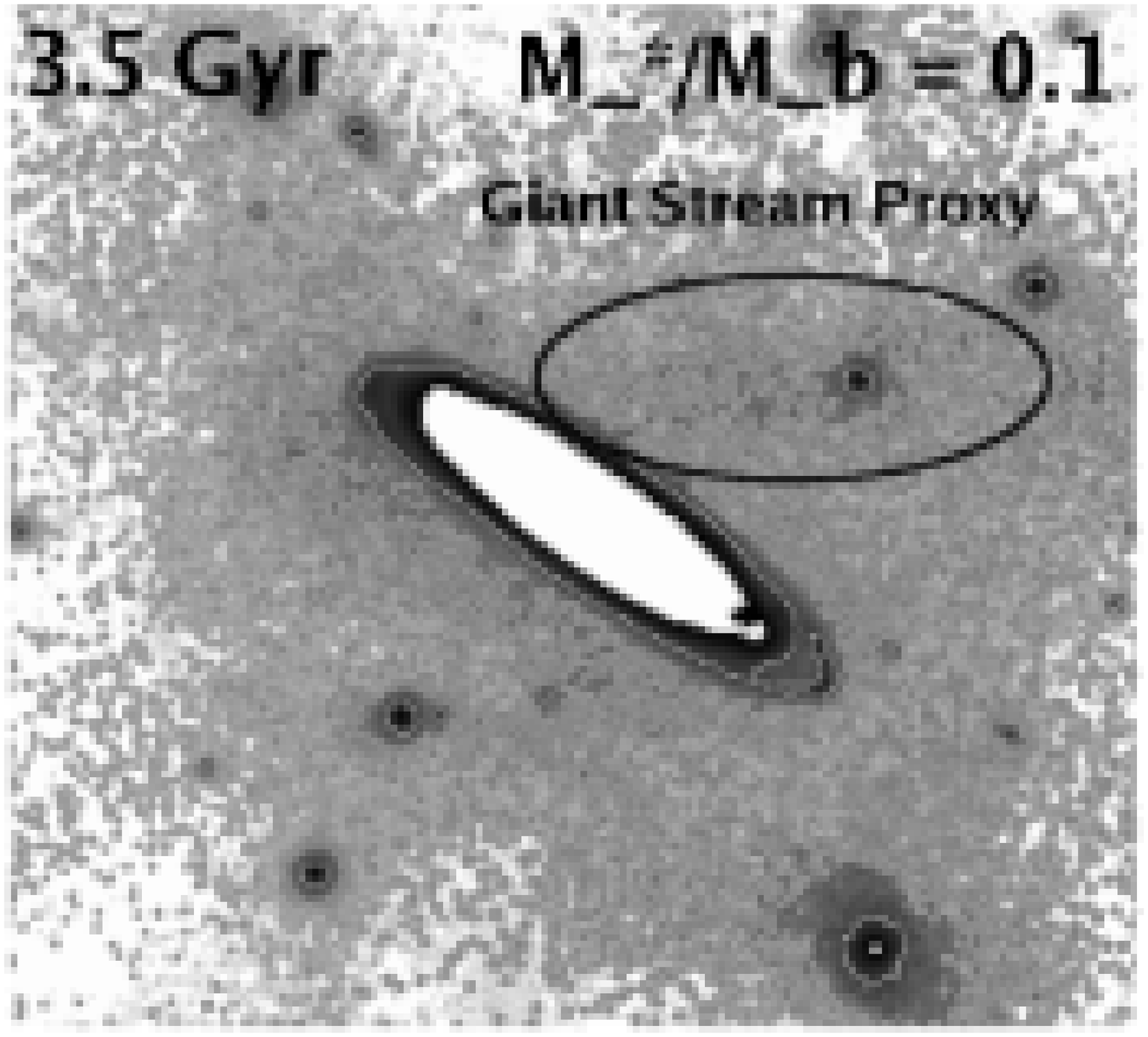}
    }
   }
  \vspace{7pt}
\centerline{\hbox{ \hspace{0.0in}
   \includegraphics[angle=0,scale=.4]{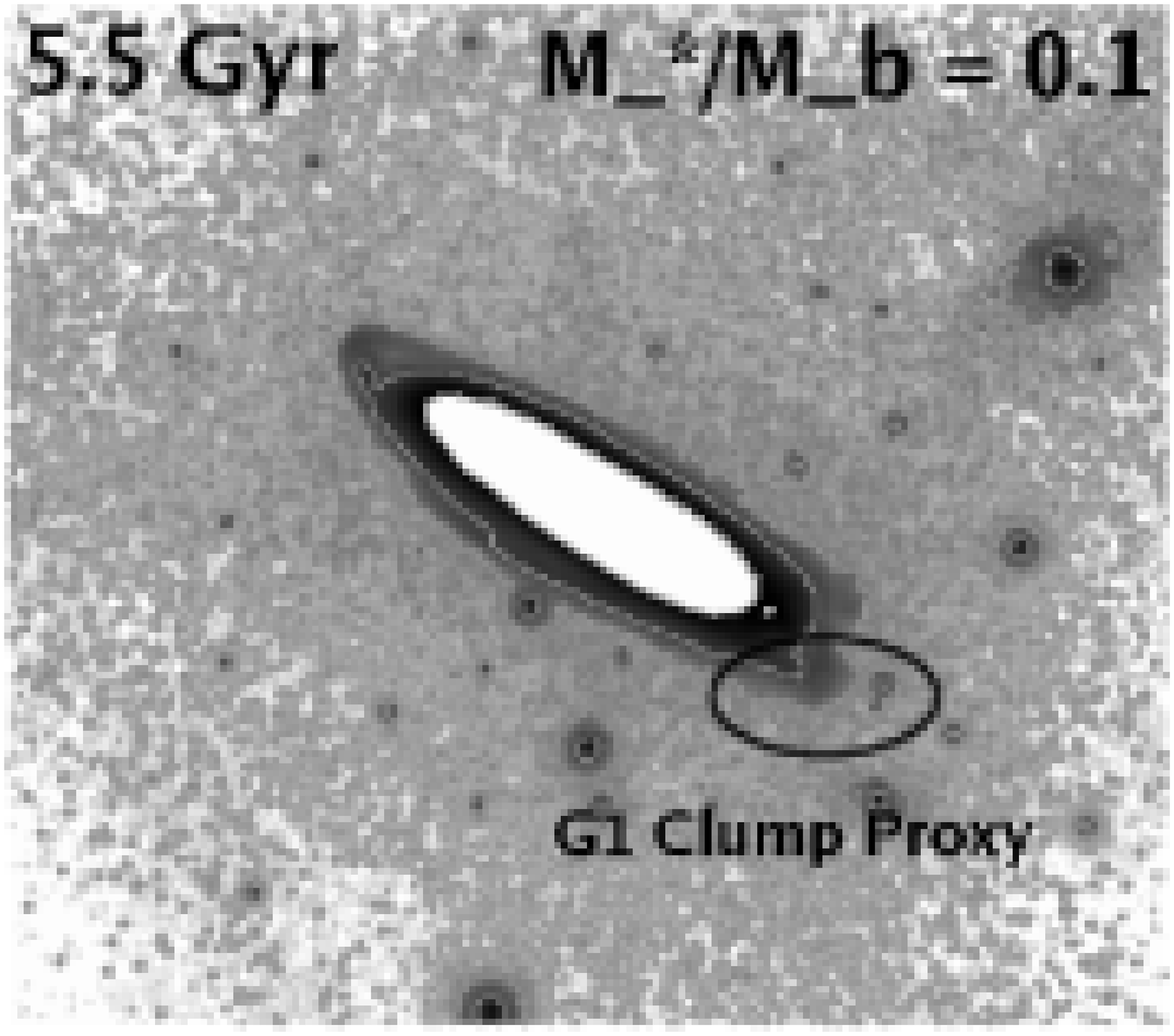}
    }
   }
 \vspace{7pt}
\caption{Close-up version of two Figure \ref{photometric_maps3} maps showing features resembling both morphologically and photometrically to the Giant Stream and G1 Clump observed 
around M31. The surface brightness of our Giant Stream is about 28.5 mag $\rm{arcsec}^{-2}$ in the B-band.(\emph{See high-resolution color version of these maps at http://www.cita.utoronto.ca/\~{}jgauthier/m31})}
\label{zoom-in}
\end{figure}

\end{document}